\documentclass[11pt]{article}

\usepackage[a4paper, total={6in, 10in}]{geometry}

\usepackage{preamble/packages}
\usepackage{preamble/commands}
\usepackage{preamble/additions_v1}
\usepackage{preamble/additions_v4}

\begin{document}

\begin{titlepage}
   \begin{center}
       \vspace*{2cm}
        \Huge
       \textbf{Small Defects at Large Charge}
       
       \vspace{0.3cm}
         \LARGE
       \vspace{0.2cm}
     
\begin{center}

{\bf
Jesse Woods
} \\
\vspace{2mm}

{\small{\it 
	Albert Einstein Center for Fundamental Physics,\\
	Institute for Theoretical Physics, University of Bern,\\
	Sidlerstrasse 5, CH-3012 Bern, Switzerland}
}
\vspace{2mm}
~\\
\texttt{jesse.woods@unibe.ch}\\
\vspace{1mm}
\end{center}
   \end{center}

   \vspace{1cm}
   \begin{abstract}
\baselineskip=14pt
We compute conformal data of cusped line defects with impurities for the abelian Higgs model in the semiclassical large-charge limit. By working perturbatively in the impurity charge $q$, we obtain the scaling dimension of the impurity to $\mathcal{O}(q^3)$ to all orders in the 't Hooft quartic and gauge couplings. Each term represents a resummation of infinitely many Feynman diagrams. Near the critical gauge coupling where conformality is lost, we resum the leading singular behaviour to all orders in $q$. We also comment on the spectrum of excited cusp states at $\mathcal{O}(q)$. Additionally, we compute the leading contributions to the Bremsstrahlung function, the coefficient of the leading irrelevant operator in the fusion limit, and argue that the Casimir energy receives no corrections semiclassically. Our results agree with known perturbative data and numerical solutions. We compare them with recent bounds on cusps. We comment on the interplay between the large rapidity limit and semiclassics.

\end{abstract}
\newpage

\vspace{5mm}
\end{titlepage}

\tableofcontents
\newpage

\section{Introduction}\label{sec:intro}
It is now quite established that the semiclassical regime of large charge is an effective tool for studying strongly coupled systems~\cite{Hellerman:2015nra,Gaume:2020bmp,Alvarez-Gaume:2016vff, Alvarez-Gaume:2019biu, Dondi:2021buw,Dondi:2022zna, Antipin:2022naw, Bersini:2025lxs}. Recently, this technique has been extended to gauge theory in~\cite{Antipin:2022hfe,Bersini:2026tpq}, for which the appropriate operators are non-local. The natural objects of study in this context are Wilson lines and their cousins. These can be interpreted as line defects~\cite{Kapustin:2005py,Drukker:2010jp}, which can host nontrivial impurities living as local operators on the line. Line operators have been much studied recently~\cite{Cuomo:2021rkm, Rodriguez-Gomez:2022gbz, Bianchi:2022sbz, Gimenez-Grau:2022czc, Aharony:2022ntz, Aharony:2023amq, Cuomo:2024psk,Iqbal:2025yop, Sun:2024qhv, Giombi:2025evu,DAlise:2025knv,Bersini:2026tpq,Maeda:2026tuw} as defects in conformal field theory~\cite{Cardy:1984bb,Cardy:1991tv,McAvity:1993ue,McAvity:1995zd,Gliozzi:2015qsa,Billo:2016cpy} (CFT) due to their relevance in describing impurities and boundaries~\cite{kondo1970theory,Wilson:1974mb,Affleck:1996mm,Billo:2013jda,Cuomo:2021cnb,Cuomo:2022xgw}. While much is known about local operators in CFTs, relatively little is known about extended operators. An interesting observable that can be computed in these contexts is the cusp anomalous dimension,
\begin{equation}\label{eq:cuspdef}
  \log\Braket{\littlecusp} = -\Gamma_{Q_1 Q_2}(\alpha_{*}) \log\frac{L}{a} + \dots,
\end{equation}
which measures the leading strength of the divergence produced by a cusp in a Wilson loop, where $L$ is the length of the line and $a$ is an ultraviolet regulator. The cusp opening angle is $\alpha_*$, with the straight line corresponding to $\alpha_*\to \pi$, and $Q_1$ and $Q_2$ represent the quantum numbers of the outgoing and incoming arms. This object has been studied in many contexts~\cite{Polyakov:1980ca,Korchemsky:1987wg,Korchemsky:1991zp,Makeenko:2006ds,Beisert:2006ez,Grozin:2015kna,Correa:2012hh,Bianchi:2018zpb,Bianchi:2019dlw,Bruser:2019auj,Henn:2019swt,Grozin:2022umo,Gromov:2015dfa,Cavaglia:2018lxi}. Contained in this quantity are several important observables - in particular, the scaling dimension of the impurity living at the cusp point, the Bremsstrahlung function at small deflection angles~\cite{Correa:2012at,Correa:2012hh}, and the Casimir fusion energy in the backtracking-line limit.

In this work, we will consider a line defect of the abelian Higgs model, or scalar quantum electrodynamics (QED), in the limit of large charge, following the precedent of~\cite{Bersini:2026tpq}. The abelian Higgs model possesses a non-trivial infrared CFT in $d=4-\epsilon$ dimensions. At criticality, it describes phase transitions in a variety of physical systems ranging from superconductors~\cite{Landau:1937obd,Herbut, Dasgupta:1981zz} to liquid crystals~\cite{Halperin:1973jh} and cosmic strings~\cite{Hindmarsh:2008dw}.

In the large-charge limit, $Q\to\infty$ with $e^2Q$ fixed, where $e$ is the gauge coupling, the path integral is dominated by a semiclassical saddle with quantum corrections suppressed by inverse powers of the charge. A natural way to access defect-changing operators at strong coupling in this regime is to perturb around the exactly solvable saddle configuration of equal charges. Writing the charges of the outgoing and incoming arms as $Q_1=(1+q)Q$ and $Q_2=Q$ and expanding in the small impurity charge $q$, the semiclassical problem becomes linear at each order and can be solved exactly in the gauge and quartic couplings. The result at each order is a closed form or an integral that resums infinitely many Feynman diagrams. Physically, the expansion in $q$ is an expansion in the number of charged quanta bound to the line. This configuration naturally describes the process in which a heavy charged object changes quantum numbers, possibly with a deflection of its worldline, or, for imaginary angles, with a boost. For example, processes like $\beta$-decay or electron capture by a heavy nucleus or ion of charge $Z$ experience a relatively small change of charge for the heavy particle, and are analogous to our setup. The cusp anomalous dimension controls the infrared suppression of such processes. Its role is analogous to that of the dimensions of boundary-condition-changing operators, which give the exponents of the Fermi-edge singularity~\cite{Affleck_1994}.

The large-charge limit in this context can be thought of as the limit of a highly charged nucleus at fixed $Za_S$, for $a_S$ the fine structure constant. Here, an expansion in $q$ is the analogue of a $1/Z$ expansion where the Coulomb field of the nucleus is treated exactly and the interactions between the bound particles perturbatively~\cite{furry1951bound,shabaev2002two}. This can be seen by examining the Feynman diagrams that are resummed by semiclassics: at $\mathcal{O}(q^n)$ we have up to $n$-body interactions, dressed by the line.

The atomic picture also illustrates the limits of this expansion.  At leading order in $q$, each quantum is a charged scalar in the Coulomb field of the line, and this Klein--Gordon Coulomb problem becomes supercritical at $e^2Q/4\pi=\frac12$, the scalar analogue of the critical charge of a nucleus~\cite{pomeranchuk1945energy,zeldovich1972electronic}. At this threshold, there are no stable orbits for the charged particle. From the point of view of the defect, the scaling dimension of the lowest charged operator on the line becomes complex, which signifies that conformality is lost and the line is screened~\cite{Aharony:2022ntz,Aharony:2023amq}. The impurity charge is the natural knob that drives the line towards this behaviour. Near criticality we can resum the singular behaviour to all orders in $q$, providing a semiclassical window around conformality loss.

These cusps are additionally a generalisation of dressed two-point correlation functions. Indeed, when one arm of the cusp is trivial, the observable is Mandelstam--Schwinger-dressed~\cite{Mandelstam:1962mi,Schwinger:1959xd,Schwinger:1962tn,Schwinger:1962tp}, which is one choice of dressing to render the correlation function gauge invariant. This has implications for the study of the superconducting phase transition~\cite{Kleinert_2003,Kleinert:2005sa,Herbut,Bersini:2026tpq,Antipin:2022hfe}. These dressings also lead to interesting infrared physics, like new memory effects~\cite{Araujo-Regado:2024dpr,Woods:softym}. Recently, such dressed correlators have also been studied in AdS~\cite{Ankur:2026ylr}.

There have been several recent bounds on the cusp anomalous dimension~\cite{Lanzetta:2026pvr,Lanzetta:2026ogn,Chandra:2026wnf,Cuomo:2026mop,Bianchi:2026dhj}. Many of these rely on the cusp being reflection symmetric, or on the two rays fusing to the identity. The first property holds only for equal charges, and in an abelian theory the latter also. Our results supply defect-changing data in an interacting, non-supersymmetric theory, exact in the coupling, which can be compared with the bounds, as well as provide input to the cusp bootstrap~\cite{Bianchi:2026dhj,Lanzetta:2026ogn}.

The outline of our results is the following. In Section \ref{sec:setup}, we establish the saddle point equations that we must solve in order to compute the anomalous dimension in the semiclassical limit. In Section \ref{sec:impurity}, by working in the straight line limit $\alpha_*\to \pi$, we compute the impurity dimension $\Delta_{1+q,1}$ through to $\mathcal{O}(q^3)$, exact in the quartic and gauge couplings. We show how the $\mathcal{O}(q)$ term resums an infinite series of Feynman diagrams with Catalan number coefficients. Also at this order, we determine the spectrum of excited states on the straight line exactly in the gauge coupling, and perturbatively at generic angles. Near the critical gauge coupling where conformality is lost, we resum the singular behaviour to all orders in $q$. In Section \ref{sec:bremandothers}, we additionally compute the Bremsstrahlung function to $\mathcal{O}(q^2)$ and show the Casimir energy is exact to this order in semiclassics. We also compute the dimension of the fused line to $\mathcal{O}(q^3)$ and identify the leading irrelevant operator on the fused line. Finally, in Section \ref{sec:discussion}, we review our results in light of recent literature. We find the results are in good agreement with numerics and with~\cite{Bersini:2026tpq} in the region of perturbative overlap as well as satisfy recent bounds on cusps. To close, we comment on the Lorentzian cusp and the large rapidity limit, where we see indications of the screening of the Maxwell saddle and a breakdown of semiclassics at a critical rapidity.

\section{Setup}\label{sec:setup}
Cusps with defect changing impurity operators correspond to the correlation function
\begin{equation}\label{eq:Zcusp}
  Z_{Q_1 Q_2}(\alpha_{*}) = \Braket{\littlecusp} = \braket{\O_{\substack{\yngrow{4}\\Q_1-Q_2}}^{*}(x_\infty) \,W_{\gamma_1}(Q_1) \, \O_{\substack{\yngrow{4}\\Q_1-Q_2}}(0) \, W_{\gamma_2}(-Q_2)} \,,
\end{equation}
where \(\gamma_1\) and \(\gamma_2\) are straight, outgoing Wilson lines from the origin to infinity, with relative cusp angle \(\alpha_{*}\). Concretely, we study this operator in the abelian Higgs model, describing the dynamics of \(N\) complex scalar fields with unit charge under a gauged \(U(1)\) symmetry. In Euclidean signature, the action is given by
\begin{equation}\label{eq:actionAH}
  S_{\text{AH}}[A, \phi] = \int d{x} \left[\frac{1}{4} F_{\mu\nu} F^{\mu\nu} + (D^{\mu} \phi_a)^{*} D_{\mu} \phi_{a} + \frac{(4 \pi)^2}{6}\lambda (\phi_a^{*} \phi_a)^2 \right] \, ,
\end{equation}
with \(D_{\mu} = \del_{\mu}{} + i e A_{\mu}\). It is well known that in \(d = 4 - \epsilon\) the model has a perturbative fixed point for \(\epsilon \to 0\), where both couplings \(\lambda\) and \(e^2\) are under perturbative control~\cite{Halperin:1973jh},
\begin{align}\label{eq:fixedpoint}
  \lambda^* \sim \frac{3\left( N +18 \pm \sqrt{(N-180) N-540}\right) }{4 N(N+4)}  \epsilon+\mathcal{O}(\epsilon^2),  \ \ \ \pqty{e^{*}}^2 \sim \frac{24 \pi ^2 }{N}  \epsilon  +\mathcal{O}(\epsilon^2).
\end{align}
At this order, the fixed points are real for $N \ge 183$, whereas modern calculations up to four loops reduce this critical value to around $N \approx 12$~\cite{Ihrig:2019kfv}. In the perturbative regime, we can identify the defect-changing operator with the symmetrised product
\begin{equation}\label{eq:operator}
  \cO_{\substack{\yngrow{4}\\Q_1-Q_2}}(x) = \phi_{( i_1} (x) \dots \phi_{ i_{Q_1-Q_2})}(x) \, .
\end{equation}
This is expected to be the lowest defect-changing operator. To perform the semiclassical calculation, we take  $\mathcal{O}(x) = \phi_1^{Q_1-Q_2}(x)$ and look for a classical profile where $\phi_{\ge2} = 0$. We drop the flavour index and refer to $\phi(x)$ henceforth.

Working in the double scaling limit, we define
\begin{align}\label{eq:doublescaling}
  q_1 &= \frac{Q_1}{Q} \, , & q_2 &= \frac{Q_2}{Q} \, , &   G &= Q e^2 , & \kappa &=  \frac{(4 \pi)^2}{6}Q \lambda\ ,
\end{align}
and take the limit $Q\to \infty$, $\epsilon\to0$ with $Q\epsilon$ fixed. In this regime, at leading order,
\begin{equation}\label{eq:kappaG}
\frac{\kappa}{G}=\frac{N+18\pm\sqrt{N^2-180N-540}}{12(N+4)},\qquad G=\frac{24\pi^2Q\epsilon}{N},
\end{equation}
and below we treat $G$ and $\kappa$ as independent, which is consistent at leading order in semiclassics where the running of couplings is suppressed. By rescaling the fields as
\begin{align}\label{eq:rescaling}
  \phi & \to \sqrt{Q} \phi, & A &\to \sqrt{Q} A \,, 
\end{align}
and absorbing the operator insertions into the action, the theory is known to be dominated by a semiclassical saddle~\cite{Bersini:2026tpq}. This is most conveniently expressed in terms of the gauge invariant fields $\rho$ and $B$, defined by
\begin{align}\label{eq:gaugeinv}
  \phi &= \frac{\rho}{\sqrt{2}}e^{i \chi}, & B =i\left( \sqrt{G} A + \dd{\chi}\right).
\end{align}
By Weyl rescaling around the cusp, flat space is mapped to the cylinder with radius $R$. The cusp anomalous dimension is then expressed in terms of the energy of the lines on the cylinder of length $T$ in the time direction, by the state operator correspondence. In this case, the scalar insertions are traded for boundary conditions and are absent from the action. We can thus write the anomalous dimension in terms of the rescaled action as
\begin{equation}\label{eq:action}
    \Gamma_{Q_1Q_2}(\alpha_*)=R E_{Q_1 Q_2}(\alpha_*)=  Q\frac{R\bar S_{\rm saddle}(q_1,q_2)}{T}+\mathcal{O}(Q^0)
\end{equation}
with
\begin{multline}\label{eq:actionsaddle}
    \frac{R\bar S_{\rm saddle}(q_1,q_2)}{T}=  \pi R^2  \int_0^{\pi }  \sin(\theta) \dd{\theta} \int_0^{\pi } \sin ^2(\alpha ) \dd{\alpha} \Bigg[\frac{1}{G}B\Laplacian_{S^3}B-\rho\Laplacian_{S^3}\rho \\
  +\left(1-(R B)^2 +\frac{\kappa}{2}(R\rho)^2\right)\rho ^2+ \left(\frac{q_1\,\delta(\alpha)}{2\pi R\sin^2\alpha}
-\frac{q_2\,\delta(\theta)\,\delta(\alpha-\alpha_*)}{\pi R\sin^2\alpha\,\sin\theta}\right)B\Bigg],
\end{multline}
where the subleading $\mathcal{O}(Q^0)$ term is the one-loop fluctuation determinant around the saddle. We focus only on the leading semiclassical term in this work. The saddle point equations are
\begin{gather}\label{eq:saddle} 
  \Laplacian_{S^3} \rho(\alpha, \theta) = \left(1-(R B(\alpha, \theta))^2+\kappa  (R \rho(\alpha, \theta)) ^2\right) \rho(\alpha, \theta)  , \\
  \label{eq:saddleB}\Laplacian_{S^3}B(\alpha, \theta) = G \left((R \rho(\alpha, \theta))^2 B(\alpha, \theta)  - \frac{q_1 \delta(\alpha)}{4 \pi R \sin ^2\alpha}+ \frac{q_2 \delta (\alpha-\alpha_* ) \delta(\theta)}{2 \pi R \sin ^2\alpha \sin\theta }\right),  \\
  \label{eq:saddlecharge}\int_0^\pi \sin(\theta) d{\theta}  \int_0^\pi \sin^2(\alpha) \dd{\alpha} \rho^2(\alpha, \theta) B(\alpha, \theta)  =\frac{q_1 - q_2 }{2 \pi R^3} \, ,
\end{gather}
where \(\Laplacian_{S^3}{}\) is the Laplacian on the three-sphere
\begin{equation}\label{eq:laplacian}
   \Laplacian_{S^3}{ } =  \tfrac{1}{\sin^2 \alpha}\del_\alpha\left(\sin^2 \alpha \del_\alpha{} \right) + \tfrac{1}{\sin^2 \alpha \sin \theta}\del_\theta\left(\sin \theta \del_\theta {} \right) + \tfrac{1}{\sin^2 \alpha \sin^2\theta}\del_\phi^2 {} \, .
\end{equation}
In~\cite{Bersini:2026tpq}, these semiclassical equations were solved exactly in the case of equal charges $q_1=q_2$, and perturbatively in $G$ for general charges. In the present work, we will consider the small defect charge limit, $q_1=1+q$, $q_2=1$, where $q\ll 1$ is treated as a perturbative parameter. To solve the saddle point equations, we may take series
\begin{equation}\label{eq:rhoexp}
    \rho=\sum_{i=0}^\infty q^{i/2}\rho^{(i/2)}(\alpha,\theta),
\end{equation}
and
\begin{equation}\label{eq:Bexp}
    B=\sum_{i=0}^\infty q^iB^{(i)}(\alpha,\theta).
\end{equation}
Immediately, we may use the equal charge solution of~\cite{Bersini:2026tpq} as a starting point for our perturbative analysis,
\begin{equation}\label{eq:B0}
    \rho^{(0)}=0,\quad B^{(0)}=\frac{G}{4 \pi ^2 R}\left((\pi -\alpha ) \cot (\alpha )-(\pi -\gamma ) \cot (\gamma )\right)+\frac{C_0(\alpha_*)}{R}, 
\end{equation}
where $\gamma$ is the geodesic angle
\begin{equation}
  \label{eq:geodesic-angle}
  \cos(\gamma) = \cos(\alpha) \cos(\alpha_{*}) + \sin(\alpha) \sin(\alpha_{*}) \cos(\theta),
\end{equation}
and $C_0(\alpha_*)$ is an integration constant to be determined. To compute the cusp anomalous dimension, we can write it as an expansion in $q$
\begin{equation}\label{eq:Gammaexp}
    \Gamma_{(1+q)Q,Q}(\alpha_*)=Q\sum_{n=0}^\infty q^n \Gamma^{(n)}_{1+q,1}(\alpha_*)+\mathcal{O}(Q^0),
\end{equation}
where
\begin{equation}\label{eq:Gammandef}
    \Gamma^{(n)}_{1+q,1}(\alpha_*)=\frac{R}{n!T}\frac{d^n \bar S_{\rm saddle}(q_1,q_2)}{dq^n}\bigg|_{q=0}.
\end{equation}
Importantly, as we are working around the classical saddle where variations in the field vanish, by the chain rule these derivatives reduce to only the explicit $q$-dependence. This is governed by the source on the first arm and so
\begin{equation}\label{eq:GnBreg}
    \Gamma^{(n)}_{1+q,1}(\alpha_*)=\frac{R}{n}\,B^{(n-1)}_{\rm reg}|_{\alpha=0}.
\end{equation}
Here, we define
\begin{equation}\label{eq:Breg}
    B_{\rm reg}|_{\alpha=0}=\lim_{\alpha\to0}\left[B(\alpha)-\frac{G}{4\pi}\frac{1+q}{R\alpha}\right],
\end{equation}
which subtracts its own Coulomb field, as in~\cite{Bersini:2026tpq}. Furthermore, by working at the saddle, it is sufficient to know the fields at order $k+1/2$ to compute $\Gamma^{(2k+1)}_{1+q,1}(\alpha_*)$. Indeed, truncating the fields at $\mathcal{O}(q^{k+1/2})$ yields the action correct up to terms of $\mathcal{O}(q^{2k+2})$. Therefore, we can express $B^{(n-1)}_{\rm reg}$ in terms of lower-order fields using the equations of motion to obtain the anomalous dimension.

\section{Scaling Dimension of the Impurity}\label{sec:impurity}
Let us begin by computing the scaling dimension $\Delta_{1+q,1}$ of the impurity, at leading order in semiclassics. This is given by the straight line limit of the cusp, $\Delta_{1+q,1}=\lim_{\alpha_*\to \pi}\Gamma_{1+q,1}(\alpha_*)$. We compute this quantity as an expansion in $q$ with $\Delta^{(n)}_{1+q,1}=\Gamma^{(n)}_{1+q,1}(\pi)$. At the leading order $\mathcal{O}(q^0)$, $\Gamma^{(0)}_{1+q,1}(\pi)$ vanishes by the normalisation convention. Thus, we begin our computation with the $\mathcal{O}(q)$ piece.

\subsection{$\mathcal{O}(q)$ Contribution}\label{sec:oq}
To determine $\Delta_{1+q,1}^{(1)}$, it suffices to know $B^{(0)}$, which requires determining the integration constant $C_0\equiv C_0(\alpha_*=\pi)$. Indeed, after the subtraction of the divergent piece near the endpoint, all that remains is this integration constant,
\begin{equation}\label{eq:G1C0}
    \Delta^{(1)}_{1+q,1}=C_0.
\end{equation}
This coefficient can be fixed at $\mathcal{O}(q)$ by determining $\rho^{(1/2)}$. For convenience, we define
\begin{equation}\label{eq:taunu}
   \tau=\frac{G}{4\pi},\quad  \nu=\frac{1}{2}+\frac{1}{2}\sqrt{1-4\tau^2},
\end{equation}
obeying the relation
\begin{equation}\label{eq:taunurel}
   \tau^2+\nu^2=\nu.
\end{equation}
To solve for $\rho^{(1/2)}$, the $\kappa$ term does not contribute, the straight line is $\theta$-independent, $\gamma=\pi-\alpha$, and so the matter equation is linear. It is convenient to work with $\varrho=\rho\sin\alpha$, in which case the matter equation reduces to 
\begin{equation}\label{eq:vrho}
    \varrho''+(\tau\cot\alpha+C_0)^2\varrho=0.
\end{equation}
This is a trigonometric Rosen--Morse equation, which has solutions in terms of hypergeometric functions
\begin{gather}\label{eq:rhohyper}
    \varrho=k_1\sin^{\nu}(\alpha) e^{-\tau\alpha}e^{i(\nu-C_0)\alpha}{}_2F_1(\nu-C_0,\nu+i\tau;1-C_0+i\tau;e^{2i\alpha})\\\label{eq:rhohyper2}+k_2\sin^{\nu}(\alpha)e^{\tau\alpha}e^{i(\nu+C_0)\alpha}{}_2F_1(\nu+C_0,\nu-i\tau;1+C_0-i\tau;e^{2i\alpha}).
\end{gather}
To fix the constants of integration, we need to impose reality, the integrated charge condition, positivity of $\rho$ and $q$, and the boundary conditions yielding finite energy of the solution near the arms which also coincide with the limit $G\to 0$ smoothly connecting to the global model~\cite{Badel:2019oxl}. These fix the 5 constants coming from the two complex $k_1$ and $k_2$ and the real $C_0$. This is done by examining the singular behaviour of the hypergeometric functions at the endpoints and tells us that we obtain a nontrivial solution with the desired boundary conditions near the arms only if the condition
\begin{equation}\label{eq:Cvals}
C_0=\pm\bigl(n+\nu\bigr),\qquad n\in\mathbb{N}
\end{equation}
is obeyed. The positive root is selected as we consider $q>0$. This also forces $k_2=0$ and leaves $k_1$ free. As $n$ generally induces nodes in the solution, we take $n=0$ in order to have $\rho\geq 0$, leaving $C_0=\nu$. In terms of $G$, this tells us
\begin{equation}\label{eq:Delta1}
    \boxed{
    \Delta_{1+q,1}^{(1)}=\frac{1}{2}+\frac{1}{2}\sqrt{1-\frac{G^2}{4\pi ^2}}.
}
\end{equation}
This is the exponent of the stable defect fixed point of \cite{Aharony:2022ntz,Aharony:2023amq}. The boundary condition is a choice of defect.
While simple, this result encodes a resummation of infinitely many Feynman diagrams. In fact, this formula is precisely the form of the generating function for the Catalan numbers,
\begin{equation}\label{eq:catalan}
    c_m=\frac{(2m)!}{m!(m+1)!},
\end{equation}
and so
\begin{equation}
\Delta_{1+q,1}^{(1)}=1-\sum_{m\ge1}c_{m-1}\Big(\frac{G}{4\pi}\Big)^{2m}=1-\frac{G^2}{16\pi^2}-\frac{G^4}{256\pi^4}-\frac{2G^6}{4096\pi^6}-\frac{5G^8}{65536\pi^8}-\dots.    
\end{equation}
We can see how these combinatorial numbers occur by considering how the Feynman diagrams arise perturbatively. In terms of the original couplings, the terms in the expansion correspond to diagrams with couplings $(Q_1-Q_2)(e^2Q)^{2m}$. A photon ending on a Wilson line contributes a factor $eQ$. At $\mathcal{O}(q)$ there is a single scalar particle interacting with the Wilson line, so there are no factors of $\kappa$ coming from exchange, nor photon exchange between scalar lines.  The relevant diagrams are thus one scalar line with any number of photons to the Wilson line. We can express this as a quantum mechanics problem, and explicitly compute matrix elements to determine the perturbative contributions. 

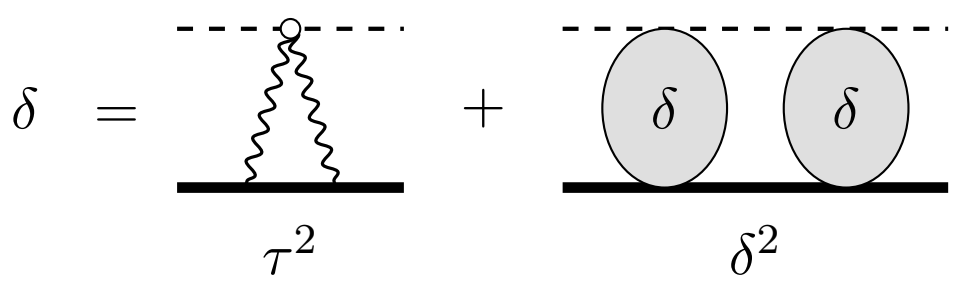
\begin{figure}[H]
{\linespread{1}\selectfont
\begin{center}
\begin{tikzpicture}[scale=0.75]
  \node at (-0.9,0.7) {$\delta\ \ =$};
  \draw[dgW] (0,0) -- (2,0); \draw[dgS] (0,1.4) -- (2,1.4);
  \draw[dgP] (1,1.4) -- (0.6,0); \draw[dgP] (1,1.4) -- (1.4,0); \node[dgSg] at (1,1.4) {};
  \node at (1,-0.55) {$\tau^2$};
  \node at (2.7,0.7) {$+$};
  \begin{scope}[xshift=3.4cm]
  \draw[dgW] (0,0) -- (3.4,0); \draw[dgS] (0,1.4) -- (3.4,1.4);
  \draw[fill=gray!25] (0.9,0.7) ellipse (0.55 and 0.7); \node at (0.9,0.7) {$\delta$};
  \draw[fill=gray!25] (2.5,0.7) ellipse (0.55 and 0.7); \node at (2.5,0.7) {$\delta$};
  \node at (1.7,-0.55) {$\delta^2$};
  \end{scope}
\end{tikzpicture}
\end{center}
\par}
    \caption{Recursive diagrams, whose nestings are counted by Catalan numbers.}
    \label{fig:catalandiag}
\end{figure}

These exchanges come in two types: single exchanges, and seagull exchanges. The single exchanges are cubic vertices coming from the derivative coupling of the gauge field to the current $A_\mu (\phi^*\del^\mu\phi-\phi\del^\mu\phi^*)$. The seagull vertex is tetravalent coming from the term $e^2A^2|\phi|^2$ in the action.
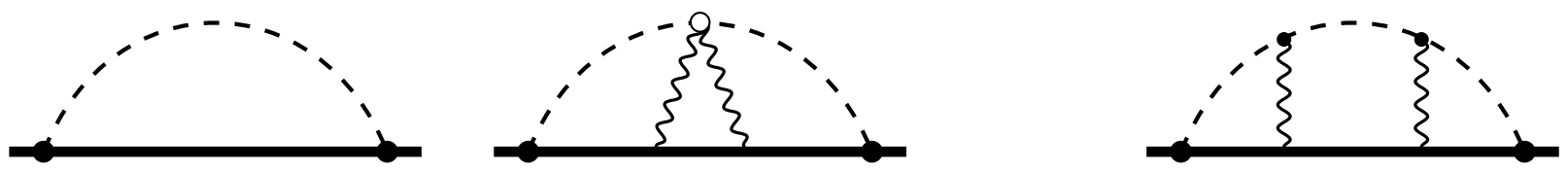
\begin{figure}
    {\linespread{1}\selectfont
\begin{center}
\begin{tabular}{ccc}
\begin{tikzpicture}[scale=0.78]\dgOne\end{tikzpicture} &
\begin{tikzpicture}[scale=0.78]\dgOne \draw[dgP] (1.5,1.13) -- (1.1,0); \draw[dgP] (1.5,1.13) -- (1.9,0); \node[dgSg] at (1.5,1.13) {};\end{tikzpicture} &
\begin{tikzpicture}[scale=0.78]\dgOne \draw[dgP] (0.9,0.98) -- (0.9,0); \node[dgVx] at (0.9,0.98) {}; \draw[dgP] (2.1,0.98) -- (2.1,0); \node[dgVx] at (2.1,0.98) {};\end{tikzpicture}\\
$\tau^0$: free, $1$ & $\tau^2$: seagull, $-\tau^2$ & $\tau^2$: two single photons, $-\tau^2C_0^2$
\end{tabular}
\end{center}
\par}
    \caption{Leading photon exchange diagrams and their contributions to $C_0^2$ on the cylinder.}
    \label{fig:Gamma1leading}
\end{figure}
Examining our equation \eqref{eq:vrho} for $\varrho$, we have an eigenvalue problem for $C_0$ that one can solve perturbatively in $\tau$,
\begin{equation}\label{eq:QM}
    \Big[\underbrace{-\partial_\alpha^2}_{\text{free}}\ \underbrace{-\,2\tau C_0\cot\alpha}_{\text{single photon}}\ \underbrace{-\,\tau^2\cot^2\alpha}_{\text{seagull}}\Big]\varrho=C_0^2\,\varrho,\qquad \varrho(0)=\varrho(\pi)=0 .
\end{equation}
This is simple quantum mechanics. 
One can compute the corrections to the eigenvalue using matrix elements via Rayleigh-Schr\"odinger perturbation theory, treating the $C_0$ in the vertex as a fixed parameter,
\begin{multline}\label{eq:RS}
C_0^2=1+V_{11}+{\sum_{n}}'\frac{|V_{1n}|^2}{1-n^2}+{\sum_{n,m}}'\frac{V_{1n}V_{nm}V_{m1}}{(1-n^2)(1-m^2)}\\+{\sum_{n,m,l}}'\frac{V_{1n}V_{nm}V_{ml}V_{l1}}{(1-n^2)(1-m^2)(1-l^2)}
-\Big(V_{11}+{\sum_{n}}'\frac{|V_{1n}|^2}{1-n^2}\Big){\sum_{n}}'\frac{|V_{1n}|^2}{(1-n^2)^2}+\mathcal{O}(\tau^6),
\end{multline}
\[
V_{nm}=-2\tau C_0\,\langle n|\cot\alpha|m\rangle-\tau^2\langle n|\cot^2\!\alpha|m\rangle,
\]
where the primes denote $n,m,l\neq1$. Finally, $C_0$ is fixed self-consistently.
At leading order $\tau^0$, one has a particle in a box with eigenstates $\ket{n}=\sqrt{2/\pi}\sin n\alpha$. By construction, $\ket{1}$ is precisely $\varrho$ at $\tau=0$. The required matrix elements for the perturbative corrections are
\begin{equation}\label{eq:cot2me}
\langle n|\cot^2\!\alpha|m\rangle=
\begin{cases}
     2\min(n,m)-\delta_{nm}, \quad \text{$n+m$ even},\\
     0 \quad \text{otherwise},
\end{cases}
\end{equation}
and
\begin{equation}\label{eq:cotme}
\langle n|\cot\alpha|m\rangle=
\begin{cases}
    \frac2\pi\sum_{\substack{k\ \text{even}\\|n-m|<k<n+m}}\frac{2k}{k^2-1}, \quad \text{$n+m$ odd},\\
       0 \quad \text{otherwise}.
\end{cases}
\end{equation}
The first order energy shift $\braket{1|V|1}$ thus comes from the seagull diagram, $-\tau^2\braket{1|\cot^2\alpha|1}=-\tau^2$. The other contributions at $\mathcal{O}(\tau^2)$ come from matrix elements $|\braket{1|V|n}|^2$ and by considering just the single photon exchanges which are of order $\tau$, yielding a $\tau^2$ squared-matrix element. The shift is thus
\begin{equation}\label{eq:tau2shift}
    \sum_{n\geq 2}\frac{4\tau^2C_0^2|\braket{1|\cot\alpha|n}|^2}{1-n^2}=-\frac{64\tau^2 C_0^2}{\pi^2}\sum_{n\in 2\mathbb{N}}\frac{n^2}{(n^2-1)^3}=-\tau^2C_0^2.
\end{equation}
This yields the relation $C_0^2=1-\tau^2-\tau^2C_0^2$ and so
\begin{equation}\label{eq:C0tau2}
    C_0=1-\tau^2+\mathcal{O}(\tau^4).
\end{equation}
This is the first Catalan term. At order $\tau^4$, we have four types of diagrams to consider, as seen in Figure \ref{fig:tau4}.
\begin{figure}
{\linespread{1}\selectfont
\begin{center}
\begin{tabular}{cccc}
\begin{tikzpicture}[scale=0.66]\dgOne \draw[dgP] (0.9,0.98) -- (0.6,0); \draw[dgP] (0.9,0.98) -- (1.2,0); \node[dgSg] at (0.9,0.98) {}; \draw[dgP] (2.1,0.98) -- (1.8,0); \draw[dgP] (2.1,0.98) -- (2.4,0); \node[dgSg] at (2.1,0.98) {};\end{tikzpicture} &
\begin{tikzpicture}[scale=0.66]\dgOne \draw[dgP] (0.6,0.82) -- (0.6,0); \node[dgVx] at (0.6,0.82) {}; \draw[dgP] (1.5,1.13) -- (1.2,0); \draw[dgP] (1.5,1.13) -- (1.8,0); \node[dgSg] at (1.5,1.13) {}; \draw[dgP] (2.4,0.82) -- (2.4,0); \node[dgVx] at (2.4,0.82) {};\end{tikzpicture} &
\begin{tikzpicture}[scale=0.66]\dgOne \foreach \x/\y in {0.5/0.75,1.15/1.05,1.85/1.05,2.5/0.75}{\draw[dgP] (\x,\y) -- (\x,0); \node[dgVx] at (\x,\y) {};}\end{tikzpicture} &
\begin{tikzpicture}[scale=0.66]\dgOne \draw[dgP] (0.9,0.98) -- (0.9,0); \node[dgVx] at (0.9,0.98) {}; \draw[dgP] (2.1,0.98) -- (2.1,0); \node[dgVx] at (2.1,0.98) {}; \node at (0.9,1.35) {\scriptsize $\times$};\end{tikzpicture}\\
two seagulls: $-\tau^4$ & seagull $+$ two singles: $-2\tau^4$ & four singles: $0$ & feedback: $+2\tau^4$
\end{tabular}
\end{center}
\par}
    \caption{$\mathcal{O}(\tau^4)$ diagrams contributing to $C_0^2$. Each value collects the terms of \eqref{eq:RS} with those vertices, including the renormalisation terms with squared denominators. The feedback diagram is from the shift of $C_0$ in the numerator.}
    \label{fig:tau4}
\end{figure}
The new ingredient is the feedback diagram, which is the same $\tau^2$ diagram, but with $C_0$ corrected by its $\mathcal{O}(\tau^2)$ correction. Overall this gives
\begin{equation}\label{eq:C0tau4}
        C_0=1-\tau^2-\tau^4+\mathcal{O}(\tau^6),
\end{equation}
which contains the next (albeit trivial) Catalan number. To understand the counting, it is advantageous to use conformal invariance and compute the exponent perturbatively for the sake of argument. We let $\delta=1-\nu$ be the shift in the dimension of the lowest charged defect operator. Near the line, we know $\varrho\sim \alpha^{\nu}$. The potential is dominated by the seagull term, which is most singular. By expanding 
\begin{equation}\label{eq:frobenius}
    \varrho=\alpha^\nu(1+c_1\alpha+c_2\alpha^2+\dots),
\end{equation}
the leading singular behaviour of the ODE tells us that the exponent obeys
\begin{equation}\label{eq:indicial}
    \nu(\nu-1)=-\tau^2.
\end{equation}
Here, only the more singular seagull term contributes to the scaling dimension, while the single photons modify the shape of the wavefunction near the line via the $c_i$ attached to the subleading powers of $\alpha$. In terms of the deficit, this becomes
\begin{equation}\label{eq:deficit}
    \delta=\tau^2+\delta^2.
\end{equation}
This equation can be expressed diagrammatically in Figure \ref{fig:catalandiag}.
By iteratively solving this relation, $\delta$ is replaced by either a seagull diagram yielding a power of $\tau^2$ or another pair of $\delta$s. At each order $\tau^{2m}$, we have a planar binary tree of $m$ seagull leaves of which there are $c_{m-1}$. The Catalan numbers count this nesting. This tells us the exponent of $\varrho$, which by conformal invariance tells us the energy. While it appears that we have only used the seagulls to compute the energy, neglecting the single photon diagrams, the truth is that these single photon diagrams are what imposes conformal invariance. Without the single photon diagrams, the equation becomes the P\"oschl--Teller problem
\begin{equation}
    -\varrho''-\frac{\tau^2}{\sin^2\alpha}\,\varrho=\big(C_0^2-\tau^2\big)\varrho,
\end{equation}
which has ground state $\sin^\nu\alpha$ (compared to $\sin^\nu\alpha e^{-\tau\alpha}$), which has the same near-arm exponent as before. But, the eigenvalue is now $C_0^2-\tau^2=\nu^2$, which yields $C_0=\sqrt{\nu}$, not $\nu$. So, while the exponent is unchanged, the energy is not equal to the exponent - without the single photon coupling, the equation is neither gauge nor conformally invariant, and thus the state operator correspondence cannot be applied. This division between seagulls and single photons holds in the Feynman gauge on the cylinder, and is gauge-dependent.

\subsection{Excited States}\label{sec:excited}

We can allow angular dependence around the straight cusp. This will lead to states of higher energy, which include both descendants and new primaries. Doing so results in a centrifugal term in the Rosen--Morse equation coming from the spherical harmonics, which can be solved analogously. The spectrum is, at leading order in $q$ and exact in $G$,
\begin{equation}\label{eq:Cnl}
    C_{nl}=n+\frac{1}{2}+\sqrt{\left(l+\frac{1}{2}\right)^2-\tau^2}\equiv \Delta_{l}+n.
\end{equation}
For $l=0$, these are the constants $C_n=n+\nu$ from \eqref{eq:Cvals} above. Here, $n$ labels the $SL(2,\mathbb R)$ descendants from derivatives along the line, which have spacing $1$ in units of $\frac{1}{R}$, and $\Delta_l$ is precisely the dimension of the defect operator of the charged scalar on the Wilson line observed in~\cite{Aharony:2022ntz,Aharony:2023amq}. These operators appear in the defect OPE of the bulk field near the line
\begin{equation}\label{eq:defectOPE}
    \rho(r,\alpha,\Omega)=\sum_{l,m}(r\alpha)^{\Delta_l-1}Y_{lm}(\Omega)\hat{\rho}_{lm}(r)+\dots.
\end{equation}
The level $C_{nl}$ is the descendant $\del_r^n\hat{\rho}_{lm}$ along the line. At $G=0$, the primaries are identified with the traceless transverse derivatives $\hat{\rho}_{lm}=\del_{(\perp i_1}\dots\del_{\perp i_l)}\rho$. A single excited quantum on top of the saddle has energy $C_{nl}-C_0$.

When introducing a small deflection to the straight line configuration, there remains only an $SO(2)$ rotation symmetry of the plane orthogonal to the cusp. In this case, at $G=0$ coupling, the scalar is free and decomposes into harmonics with levels precisely $C_{N}=N+1$, with $(N+1)^2$ degenerate states labelled by $m=-N,\dots,N$ with multiplicities $N+1-|m|$ from the $SO(4)$ symmetry. Introducing a small coupling, on the straight line, $SO(3)$ splits level $N$ into levels $l=0,\dots,N$ with $2l+1$ states in each. For the bent line, only an $SO(2)$ subgroup remains leaving for $l\geq|m|$ just one state of charge $m$. Using degenerate perturbation theory, one obtains at $\mathcal{O}(G)$ for the $N=1$ levels
\begin{equation}\label{eq:CN1}
    C_{m=0}=2\mp \frac{G}{4\pi^2}\sin\alpha_*, \quad C_{m=\pm1}=2.
\end{equation}
These correspond to modes leaning towards the arms, away from the arms, and perpendicular to the arms respectively.  More generally, at order $\mathcal{O}(q)$, the splitting of the $N$th level takes the form
\begin{equation}\label{eq:CNm}
    C_{Nm}=N+1-\frac{G}{4\pi^2}\sigma_{Nm}+\mathcal{O}(G^2).
\end{equation}
The eigenvalues come in pairs due to reflections of the cusp.  The first few values can be seen in Table \ref{tab:descsplit}. Each entry with $|m|\ge1$ is doubly degenerate ($\pm m$) due to reflections. 

For the straight line, the splitting enters only at $\mathcal{O}(G^2)$ from $\Delta_1-\Delta_0-1=\frac{2}{3}\tau^2+\dots$. Bending lifts the degeneracy at $\mathcal{O}(\varphi G)$, for a deflection angle $\varphi=\pi-\alpha_*$, by inducing a change in $B^{(0)}$ by $\varphi\frac{\tau}{\pi}\left(\cot\alpha-\frac{\alpha}{\sin^2\alpha}\right)\cos\theta$, which is the potential of the bent arm minus that of the straight arm. This corresponds to the insertion of the displacement operator
\begin{equation}\label{eq:displacement}
    \mathcal{D}_i=ieQ_2 F_{\parallel i},
\end{equation}
which deforms Wilson lines $W=\exp\left(ie Q_2\int_{\gamma(s)} A\right)$ in the transverse directions $\delta x_\perp=\varphi s$ by $\delta W=-\varphi\int_{\gamma}ds\,s\,\mathcal{D}_{\perp}(s)\cdot W$. Under $SO(3)$, $\mathcal{D}_\perp$ transforms as an $l=1,m=0$ operator, so near the straight line it mixes between modes $m'=m$ and $l'=l\pm 1$. Interactions which enter at $\mathcal{O}(q^2)$ further modify these splittings.

\begin{table}[H]
    \begin{center}
\renewcommand{\arraystretch}{1.6}
\begin{tabular}{c|c|l}
$N$ & $|m|$ & $\sigma_{N|m|}$\\ \hline
$1$ & $1$ & $0$\\
    & $0$ & $\pm\sin\alpha_*$\\ \hline
$2$ & $2$ & $0$\\
    & $1$ & $\pm\frac12\sin\alpha_*$\\
    & $0$ & $0,\ \ \pm\sin\alpha_*\sqrt{2+\cos^2\alpha_*}$\\ \hline
$3$ & $3$ & $0$\\
    & $2$ & $\pm\frac13\sin\alpha_*$\\
    & $1$ & $0,\ \ \pm\frac13\sin\alpha_*\sqrt{6+\cos^2\alpha_*}$\\
     & $0$ & $\pm\frac13\sin\alpha_*\Big(\sqrt{4\cos^4\alpha_*+4\cos^2\alpha_*+10}\pm\sqrt{4\cos^4\alpha_*+\cos^2\alpha_*+4}\Big)$
\end{tabular}
\end{center}
    \caption{Leading order in $G$ splitting of the levels for the first few $N$.}
    \label{tab:descsplit}
\end{table}

\begin{figure}
   \begin{center}
\includegraphics[width=\textwidth]{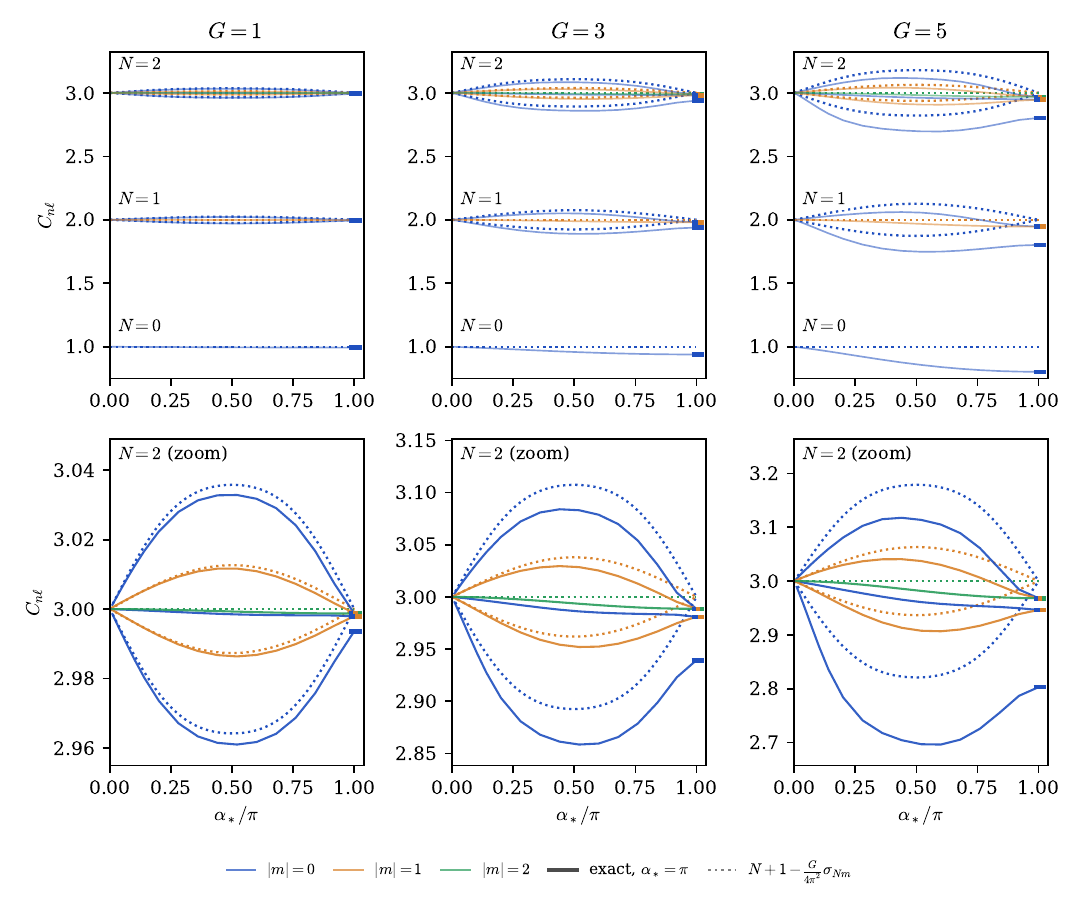}
\end{center}
    \caption{Numerical level splitting of the lowest levels at $\mathcal{O}(q)$. The  ticks at $\alpha_*/\pi=1$ are the exact values from the straight line at all orders in $G$. The dotted line is the leading order in $G$ correction at each level.}
    \label{fig:excited}
\end{figure}

In Figure \ref{fig:excited} we see numerically the level splitting for arbitrary $G$ as a function of cusp angle compared to the leading $\mathcal{O}(G)$ analytic result along with markings for the exact results for the straight line, at leading order in $q$.
For the straight line, the numerics coincide exactly with the analytically derived $C_{nl}$. For small cusp angles, we see each level approach the value $N+1$. This coincides with the fusion limit for which the cloud should see no net source. We see that even for moderate $G$, our small $G$ formula still agrees reasonably well with the numerics - indeed for $N=1$ at $\alpha_*=\pi/2$, our formula gives the gap between the two $m=0$ levels to within $0.6\%$ for $G=1$ and about $18\%$ for $G=5$.

The numerical curves sit slightly lower than the analytic formulas. Majority of this shift is explainable by the next order $-\tau^2/(2l+1)$ shift which is already present in the exact result for the straight line, for which we see excellent agreement.

We can observe also that the splitting is not symmetric numerically, which is a higher order effect in $G$. The equal level spacing is also lost for the bent line. This was anticipated on general grounds in~\cite{Bianchi:2026dhj}. There is also no level crossing at any $0<\alpha_*<\pi$ at $\mathcal{O}(q)$ for the checked values of $N\leq 2$. Levels with equal $m$ are generally not expected to cross. What is more interesting is that we do not observe crossings of levels with different $m$ numerically.

In~\cite{Bersini:2026tpq}, it was observed that for generic charges the spectrum of fluctuations around the semiclassical saddle is massive, in contrast to the model with only global symmetries. The usual type-I Goldstone, whose spectrum would contain the conformal descendants, gains a mass via the Higgs mechanism. Here, the condensate gives a mass to the photon proportional to $Gq$, which should shift the energy levels for the bent line at the next order in $q$.  On the straight line, the $SL(2,\mathbb R)$ of the line is unbroken and its descendants appear exactly, as $C_{nl}=\Delta_l+n$, and are protected. Away from the straight line there is no translation along the line, the levels are no longer integer-spaced~\cite{Bianchi:2026dhj}, and, as in~\cite{Bersini:2026tpq}, no unit-energy descendants are expected.

\subsection{$\mathcal{O}(q^2)$ Contribution}\label{sec:oq2}
To determine the next correction to the scaling dimension, we are required to fix the normalisation of $\rho^{(1/2)}$
\begin{equation}\label{eq:rhohalf}
    \rho^{(1/2)}=k_1\sin^{\nu-1}(\alpha)\,e^{-\tau\alpha},
\end{equation}
in order to determine the $\mathcal{O}(q)$ correction to $C_0$, which we denote $C_0^{(1)}$, and hence $B^{(1)}$ near the arm of the cusp.

One fixes $k_1\in \mathbb{R}$ by using the charge condition, which boils down to a master integral of the form 
\begin{equation}\label{eq:masterint}
    \int_0^\pi \sin^\sigma(\alpha)e^{\beta\alpha}d\alpha=\frac{\pi e^{\beta\pi/2}\Gamma(\sigma+1)}{2^\sigma\Gamma\left(1+\frac{\sigma+i\beta}{2}\right)\Gamma\left(1+\frac{\sigma-i\beta}{2}\right)},
\end{equation}
which yields
            \begin{equation}\label{eq:C1}
    k_1=\frac{2^{\nu }e^{\pi\tau/2}|\Gamma (1+\nu +i \tau)|}{2\pi R\sqrt{\Gamma(1+2\nu)}}.
    \end{equation}
Then, we must solve for $B^{(1)}$ via the Maxwell equation
\begin{equation}\label{eq:gaussB1}
\Laplacian_{S^3}B^{(1)}=GR^2(\rho^{(1/2)})^2B^{(0)}-\frac{G}{R}\left(\frac{\delta(\alpha)}{4 \pi \sin^2\alpha}\right).
\end{equation}
This can be integrated explicitly. Near the arms of the cusp, we have
\begin{equation}\label{eq:B1arm}
    R B^{(1)}=\frac{\tau}{\alpha}-\frac{2\tau^2}{2\nu-1}+C_{0}^{(1)}+\mathcal{O}(\alpha^{2\nu-1}).
\end{equation}
To determine $C_0^{(1)}$ requires discerning the consistency conditions on $\rho^{(3/2)}$ through the matter equation at $\mathcal{O}(q^{3/2})$ such that it is regular and has the correct charge, in the same way $C_0$ was fixed. This is outlined in Appendix \ref{sec:appA}. These manipulations allow $C_0^{(1)}$ to be isolated, and expressed in terms of integrals of the form \eqref{eq:masterint}. Evaluating these gives
\begin{equation}\label{eq:C01}
\boxed{\;C_0^{(1)}=\frac{\Gamma\!\left(2\nu-\tfrac12\right)\,|\Gamma(\nu+i\tau)|^2}{4\sqrt{\pi}\;\Gamma(2\nu)\,\left|\Gamma\!\left(\nu+\tfrac12+i\tau\right)\right|^2}\left[\frac{\kappa}{2\pi}-\frac{\tau\,(\nu+\tfrac12)}{\nu-\tfrac12}\right].\;}
\end{equation}
Thus, we obtain
\begin{equation}\label{eq:G2C01}
    \Gamma^{(2)}_{1+q,1}(\pi)=\frac{1}{2}\left(C_0^{(1)}-\frac{\tau^2}{\nu-\frac{1}{2}}\right),
\end{equation}
which, when written in terms of $G$, becomes
\begin{equation}\label{eq:G2}
\boxed{\begin{aligned}
\Gamma^{(2)}_{1+q,1}(\pi)={}&\frac{\Gamma\!\left(\tfrac12+\sqrt{1-\tfrac{G^2}{4\pi^2}}\right)\left|\Gamma\!\left(\tfrac12+\tfrac12\sqrt{1-\tfrac{G^2}{4\pi^2}}+\tfrac{iG}{4\pi}\right)\right|^2}{4\sqrt{\pi}\;\Gamma\!\left(1+\sqrt{1-\tfrac{G^2}{4\pi^2}}\right)\left|\Gamma\!\left(1+\tfrac12\sqrt{1-\tfrac{G^2}{4\pi^2}}+\tfrac{iG}{4\pi}\right)\right|^2}\\
&\times\left[\frac{\kappa}{4\pi}-\frac{G\left(2+\sqrt{1-\tfrac{G^2}{4\pi^2}}\right)}{8\pi\sqrt{1-\tfrac{G^2}{4\pi^2}}}\right]
-\frac{G^2}{16\pi^2\sqrt{1-\tfrac{G^2}{4\pi^2}}}.
\end{aligned}}
\end{equation}
Note that this term is well-behaved except at $G=2\pi$. Indeed, near this value, we have
\begin{equation}\label{eq:G2crit}
    \Gamma^{(2)}_{1+q,1}(\pi)\sim -\frac{1}{2(1+e^{-\pi})}\sqrt{\frac{\pi}{2\pi-G}},\qquad G\to2\pi,
\end{equation}
which is the expected singular square-root behaviour. To see why this singular behaviour arises, we can examine more closely our perturbative expansion. At finite $q$, the scalar near the arm with charge $1+q$ has exponent
\begin{equation}\label{eq:muq}
    \mu(q)=\sqrt{\tfrac14-\tau^2(1+q)^2},
\end{equation}
through which the cusp obtains $q$-dependence via $C_0$. Taking $\mu\equiv\mu(0)$, the branch point sits precisely at 
 \begin{equation}\label{eq:qstar}
     q_*=\frac{1}{2\tau}-1=\frac{1}{\sqrt{1-4\mu^2}}-1=2\mu^2+6\mu^4+\dots,
 \end{equation} 
for which the arm becomes critical. As $\tau\to1/2$, the region of valid $q$ shrinks to zero. Thus, the limit $G\to2\pi$ should be taken as $q\to 0$. This suggests that we should consider $\frac{q}{q_*}$ fixed. Heuristically, $q_*\sim 2\mu^2$, and at leading order in $\mu$,
 \begin{equation}\label{eq:muqscaling}
     \mu(q)\sim \mu\sqrt{1-\frac{q}{2\mu^2}}.
 \end{equation}
Thus, in an expansion in small $q\ll \mu^2$, we get powers
\begin{equation}\label{eq:muqseries}
    \mu(q)\sim \mu-\frac{q}{4\mu}-\frac{q^2}{32\mu^3}-\frac{q^3}{128\mu^5}+\mathcal{O}(q^4).
\end{equation}
Each order in $q$ adds a factor of $\frac{1}{\mu^2}$, so we expect singular behaviour $\Gamma^{(n)}_{1+q,1}(\pi)\sim \mu^{3-2n}$. We can compute this behaviour exactly. Instead of holding $\mu$ fixed and expanding in small $q\ll \mu^2$, we can consider the regime where $x=\frac{q}{\mu^2}$ is fixed and both $q\to 0$ and $\mu\to 0$. 

By using the method of matched asymptotics, we can compute the anomalous dimension resummed to all orders in $x$ at leading order in $\mu$.

Taking $s=\log\tan(\alpha/2)$ and $\rho=\sqrt{\cosh s}\,\Psi$, the arms sit at $s=\mp\infty$ and the matter equation becomes
\begin{equation}\label{eq:Psieq}
    \Psi''=U\Psi,\qquad U=\frac14+\frac{\frac14-(RB)^2}{\cosh^2s}+\frac{\kappa R^2\Psi^2}{\cosh s},
\end{equation}
with tails $\Psi\propto e^{\mu(q)s}$ as $s\to-\infty$ and $\Psi\propto e^{-\mu s}$ as $s\to\infty$. At $G=2\pi$ and $q=0$ the solutions are $RB_c=\frac12(1-\sinh s)$ and $\Psi_c=e^{-\alpha/2}$, obeying
\begin{equation}\label{eq:critint}
    \int_{-\infty}^{\infty}\Psi_c^2\,RB_c\,\frac{ds}{\cosh^2s}=\frac{1+e^{-\pi}}{2}.
\end{equation}
At leading order in $\mu$, the core is this critical solution, with $\Psi=A\Psi_c$ and $RB=RB_c+\delta C$, where $\delta C=\bar{C_0}(q)-\frac12$, where $\bar{C_0}(q)$ is the constant in the core with $\bar{C}_0(0)=C_0$. The quartic term is subleading. Since the charge density carries a factor $\cosh^{-2}s$, the charge sits in the core, so that \eqref{eq:critint} fixes
\begin{equation}\label{eq:Anorm}
    A^2=\frac{q}{2\pi(1+e^{-\pi})R^2}.
\end{equation}
To find $\delta C$, we integrate $(\Psi_c\Psi'-\Psi\Psi_c')'=(U-U_c)\Psi\Psi_c$ over all $s$. The boundary terms vanish, so $\int ds\,(U-U_c)\Psi\Psi_c=0$. In the tails $U-U_c\to\mu(q)^2$ and $\mu^2$, which contribute $A\mu(q)$ and $Ae^{-\pi}\mu$, while in the core $U-U_c\simeq-2RB_c\delta C/\cosh^2s$, which contributes $-A(1+e^{-\pi})\delta C$ by \eqref{eq:critint}. Thus,
\begin{equation}\label{eq:C0crit}
    \bar{C_0}(q)=\frac12+\frac{\mu(q)+e^{-\pi}\mu}{1+e^{-\pi}}+\mathcal O(\mu^2).
\end{equation}
Finally, note that $\frac{d\Gamma_{1+q,1}(\pi)}{dq}=RB_{\rm reg}|_{\alpha=0}$ differs from $\bar{C}_0(q)$, the constant in the core, since the long tail carries charge close to the arm. By Gauss' law, this lowers the regular part by $\frac{\pi R^2A^2}{2\mu(q)}$, and so
\begin{equation}\label{eq:dGdqcrit}
    \frac{d\Gamma_{1+q,1}(\pi)}{dq}\sim\frac12+\frac{\mu(q)+e^{-\pi}\mu}{1+e^{-\pi}}-\frac{q}{4(1+e^{-\pi})\mu(q)}.
\end{equation}
Integrating this yields
\begin{equation}\label{eq:Gcrit}
    \Gamma_{1+q,1}(\pi)=\frac q2+\frac{q}{1+e^{-\pi}}\big[\mu(q)+e^{-\pi}\mu\big]+\mathcal O(\mu^4),
\end{equation}
which, expressed in terms of $x$, gives us the resummed result
\begin{equation}\label{eq:critresum}
   \boxed{ \Gamma_{1+q,1}(\pi)=\frac{q}{2}+\mu^3\frac{x(\sqrt{1-x/2}+e^{-\pi})}{1+e^{-\pi}}+\mathcal{O}(\mu^4).}
\end{equation}
The formula is a function of arbitrary $0\leq x\leq 2$. Expanding in $x\ll1$, this yields explicitly the Catalan numbers at each leading order in $\mu$, which are those visible in the small $q$ expansion:
\begin{align}\label{eq:critgamma}
\Gamma^{(1)}_{1+q,1}(\pi)=\frac12+\mu=\nu,\qquad
\Gamma^{(2)}_{1+q,1}(\pi)\simeq-\frac{1}{4(1+e^{-\pi})\,\mu},\\
\label{eq:critgamman}\Gamma^{(n)}_{1+q,1}(\pi)\simeq-\frac{c_{n-2}}{2^{3n-4}(1+e^{-\pi})}\,\mu^{3-2n}\quad(n\ge2).
\end{align}
Less singular pieces are of higher order in $\mu$ and are not captured. Away from $G\sim 2\pi$, this says nothing. This leading singular behaviour is in perfect agreement for $\Gamma^{(1)}_{1+q,1}(\pi)$ and $\Gamma^{(2)}_{1+q,1}(\pi)$ as calculated in the previous sections above, and matches the pole in the value of $\Gamma^{(3)}_{1+q,1}(\pi)$ which we compute in the next section. Note that this is a large $Q$ result: as $qQ\gg 1$, we need $Q\mu^2\gg1$ and so describes the semiclassical approach to the loss of conformality.

To verify the result of \eqref{eq:G2}, if one expands in small $G$ we obtain
\begin{equation}\label{eq:G2series}
    \Gamma^{(2)}_{1+q,1}(\pi)=\frac{\kappa}{8\pi^2}-\frac{3G}{16\pi^2}+G^2\left(\frac{(\pi^2-6)\kappa}{384\pi^4}-\frac{1}{16\pi^2}\right)-\frac{(\pi^2-2)G^3}{256\pi^4}+\mathcal{O}(G^4).
\end{equation}
Through to $\mathcal{O}(G^2)$ this is exactly the $q^2$ coefficient of the result of~\cite{Bersini:2026tpq} by taking $\alpha_*=\pi$ and $q_1=1+q$ and $q_2=1$. This establishes an extension of the results of~\cite{Bersini:2026tpq} with predictions for the $G^3$ contributions and beyond, at this order in $q$.

\begin{figure}[H]
    {\linespread{1}\selectfont
\begin{center}
\renewcommand{\arraystretch}{1.2}
\begin{tabular}{c}
\textbf{$G^0$: the contact}\\
\begin{tikzpicture}[scale=0.8]\dgBase \draw[dgS] (0,0) .. controls (0.5,1.4) and (1.1,1.4) .. (1.5,1.0); \draw[dgS] (1.5,1.0) .. controls (1.9,1.4) and (2.5,1.4) .. (3,0); \draw[dgS] (0,0) .. controls (0.7,0.5) and (1.1,0.7) .. (1.5,1.0); \draw[dgS] (1.5,1.0) .. controls (1.9,0.7) and (2.3,0.5) .. (3,0); \node[dgK] at (1.5,1.0) {};\end{tikzpicture}\\
\shortstack{(a) contact\\[2pt] $\displaystyle\frac{\kappa}{8\pi^2}$}
\end{tabular}

\vspace{6pt}
\begin{tabular}{ccc}
\multicolumn{3}{c}{\textbf{$G^1$: one-photon exchange}}\\
\begin{tikzpicture}[scale=0.8]\dgBase \draw[dgS] (0,0) .. controls (0.45,1.9) and (2.55,1.9) .. (3,0); \draw[dgS] (0,0) .. controls (0.8,1.0) and (2.2,1.0) .. (3,0); \draw[dgXP] (1.5,1.42) -- (1.5,0.75); \node[dgVx] at (1.5,1.42) {}; \node[dgVx] at (1.5,0.75) {};\end{tikzpicture} &
\begin{tikzpicture}[scale=0.8]\dgBase \draw[dgS] (0,0) .. controls (0.6,1.5) and (2.4,1.5) .. (3,0); \draw[dgXP] (1.5,1.13) -- (1.5,0.08); \node[dgVx] at (1.5,1.13) {};\end{tikzpicture} &
\begin{tikzpicture}[scale=0.8]\dgBase \draw[dgXP] (0.7,0.08) .. controls (1.1,1.0) and (1.9,1.0) .. (2.3,0.08);\end{tikzpicture}\\
\shortstack{(b1) photon exchange\\[2pt] $\displaystyle+\frac{G}{16\pi^2}$} & \shortstack{(b2) scalar--charge\\[2pt] $\displaystyle-\frac{G}{8\pi^2}$} & \shortstack{(b3)  charge--charge\\[2pt] $\displaystyle-\frac{G}{8\pi^2}$}
\end{tabular}

\vspace{6pt}
\begin{tabular}{ccc}
\multicolumn{3}{c}{\textbf{$G^2$, $\kappa^0$: the arm}}\\
\begin{tikzpicture}[scale=0.8]\dgBase \draw[dgS] (0,0) .. controls (0.6,1.5) and (2.4,1.5) .. (3,0); \draw[dgXP] (1.0,1.04) -- (1.0,0.08); \node[dgVx] at (1.0,1.04) {}; \draw[dgP] (2.0,1.04) -- (2.0,0); \node[dgVx] at (2.0,1.04) {};\end{tikzpicture} &
\begin{tikzpicture}[scale=0.8]\dgBase \draw[dgS] (0,0) .. controls (0.6,1.5) and (2.4,1.5) .. (3,0); \draw[dgXP] (1.5,1.13) -- (1.15,0.08); \draw[dgP] (1.5,1.13) -- (1.85,0); \node[dgSg] at (1.5,1.13) {};\end{tikzpicture} & \\
\shortstack{(c1) two single photons\\[2pt] $\displaystyle-\frac{G^2}{32\pi^2}$} & \shortstack{(c2) seagull\\[2pt] $\displaystyle-\frac{G^2}{32\pi^2}$} &
\end{tabular}

\vspace{6pt}
\begin{tabular}{cc}
\multicolumn{2}{c}{\textbf{$\kappa G^2$: the contact with line photons}}\\
\begin{tikzpicture}[scale=0.8]\dgBase \draw[dgS] (0,0) .. controls (0.5,1.4) and (1.1,1.4) .. (1.5,1.0); \draw[dgS] (1.5,1.0) .. controls (1.9,1.4) and (2.5,1.4) .. (3,0); \draw[dgS] (0,0) .. controls (0.7,0.5) and (1.1,0.7) .. (1.5,1.0); \draw[dgS] (1.5,1.0) .. controls (1.9,0.7) and (2.3,0.5) .. (3,0); \node[dgK] at (1.5,1.0) {}; \draw[dgP] (0.75,1.18) -- (0.75,0); \node[dgVx] at (0.75,1.18) {}; \draw[dgP] (2.25,0.62) -- (2.25,0); \node[dgVx] at (2.25,0.62) {};\end{tikzpicture} &
\begin{tikzpicture}[scale=0.8]\dgBase \draw[dgS] (0,0) .. controls (0.5,1.4) and (1.1,1.4) .. (1.5,1.0); \draw[dgS] (1.5,1.0) .. controls (1.9,1.4) and (2.5,1.4) .. (3,0); \draw[dgS] (0,0) .. controls (0.7,0.5) and (1.1,0.7) .. (1.5,1.0); \draw[dgS] (1.5,1.0) .. controls (1.9,0.7) and (2.3,0.5) .. (3,0); \node[dgK] at (1.5,1.0) {}; \draw[dgP] (0.75,1.18) -- (0.45,0); \draw[dgP] (0.75,1.18) -- (1.05,0); \node[dgSg] at (0.75,1.18) {};\end{tikzpicture}\\
\shortstack{(d1) single photons\\[2pt] $\displaystyle\frac{(\pi^2-9)\,\kappa G^2}{384\pi^4}$} & \shortstack{(d2) seagull\\[2pt] $\displaystyle\frac{\kappa G^2}{128\pi^4}$}
\end{tabular}
\end{center}
\par}
    \caption{Feynman diagrams contributing to $\Gamma^{(2)}_{1+q,1}(\pi)$ to $\mathcal{O}(G^2)$. Each picture represents a sum over all placements and orderings of the vertices. Values are in the Feynman gauge on the cylinder; only sums are gauge-invariant.}
    \label{fig:q2diagrams}
\end{figure}
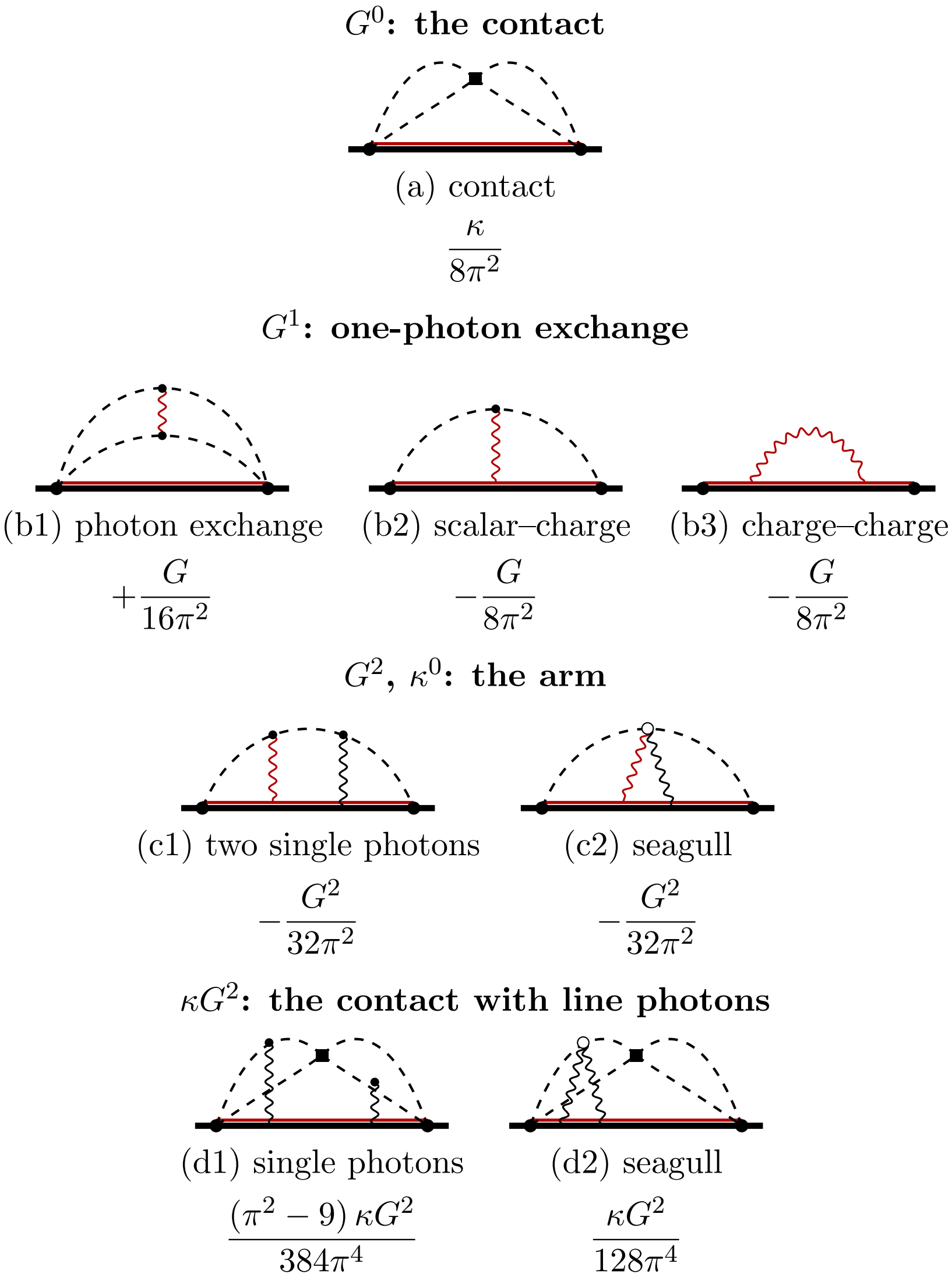

Again, the full expression $\Gamma^{(2)}_{1+q,1}(\pi)$ is effectively the sum of infinitely many Feynman diagrams. By counting the powers of $\kappa, G$, and $q$, in terms of the original couplings these terms are of the form $(Q_1-Q_2)^2(e^2Q)^i(\lambda Q)^j/Q$. This means that each term is an interaction between two quanta of the impurity,  between a quantum and the extra charge, or of the extra charge with itself, dressed by interactions with the Wilson line or between scalar lines, which yields powers $e^2Q$ or $\lambda Q$ depending on whether a photon or scalar is exchanged. The closed form can be organised into three families of diagrams,
\begin{equation}\label{eq:Gamma2split}
\Gamma^{(2)}_{1+q,1}(\pi)=\underbrace{-\frac{\tau^2}{2\mu}}_{\text{(i) arm}}\;+\;\underbrace{\frac{\kappa}{4\pi}\,P}_{\text{(ii) contact}}\;\underbrace{-\;\frac{\tau(\nu+\frac12)}{2\mu}\,P}_{\text{(iii) exchange}},\qquad P=\frac{\Gamma\!\left(2\nu-\tfrac12\right)|\Gamma(\nu+i\tau)|^2}{4\sqrt\pi\,\Gamma(2\nu)\,|\Gamma(\nu+\tfrac12+i\tau)|^2}.
\end{equation}
The first is the arm family, which is the change of the $\mathcal{O}(q)$ result when the charge of the arm is raised to $1+q$, which gives the derivative of the Catalan series of the previous order. This series consists of precisely the double photon and seagull diagrams seen at the previous order, with the difference being that one of the photons ends on the original Wilson line and the other on the extra-charged part of the line. The contact family comes from a single quartic vertex, with various attached photon lines appropriate to the order. The exchange family comes from a photon exchange between two scalar lines, a scalar line and the impurity, or between the impurity and itself. As a series in $G$, the first few diagrams can be seen in Figure \ref{fig:q2diagrams}. Here we colour photons red when they come with a factor of $1/Q$. This includes the photon connecting to the shifted charge-$q$ part of the Wilson line, as opposed to the black that connect to the unshifted charge-$1$ line. Exchange between the two scalar lines is also red. Near the critical point, the arm and exchange terms together produce the scaling result seen in \eqref{eq:critgamma}, while the contact series is subleading.
The semiclassical method immediately computes the result of summing these diagrams, without having to compute their contributions and multiplicities explicitly.

\subsection{$\mathcal{O}(q^3)$}\label{sec:oq3}
To compute $\Gamma^{(3)}_{1+q,1}(\pi)$ we follow a similar recipe as to the previous orders. We need enough knowledge of $\rho^{(3/2)}$ to eliminate it in favour of $\rho^{(1/2)}$, $B^{(0)}$ and $B^{(1)}$ by using the solvability of the matter equation at $\mathcal{O}(q^{5/2})$. In this vein, it is more convenient to work with the ratio,
\begin{equation}\label{eq:ydef}
    y=\frac{\rho^{(3/2)}}{\rho^{(1/2)}}.
\end{equation}
In fact, after some manipulation, as shown in Appendix \ref{sec:appB}, we can eliminate $y$ and express the anomalous dimension in terms of $y'$,
\begin{equation}\label{eq:gamm3}
\boxed{\;\Gamma^{(3)}_{1+q,1}(\pi)=-2\pi R^2\int_0^\pi\varrho^2\Big[\big(y'\big)^2+\big(RB^{(1)}\big)^2\Big]d\alpha.\;}
\end{equation}
This is a more tractable form for evaluation, as although $y$ is not known (as it requires the explicit form of $\rho^{(3/2)}$), the combination $\varrho^2y'$ is:
\begin{equation}\label{eq:yprime}
    \varrho^2y'(\alpha)=\int_0^\alpha\Big[\kappa R^2k_1^4\sin^{4\nu-2}a\;e^{-4\tau a}-2(\tau\cot a+\nu)\,RB^{(1)}(a)\,k_1^2\sin^{2\nu}a\;e^{-2\tau a}\Big]da .
\end{equation}
Thus, $\Gamma^{(3)}_{1+q,1}(\pi)$ reduces to a nested integral of elementary functions and an incomplete integral. Note that this formula splits up into three parts, each with a different $\kappa$-dependence:
\begin{equation}\label{eq:G3kappa}
    \Gamma^{(3)}_{1+q,1}(\pi)=\mathcal{A}_0+\mathcal{A}_1\kappa+\mathcal{A}_2\kappa^2,
\end{equation}
where
\begin{align}\label{eq:A012}
\mathcal A_0&=-\frac{1}{2\pi R^2}\int_0^\pi\frac{\big(RB^{(1)}|_{\kappa=0}\,F+\tau\int_0^\alpha\frac{F(1-F)}{\sin^2a}da\big)^2}{\varrho^2}\,d\alpha-2\pi R^2\int_0^\pi\varrho^2\big(RB^{(1)}|_{\kappa=0}\big)^2d\alpha, \\
\label{eq:A1}\mathcal A_1&=\frac{P}{2\pi^2R^2}\int_0^\pi\frac{(F_4-F)\big(RB^{(1)}|_{\kappa=0}\,F+\tau\int_0^\alpha\frac{F(1-F)}{\sin^2a}da\big)}{\varrho^2}\,d\alpha-2PR^2\int_0^\pi\varrho^2\,RB^{(1)}\big|_{\kappa=0}\,d\alpha,\\
\label{eq:A2}\mathcal A_2&=-\frac{P^2}{8\pi^2}-\frac{P^2}{8\pi^3R^2}\int_0^\pi\frac{(F_4-F)^2}{\varrho^2}\,d\alpha,
\end{align}
where $F$ is the charge fraction %of the previous section
\begin{gather}\label{eq:Fdef}
    F(\alpha)
    =4\pi R^2k_1^2\Big[\frac{\tau}{2\nu}\sin^{2\nu}\!\alpha\,e^{-2\tau\alpha}+\int_0^\alpha\sin^{2\nu}\!a\,e^{-2\tau a}\,da\Big],
\end{gather}
$F_4$ the corresponding fraction of the density $\big(\rho^{(1/2)}\big)^4$
\begin{equation}\label{eq:F4def}
         F_4(\alpha)=\frac{4\pi^2R^4k_1^4}{P}\int_0^\alpha\sin^{4\nu-2}\!a\,e^{-4\tau a}\,da,
\end{equation}
and $P$ is the same constant from \eqref{eq:Gamma2split}. From \eqref{eq:gamm3}, we can see $\Gamma^{(3)}_{1+q,1}(\pi)\le0$ manifestly. Furthermore, the integral converges for $G<2\pi$. While it is explicit, the integral likely cannot be evaluated analytically. It, however, can be expanded to arbitrary orders in small $G$, for which we can evaluate each term to get
\begin{multline}\label{eq:G3series}
\Gamma^{(3)}_{1+q,1}(\pi)=-\frac{\kappa^2}{32\pi^4}
+G^2\Big[\frac{3-8\pi^2}{384\pi^4}+\frac{(\pi^2-6)\kappa}{384\pi^4}+\frac{(\pi^2-18)\kappa^2}{1536\pi^6}\Big]\\
-G^3\Big[\frac{\pi^2-2}{256\pi^4}+\frac{\big(\pi^4+60\pi^2-480-210\zeta(3)\big)\kappa}{7680\pi^6}\Big]\\
+G^4\Big[\frac{6\pi^4-85\pi^2-870-1260\zeta(3)}{30720\pi^6}
-\frac{\big(\pi^4+15\pi^2+270-540\zeta(3)\big)\kappa}{46080\pi^6}\\
+\frac{\big(180(22+12\zeta(3)-15\zeta(5))+15\pi^2(12\zeta(3)-41)-2\pi^4\big)\kappa^2}{368640\pi^8}\Big]+\mathcal O(G^5).
\end{multline}
Note there is no $G^1$ term. This agrees with the observation in~\cite{Bersini:2026tpq}, which computes up to $\mathcal{O}(G^2)$. Note that at higher orders in $G$ than displayed here, the multiple zeta-value $\zeta(5,3)$ occurs, with the first instance being in the term
\begin{equation}
    -\frac{9\zeta(5,3)}{65536\pi^{12}}\kappa^2 G^8.
\end{equation}
 This suggests that indeed there does not exist a closed form for the integral in terms of $\Gamma$-functions and their derivatives, as it is conjectured that $\zeta(5,3)$ is algebraically independent from single $\zeta$-values. Other multiple $\zeta$ values also occur at higher orders.

Again, this result is a summation of infinitely many Feynman diagrams. In terms of the original couplings, we are allowed diagrams with couplings $(Q_1-Q_2)^3(e^2Q)^i(\lambda Q)^j/Q^2$. This implies there are three scalar lines, or two scalars with a photon to the extra charge, or a single scalar with two photons to the extra charge. Photons from the scalar to the Wilson line gives $e^2Q=G$ for each photon, so we are allowed any number of such photons. Contact interactions come with powers of $\lambda\propto\frac{\kappa}{Q}$, scalar-scalar photon exchanges with $e^2=G/Q$, and photons to the extra charge with $G(Q_1-Q_2)/Q$. This means we can have at most two quartic vertices, and is why we get a quadratic polynomial in $\kappa$. When there are no quartic vertices, there must be at least two photons coming from the scalar-scalar photons or to the extra charge and so the series for $\mathcal{A}_0$ starts at $G^2$. Likewise, for $\mathcal{A}_1$, there is a single quartic vertex, and one scalar-scalar photon exchange or to the extra charge is needed to make up the factors of $1/Q$, and thus the series can start at $G^1$ - however the $G^1$ term comes from the Coulomb energy which has charge exactly $q$, so this term vanishes and the series starts at $\mathcal{O}(G^2)$. For two quartic vertices, no photons are necessary and so $\mathcal{A}_2$ starts at $G^0$. The possible diagrams to order $\mathcal{O}(G^2)$ can be seen in Figure \ref{fig:q3diagrams}.

\providecommand{\qTwoC}{\dgBase
\draw[dgS] (0,0) .. controls (0.3,0.8) and (0.7,1.05) .. (1.0,1.25) .. controls (1.4,1.55) and (2.6,1.9) .. (3,0);
\draw[dgS] (0,0) .. controls (0.2,1.5) and (0.7,1.45) .. (1.0,1.25) .. controls (1.3,1.05) and (1.7,0.95) .. (2.0,0.75) .. controls (2.3,0.55) and (2.8,0.5) .. (3,0);
\draw[dgS] (0,0) .. controls (0.5,0.5) and (1.6,0.5) .. (2.0,0.75) .. controls (2.4,1.0) and (2.9,1.0) .. (3,0);
\node[dgK] at (1.0,1.25) {}; \node[dgK] at (2.0,0.75) {};}
\providecommand{\qOneC}{\dgBase
\draw[dgS] (0,0) .. controls (0.3,0.4) and (1.0,0.75) .. (1.5,1.1) .. controls (2.0,1.45) and (2.8,1.9) .. (3,0);
\draw[dgS] (0,0) .. controls (0.2,1.9) and (1.0,1.45) .. (1.5,1.1) .. controls (2.0,0.75) and (2.7,0.4) .. (3,0);
\node[dgK] at (1.5,1.1) {};}
\providecommand{\qOne}{\dgBase \draw[dgS] (0,0) .. controls (0.6,1.5) and (2.4,1.5) .. (3,0);}
\providecommand{\qTwo}{\dgBase \draw[dgS] (0,0) .. controls (0.45,1.9) and (2.55,1.9) .. (3,0); \draw[dgS] (0,0) .. controls (0.8,1.0) and (2.2,1.0) .. (3,0);}
\providecommand{\qThree}{\dgBase \draw[dgS] (0,0) .. controls (0.4,2.3) and (2.6,2.3) .. (3,0); \draw[dgS] (0,0) .. controls (0.6,1.6) and (2.4,1.6) .. (3,0); \draw[dgS] (0,0) .. controls (0.85,0.9) and (2.15,0.9) .. (3,0);}
\providecommand{\qLab}[2]{\shortstack{#1\\[2pt] #2}}
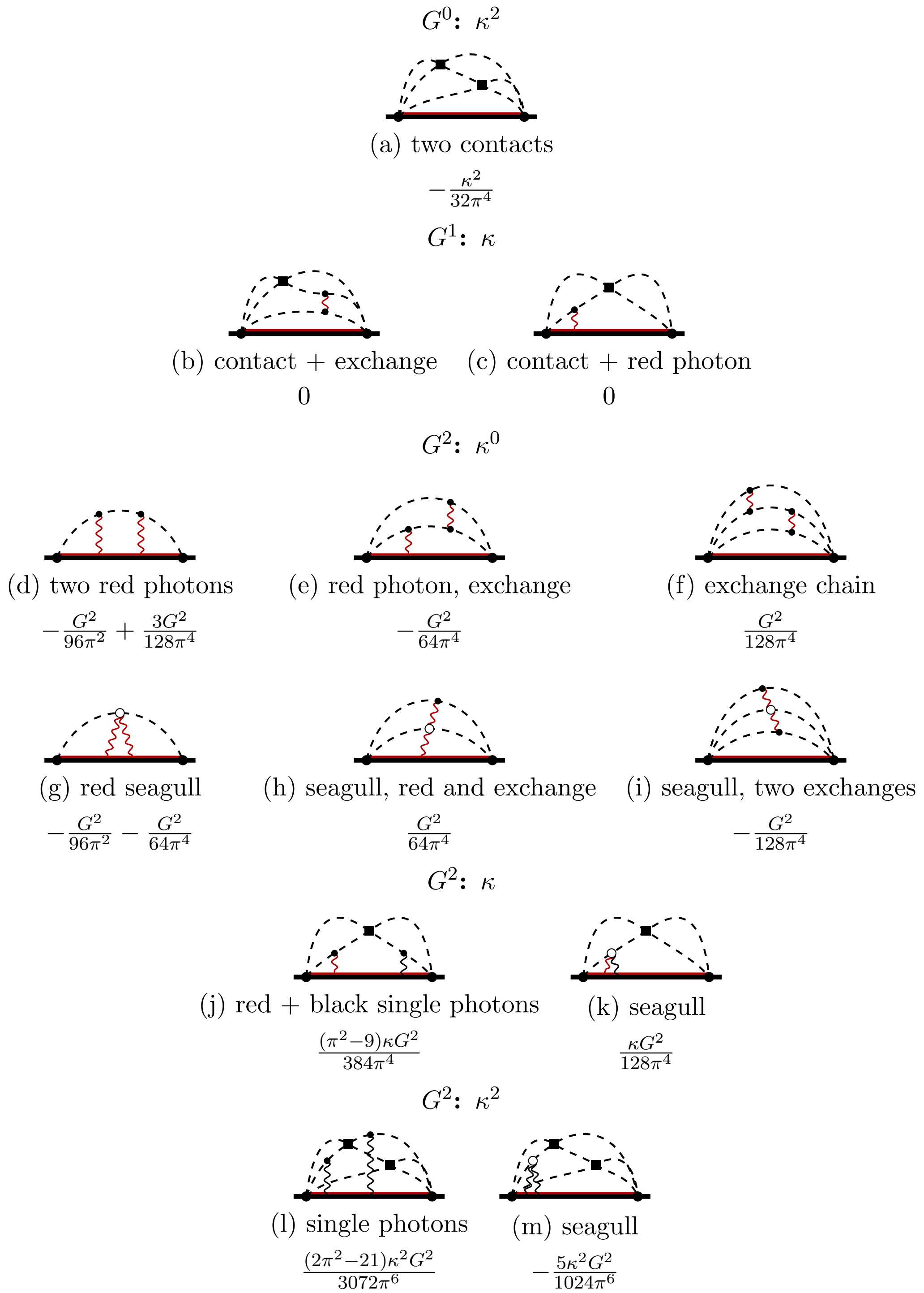
\begin{figure}[H]
{\linespread{1}\selectfont
\begin{center}
\renewcommand{\arraystretch}{1.1}
\textbf{$G^0$: $\kappa^2$}\\[2pt]
\begin{tabular}{c}
\begin{tikzpicture}[scale=0.62,color=black]\qTwoC\end{tikzpicture}\\
\qLab{(a) two contacts}{$-\frac{\kappa^2}{32\pi^4}$}
\end{tabular}

\vspace{6pt}
\textbf{$G^1$: $\kappa$}\\[2pt]
\begin{tabular}{cc}
\begin{tikzpicture}[scale=0.62,color=black]\dgBase
\draw[dgS] (0,0) .. controls (0.3,0.8) and (0.7,1.05) .. (1.0,1.25) .. controls (1.4,1.55) and (2.6,1.9) .. (3,0);
\draw[dgS] (0,0) .. controls (0.2,1.5) and (0.7,1.45) .. (1.0,1.25) .. controls (1.3,1.05) and (1.6,0.95) .. (2.0,0.95) .. controls (2.5,0.95) and (2.85,0.8) .. (3,0);
\draw[dgS] (0,0) .. controls (0.5,0.7) and (2.5,0.7) .. (3,0);
\node[dgK] at (1.0,1.25) {};
\draw[dgXP] (2.0,0.95) -- (2.0,0.52); \node[dgVx] at (2.0,0.95) {}; \node[dgVx] at (2.0,0.52) {};\end{tikzpicture} &
\begin{tikzpicture}[scale=0.62,color=black]\qOneC
\draw[dgXP] (0.675,0.569) -- (0.675,0.08); \node[dgVx] at (0.675,0.569) {};\end{tikzpicture}\\
\qLab{(b) contact $+$ exchange}{$0$} & \qLab{(c) contact $+$ red photon}{$0$}
\end{tabular}

\vspace{6pt}
\textbf{$G^2$: $\kappa^0$}\\[2pt]
\begin{tabular}{ccc}
\begin{tikzpicture}[scale=0.62,color=black]\qOne
\draw[dgXP] (1.0,1.04) -- (1.0,0.08); \node[dgVx] at (1.0,1.04) {};
\draw[dgXP] (2.0,1.04) -- (2.0,0.08); \node[dgVx] at (2.0,1.04) {};\end{tikzpicture} &
\begin{tikzpicture}[scale=0.62,color=black]\qTwo
\draw[dgXP] (1.0,0.68) -- (1.0,0.08); \node[dgVx] at (1.0,0.68) {};
\draw[dgXP] (2.0,0.68) -- (2.0,1.325); \node[dgVx] at (2.0,0.68) {}; \node[dgVx] at (2.0,1.325) {};\end{tikzpicture} &
\begin{tikzpicture}[scale=0.62,color=black]\qThree
\draw[dgXP] (1.0,1.608) -- (1.0,1.105); \node[dgVx] at (1.0,1.608) {}; \node[dgVx] at (1.0,1.105) {};
\draw[dgXP] (2.0,1.105) -- (2.0,0.609); \node[dgVx] at (2.0,1.105) {}; \node[dgVx] at (2.0,0.609) {};\end{tikzpicture}\\
\qLab{(d) two red photons}{$-\frac{G^2}{96\pi^2}+\frac{3G^2}{128\pi^4}$} & \qLab{(e) red photon, exchange}{$-\frac{G^2}{64\pi^4}$} & \qLab{(f) exchange chain}{$\frac{G^2}{128\pi^4}$}\\[6pt]
\begin{tikzpicture}[scale=0.62,color=black]\qOne
\draw[dgXP] (1.5,1.125) -- (1.2,0.08); \draw[dgXP] (1.5,1.125) -- (1.8,0.08); \node[dgSg] at (1.5,1.125) {};\end{tikzpicture} &
\begin{tikzpicture}[scale=0.62,color=black]\qTwo
\draw[dgXP] (1.5,0.75) -- (1.3,0.08); \draw[dgXP] (1.5,0.75) -- (1.7,1.409); \node[dgVx] at (1.7,1.409) {}; \node[dgSg] at (1.5,0.75) {};\end{tikzpicture} &
\begin{tikzpicture}[scale=0.62,color=black]\qThree
\draw[dgXP] (1.5,1.2) -- (1.3,1.707); \node[dgVx] at (1.3,1.707) {}; \draw[dgXP] (1.5,1.2) -- (1.7,0.665); \node[dgVx] at (1.7,0.665) {}; \node[dgSg] at (1.5,1.2) {};\end{tikzpicture}\\
\qLab{(g) red seagull}{$-\frac{G^2}{96\pi^2}-\frac{G^2}{64\pi^4}$} & \qLab{(h) seagull, red and exchange}{$\frac{G^2}{64\pi^4}$} & \qLab{(i) seagull, two exchanges}{$-\frac{G^2}{128\pi^4}$}
\end{tabular}

\vspace{6pt}
\textbf{$G^2$: $\kappa$}\\[2pt]
\begin{tabular}{cc}
\begin{tikzpicture}[scale=0.62,color=black]\qOneC
\draw[dgXP] (0.675,0.569) -- (0.675,0.08); \node[dgVx] at (0.675,0.569) {};
\draw[dgP] (2.325,0.569) -- (2.325,0); \node[dgVx] at (2.325,0.569) {};\end{tikzpicture} &
\begin{tikzpicture}[scale=0.62,color=black]\qOneC
\draw[dgXP] (0.675,0.569) -- (0.5,0.08); \draw[dgP] (0.675,0.569) -- (0.85,0); \node[dgSg] at (0.675,0.569) {};\end{tikzpicture}\\
\qLab{(j) red $+$ black single photons}{$\frac{(\pi^2-9)\kappa G^2}{384\pi^4}$} & \qLab{(k) seagull}{$\frac{\kappa G^2}{128\pi^4}$}
\end{tabular}

\vspace{6pt}
\textbf{$G^2$: $\kappa^2$}\\[2pt]
\begin{tabular}{cc}
\begin{tikzpicture}[scale=0.62,color=black]\qTwoC
\draw[dgP] (0.5,0.85) -- (0.5,0); \node[dgVx] at (0.5,0.85) {};
\draw[dgP] (1.53,1.47) -- (1.53,0); \node[dgVx] at (1.53,1.47) {};\end{tikzpicture} &
\begin{tikzpicture}[scale=0.62,color=black]\qTwoC
\draw[dgP] (0.5,0.85) -- (0.35,0); \draw[dgP] (0.5,0.85) -- (0.65,0); \node[dgSg] at (0.5,0.85) {};\end{tikzpicture}\\
\qLab{(l) single photons}{$\frac{(2\pi^2-21)\kappa^2G^2}{3072\pi^6}$} & \qLab{(m) seagull}{$-\frac{5\kappa^2G^2}{1024\pi^6}$}
\end{tabular}
\end{center}
\par}
    \caption{All Feynman diagrams contributing to $\Gamma^{(3)}_{1+q,1}(\pi)$ up to $\mathcal O(G^2)$. Note that the $\kappa^2G$ and $\kappa^0G$ terms vanish. Each picture is the sum over all placements and orderings of its vertices. Values are in the Feynman gauge on the cylinder; only sums are gauge-invariant.}
    \label{fig:q3diagrams}
\end{figure}

Note that diagrams $(d)$ and $(g)$ are the analogues of $(c1)$ and $(c2)$ of Figure \ref{fig:q2diagrams}, with the black photon replaced by a second red one. They contain the arm family, which is the part $\propto \frac{1}{\pi^2}$, which to all orders in $G$ is $\frac16\frac{d^2}{dq^2}\nu\big(\tau(1+q)\big)\big|_{q=0}=-\frac{\tau^2}{24\mu^3}$. Diagrams $(j)$ and $(k)$ are the correction to the contact family of the previous order to account for the extra charge, which replaces $\tau$ by $\tau(1+\frac q2)$ in $\frac{\kappa}{4\pi}P$. At this order they equal $(d1)$ and $(d2)$ of Figure \ref{fig:q2diagrams}. Diagram $(a)$ is the second-order contact interaction of three quanta, and $(l)$ and $(m)$ are its photon dressing.

\section{Conformal Data from Deflections and Fusion}\label{sec:bremandothers}
In this section, we will compute the conformal data that can be extracted from small deflections away from the straight line, and the fusion limit $\alpha_*\to 0$. The key observables in these cases are the Bremsstrahlung function, which describes the energy radiated due to an acceleration of a charge for equal charges in the former case; and for the latter, the dimension of the fused line, Casimir energy, and the dimension of the leading irrelevant operator and its coefficient.
\subsection{The Bremsstrahlung Function}\label{sec:brem}
In addition to the impurity scaling dimension, the cusp anomalous dimension contains other interesting quantities. If we expand in small deflection angles $\varphi=\pi-\alpha_*$, we know the Bremsstrahlung function\footnote{For $q_1=q_2$, the coefficient $B=-\frac{1}{2}\beta$ is what is usually referred to as the Bremsstrahlung function. We adopt this name at general charges for convenience.} is defined as the leading angular coefficient
\begin{equation}\label{eq:bremsdef}
    \Gamma_{q_1q_2}(\pi-\varphi)=\Delta_{q_1q_2}+\frac{1}{2}\beta_{q_1q_2}\varphi^2+\mathcal{O}(\varphi^4).
\end{equation}
This object is intrinsically tied to the displacement operator \eqref{eq:displacement}. Deflecting an arm inserts this operator along the line, so that $\beta_{q_1q_2}$ is an integrated two-point function of displacement operators in the presence of the defect-changing operator. For $q_1=q_2$ this two-point function is fixed by conformal symmetry to be $\langle\mathcal{D}_i(s)\mathcal{D}_j(0)\rangle=C_D\delta_{ij}/s^4$, so that $\beta_{q_1q_1}=-C_D/6$, which relates it to the energy radiated by an accelerated charge~\cite{Correa:2012at}. At $q_1\neq q_2$ this interpretation is less clear. 

While solving for a general cusp angle is tractable perturbatively to the first few orders in the gauge coupling $G$, when perturbing in the impurity charge $q$ the leading solutions for the fields are no longer homogeneous, and one can no longer apply linearity of the sources to solve the required differential equations. Thus, obtaining the Bremsstrahlung function in our present regime is more difficult than that of~\cite{Bersini:2026tpq}. 
\subsubsection{$\mathcal{O}(q)$}\label{sec:bremq}
At $\mathcal{O}(q)$, we need the angular correction to the integration constant,
\begin{equation}\label{eq:Cphi}
    C(\varphi)=C_0+c(G)\varphi^2.
\end{equation}
This then (combined with angular expansion of the equal-charges solution) will yield
\begin{equation}\label{eq:betafirstorder}
    \beta_{1+q,1}=-\frac{G}{6\pi^2}+\Big(2c(G)-\frac{G}{6\pi^2}\Big)q+\mathcal{O}(q^2).
\end{equation}

To compute this, we can expand the gauge field in terms of $\varphi$ and consider the ratio of the full matter solution to the straight line solution. This allows $c(G)$ to be isolated, and yields the result
\begin{equation}
    \boxed{\begin{aligned}\label{eq:c(G)}
c(G)={}&\frac{2\tau}{3\pi^2}\,\mathrm{Im}\,\psi(\nu+i\tau)-\frac{\tau^2}{6\pi^2}\bigg[\frac{3-5\nu}{\nu}\\
&+\frac{\Gamma\big(\frac32+\nu-\rt\big)\,\Gamma\big(\frac32+\nu+\rt\big)}{\Gamma(1+2\nu)}\\
&\qquad\times\sum_{n=0}^\infty\Big(f(n)-f\big(n+\rt-\nu-\tfrac12\big)\Big)\bigg],
\end{aligned}}
\end{equation}
where
\begin{equation}\label{eq:fx}
f(x)=\frac{\Gamma(x+1)\,\Gamma(x+1+2\nu)}{\Gamma\big(x+\frac32+\nu-\rt\big)\,\Gamma\big(x+\frac32+\nu+\rt\big)}\bigg[1+\frac{\nu\,(x+2)}{(x+1)(x+1+2\nu)+\nu}\bigg].    
\end{equation}

Explicitly, one may expand $c(G)$ in $G$ to obtain 
\begin{equation}\label{eq:beta2expan}
    \boxed{\begin{aligned}\beta_{1+q,1}={}&-\frac{G}{6\pi^2}+q\bigg[-\frac{G}{6\pi^2}+\frac{(\pi^2+3)G^2}{72\pi^4}\\
&\qquad-\frac{\big(45+15\pi^2+2\pi^4-360\zeta(3)\big)G^4}{34560\,\pi^6}+\dots\bigg]+\mathcal O(q^2).\end{aligned}}
\end{equation}

\subsubsection{$\mathcal{O}(q^2)$}\label{sec:bremq2}
To obtain the $\mathcal{O}(q^2)$ correction, we proceed much in the same way by expanding in $\varphi$, eliminating the ratio between the bent scalar and straight scalar solution, and evaluating the resulting integrals. The result is in terms of integrals which can be expanded in $\tau$ and integrated to yield (multiple) zeta values. The full expression can be found in Appendix \ref{sec:appE}. To the first few orders in $G$, we obtain
\begin{multline}\label{eq:beta2series}
\beta^{(2)}_{1+q,1}=\frac{(3+\pi^2)\,G^2}{72\pi^4}+\frac{\big(135-30\pi^2-8\pi^4+270\,\zeta(3)\big)G^3}{8640\,\pi^6}\\
-\frac{\big(45+15\pi^2+2\pi^4-360\,\zeta(3)\big)G^4}{17280\,\pi^6}\\
+\kappa\Big[-\frac{G^2}{32\pi^6}+\frac{\big(11\pi^4-360\pi^2+3330+240\pi^2\zeta(3)-3600\,\zeta(5)\big)G^4}{138240\,\pi^8}\Big]+\mathcal{O}(G^5).
\end{multline}
Some interesting points to note are as follows. The even in $G$, $\kappa$-independent part equals $G\,c'(G)$. Equivalently, through $\mathcal{O}(q^2)$, the scalar cloud contribution to $\beta_{1+q,1}$ is $2q\,c\big(G(1+\frac q2)\big)$: the cloud sees the mean charge of the two lines. The new information at this order sits in the odd powers of $G$ and in the $\kappa$-dependent terms. It is again likely that this expression has no closed form, as the double zeta value $\zeta(5,3)$ appears at higher orders in the expansion.

\subsection{Casimir Energy of Fusion}\label{sec:casimir}
In the opposite limit, $\alpha_*\to0$, the two antiparallel half-lines fuse into a single line of charge $q_1-q_2$. The cusp anomalous dimension then takes the fusion form~\cite{Diatlyk:2024zkk}
\begin{equation}\label{eq:fusion}
\Gamma_{q_1q_2}(\alpha_*)=\frac{\mathcal{C}_{q_1(-q_2)(\overline{q_1-q_2})}}{\alpha_*}+\Delta_{(q_1-q_2)0}+\alpha_{q_1q_2}\,\alpha_*^{\Delta_{\rm irr}-1}+\dots.
\end{equation}
Its three leading terms are the Casimir energy of the pair, the dimension of the fused line, and the contribution of the leading irrelevant operator on the fused line. The contribution from the curvature also appears at order $\alpha_*$. We may also compute these in our regime.

Firstly, we can compute the Casimir energy. This can be done exactly, at leading order in semiclassics. The $\alpha_*^{-1}$ contribution to the energy comes from the behaviour near the lines. Letting $r_1=\alpha/\alpha_*$ and $r_2=\gamma/\alpha_*$ be the rescaled distances to the lines, their Coulomb field is
\begin{equation}\label{eq:B0near}
    RB_C\sim \frac{\tau}{\alpha_*}\left(\frac{q_1}{r_1}-\frac{q_2}{r_2}\right).
\end{equation}
This extremises the action in the absence of $\rho$. The remaining part of the action is extremised for this field if $\rho$ obeys
\begin{equation}\label{eq:near}
-\nabla^2\rho-\tau^2\Big(\frac{q_1}{r_1}-\frac{q_2}{r_2}\Big)^2\rho\equiv (-\nabla^2+V)\rho=0.
\end{equation}
This is a Schrödinger equation with an attractive potential for the scalar, in a zero-energy state. If it has no bound states, then the Casimir energy is bounded from below by exactly this Coulomb energy. Indeed, at order $\alpha_*^{-1}$ we may work with the rescaled action on the near-zone $\mathbb R^3$, with $B$ vanishing at infinity, since a constant part of $RB$ is suppressed by $\alpha_*$ there. This action is convex in $\rho^2$ for $\kappa>0$ and concave in $B$. For any $B$, $\min_{\rho^2}\bar S[\rho^2,B]\le\bar S[0,B]\le\bar S[0,B_C]$. If the operator in \eqref{eq:near} is non-negative, then $\min_{\rho^2}\bar S[\rho^2,B_C]=\bar S[0,B_C]$, so the bound is saturated and the coefficient of $\alpha_*^{-1}$ is the Coulomb value.

For a single arm $q_i$, it is known that there are no bound states for $\tau q_i<\frac{1}{2}$~\cite{Aharony:2022ntz,Aharony:2023amq}. To see that our operator is positive (i.e.\ $\braket{\rho,(-\nabla^2+V)\rho}\geq 0$, $\forall\rho$), note that for any vector field $\vec{v}$, expanding $\int|\nabla\rho+\vec{v}\rho|^2\geq0$ gives
\begin{equation}\label{eq:hardyF}
    \int |\nabla\rho|^2\geq \int\big(\nabla\cdot\vec{v}-|\vec{v}|^2\big)\rho^2 .
\end{equation}
In the near zone, choosing $\vec{v}=\frac{a_1}{r_1}\hat{r}_1+\frac{a_2}{r_2}\hat{r}_2$, with $a_i=\frac12\big(1-\sqrt{1-4\tau^2q_i^2}\big)$, yields
\begin{equation}\label{eq:hardy}
    \int |\nabla\rho|^2\geq \int\left(\frac{\tau^2q_1^2}{r_1^2}+\frac{\tau^2q_2^2}{r_2^2}-\frac{2a_1a_2\,\hat{{r}}_1\cdot\hat{{r}}_2}{r_1r_2}\right)\rho^2
    \geq \tau^2\int \left(\frac{q_1}{r_1}-\frac{q_2}{r_2}\right)^2\rho^2,
\end{equation}
since $\hat{{r}}_1\cdot\hat{{r}}_2\leq1$ and $a_1a_2\leq\tau^2q_1q_2$. Therefore
\begin{equation}\label{eq:positivity}
   \braket{\rho,(-\nabla^2+V)\rho}= \int \rho\left(-\nabla^2\rho-\tau^2\Big(\frac{q_1}{r_1}-\frac{q_2}{r_2}\Big)^2\rho\right)
    =\int \left((\nabla\rho)^2-\tau^2\Big(\frac{q_1}{r_1}-\frac{q_2}{r_2}\Big)^2\rho^2\right)\geq 0 .
\end{equation}
This holds as long as $Gq_i<2\pi$ for both arms, i.e.\ as long as each arm is separately subcritical. Therefore, there is no bound state, and no corrections to the Casimir energy in this regime. Thus, the Casimir energy is exactly
\begin{equation}\label{eq:casimir}
  \boxed{\mathcal{C}_{q_1(-q_2)(\overline{q_1-q_2})}=-\frac{Gq_1q_2}{4\pi}}
\end{equation}
at leading order in semiclassics. This agrees with the perturbative result of~\cite{Bersini:2026tpq}. 

The Casimir energy also constrains the cusp states. In~\cite{Bianchi:2026dhj}, the fusion limit of a loop with four cusps is used to show that the weighted density of cusp states of equal charges grows at large dimension with an exponent set by the Casimir energy. If this extends to loops whose arcs are only pairwise equal, as suggested there, the exactness of $\mathcal{C}_{q_1(-q_2)(\overline{q_1-q_2})}$ fixes this growth at leading order in semiclassics for our impurity, which then enters only through $q_1q_2$.

\subsection{Dimension of the Fused Line}\label{sec:fused}
The constant term of the fusion form is determined by the far zone, as at distances much larger than $\alpha_*$, the pair of lines is indistinguishable from a single line of charge $q_1-q_2=q$, so $\Delta_{q0}$ is the dimension of a line of charge $q$ ending on the trivial line. Diagrammatically, $\Delta_{q0}$ is obtained from the straight-line result by deleting every diagram with a photon ending on the charge-$1$ line. In terms of our previous results, the equations of motion are easier to solve, and the energy integrals become elementary. We can use these to compute
\begin{equation}\label{eq:fuseddim}
\Delta_{q0}=q+\Big(\frac{\kappa}{8\pi^2}-\frac{3G}{16\pi^2}\Big)q^2+\Big(\frac{(3-8\pi^2)G^2}{384\pi^4}-\frac{\kappa^2}{32\pi^4}\Big)q^3+\mathcal O(q^4).
\end{equation}

These coefficients are exact in $G$. Since the line ends with a charge $q$, each photon carries a factor of $qG$, and each quartic a factor $\kappa q$. Thus, at $\mathcal{O}(q^n)$ each term is a polynomial in $G$ and $\kappa$ with powers up to $n-1$. This is unlike the cusp, where photons between the scalars and the charge-$1$ line cost $G$ with no power of $q$ and yield non-trivial, non-polynomial coefficients. The same coefficients are obtained in~\cite{Bersini:2026tpq}, as the $\alpha_*\to0$ constant of their cusp.

\subsection{Leading Irrelevant Operator}\label{sec:irrelevant}
Here, we aim to determine the subleading angular behaviour of \eqref{eq:fusion} with the goal of determining the dimension of the leading irrelevant operator. At leading order in $q$, this is fixed by the scattering length. To identify the dimension of the leading irrelevant operator, we must however work beyond this order, as we will see.

We consider the far-distance limit. Around $q=0$, at small $\alpha_*$, $r_2\sim r_1-\cos\theta$ from the expression for the geodesic angle, and so the potential \eqref{eq:near} felt by the scalar tends to
\begin{equation}\label{eq:Vfar}
   V=-\tau^2\left(\frac{1}{r_1}-\frac{1}{r_2}\right)^2 \sim -\tau^2\frac{\cos^2\theta}{r_1^4}.
\end{equation}
This is the negligible square of a dipole potential, which at long distances yields $\nabla^2\rho=0$ and so suggests $\rho$ tends to a harmonic function. Up to normalisation, we can thus use a multipole expansion to get
\begin{equation}\label{eq:rhofar}
    \rho\sim 1-\frac{a(G)}{r_1}+\mathcal{O}(r^{-2}_1),
\end{equation}
where $a(G)$ is the scattering length monopole term, which must be finite by the positivity of the operator. This can be computed order by order to be
\begin{equation}\label{eq:scattering}
\boxed{
    a(G)=-\frac{G^2}{16\pi^2}-\frac{(16-\pi^2)\,G^4}{1024\pi^4}-\frac{\big(96-4\pi^2-63\zeta(3)+6\pi^2\log2\big)\,G^6}{24576\pi^6}+\mathcal O(G^8).}
\end{equation}
Note that the first term has a simple physical origin - it is the Yukawa screening term with Higgsed photon mass $m^2=Gq/(2\pi^2R^2)$ which enters through the massive $S^3$ Green's function.

To obtain $a(G)$, we start with the matter equation where we rescale the potential $V=-\tau^2\mathcal{V}$,
\begin{equation}\label{eq:matternear}
    \nabla^2\rho=-\tau^2 \mathcal{V}\rho,
\end{equation}
and, using the multipole expansion, one may integrate the equation to extract
\begin{equation}\label{eq:aint}
    a(G)=-\frac{\tau^2}{4\pi}\int \mathcal{V} \rho\,d^3x,
\end{equation}
which is finite by the decay of the integrand. To compute it, we can invert the Laplacian and iterate. In doing so, we can write
\begin{equation}\label{eq:rhoinv}
    \rho=1+\tau^2(-\nabla^2)^{-1}(\mathcal{V}\rho).
\end{equation}
The Born series is the iteration of this,
\begin{equation}\label{eq:born}
    \rho=1+\tau^2 L_1+\tau^4L_2+\dots=\sum_{j\geq 0}\tau^{2j}L_j,
\end{equation}
where $L_{i}=[(-\nabla^2)^{-1}\circ \mathcal{V}]^{i}$. Inserting this expression for $\rho$ into the expression \eqref{eq:aint} for $a(G)$ gives
\begin{equation}\label{eq:aseries}
    a(G)=-\sum_{j\geq 1}A_j \tau^{2j}, \quad A_j=\frac{1}{4\pi}\int \mathcal{V} L_{j-1}\,d^3x.
\end{equation}
Each $A_j$ is positive, so $a(G)$ is negative, as expected for an attractive potential. The Born series converges for $G<2\pi$, as it is a geometric series with operator norm $4\tau^2<1$. We can compute the first few terms in closed form. We may write
\begin{equation}\label{eq:W}
    \mathcal{V}=-\nabla^2 W,\quad W=\log\frac{(r_1+r_2+1)^2}{4r_1r_2}.
\end{equation}
Then, for $j\leq 3$, one can use the symmetry of the Green's function to write
\begin{equation}\label{eq:AjW}
    A_j=\frac{1}{4\pi}\int \mathcal{V}W^{j-1}d^3x.
\end{equation}
After a spheroidal change of coordinates one can reduce these further to integrals
\begin{equation}\label{eq:Ajint}
    A_j=2\int_0^1 dy\frac{y^2}{1-y^2}\int_y^1\frac{dx}{x^3}\left(2\log(1+x)-\log(1-y^2)\right)^{j-1}.
\end{equation}
These are solvable integrals, giving
\begin{equation}\label{eq:Ajvalues}
    A_1=1,\quad A_2=4-\frac{\pi^2}{4},\quad A_3=16-\frac{2\pi^2}{3}-\frac{21}{2}\zeta(3)+\pi^2\log 2.
\end{equation}
The remaining $j>3$ must be evaluated numerically. For example, $A_4=9.97\pm 0.01$. This yields the series in \eqref{eq:scattering}.

At $\mathcal{O}(q)$, we know
\begin{equation}\label{eq:G1alpha}
    \Gamma^{(1)}_{1+q,1}(\alpha_*)=C_0(\alpha_*)-\frac{G}{4\pi^2}[1+(\pi-\alpha_*)\cot\alpha_*],
\end{equation}
which reduces to $\Gamma^{(1)}_{1+q,1}(\pi)=C_0$ on the straight line. For distances $\alpha_*\ll\alpha\ll1$, $\rho^{(1/2)}\simeq\rho_0\big(1-\frac{a(G)\alpha_*}{\alpha}\big)$, where $\rho_0$ is the normalisation, evaluating the equation of motion on this and integrating over $S^3$ outside the radius $\alpha$ gives us
\begin{equation}\label{eq:C2a}
    C_0(\alpha_*)^2-1=\frac{2a(G)\alpha_*}{\pi}.
\end{equation}
At small cusp angles, this can be expanded to give
\begin{equation}\label{eq:Calpha}
    C_0(\alpha_*)=1+\frac{a(G)}{\pi}\alpha_*+\dots.
\end{equation}
Combined with the energy of the line at small $\alpha_*$, this yields
\begin{equation}\label{eq:G1small}
    \Gamma^{(1)}_{1+q,1}(\alpha_*)=-\frac{G}{4\pi\alpha_*}+1+\Big(\frac{a(G)}{\pi}+\frac{G}{12\pi}\Big)\alpha_*+\dots,
\end{equation}
which are the leading contributions to the Casimir energy and fused line as computed above, and linear corrections. The second piece at $\mathcal{O}(\alpha_*)$, $\frac{G}{12\pi}\alpha_*$, comes from the photon exchange between the two lines. This appears in the effective action of the fused line as a constant with no operator proportional to the separation times the curvature on $S^3$~\cite{Kravchuk:2024qoh}. This enters with power of $\alpha_*$ exactly one. The full term, $\frac{Gq_1q_2}{12\pi}\alpha_*$, is present even at $q=0$, where the fused line is trivial, and thus is not a contribution of an irrelevant operator.

The $\frac{a(G)}{\pi}\alpha_*$ term comes from the scalar. The leading order in $\mathcal{O}(q)$ result would suggest that the leading irrelevant operator has dimension determined by $\alpha_*^{\Delta_{\rm irr}-1}=\alpha_*\implies \Delta_{\rm irr}=2$. However, this is a perturbative artefact. 

Indeed, at distances $\alpha\gg \alpha_*$, the pair of lines looks like a single line of charge $q$, perturbed by the leading irrelevant operator $\delta S=\lambda\int dt\,\mathcal O_{\rm irr}$ with $\lambda\propto \alpha_*^{\Delta_{\rm irr}-1}$. The defect operators of the fused line are read off from the behaviour of the bulk fields near it. In this case, the matter equation is
\begin{equation}\label{eq:farmatter}
    \rho''+\frac{2}{\alpha}\rho'+\frac{\tau^2q^2}{\alpha^2}\rho=0,
\end{equation}
which has leading behaviour $\rho\propto \alpha^{\hat{\Delta}-1}$ with
\begin{equation}\label{eq:Dhat}
    \hat{\Delta}=\frac{1}{2}+\sqrt{\frac{1}{4}-\tau^2q^2}.
\end{equation}
Since the bulk scalar has dimension $1$, $\rho\propto \alpha^{\hat\Delta-1}$ indicates the bulk-to-defect OPE onto a defect operator $\hat\phi$ of dimension $\hat\Delta$. We can consider the neutral operator on the line $\hat{\phi}^\dagger\hat{\phi}$, which has dimension $2\hat{\Delta}$. The scattering length fixes the coefficient of this term in the angular expansion of the cusp. In the intermediate zone $\alpha_*\ll \alpha\ll 1$, the general solution is $\rho\propto \alpha^{\hat\Delta-1}+c\,\alpha^{-\hat\Delta}$. As $q\to0$ it must reduce to the near-zone tail $1-\frac{a(G)\alpha_*}{\alpha}$. Since $\alpha_*$ is the only scale,
\begin{equation}\label{eq:rhomatch}
    \rho\propto \alpha^{\hat\Delta-1}\Big[1-a(G)\,\big(1+\mathcal O(q)\big)\Big(\frac{\alpha_*}{\alpha}\Big)^{2\hat\Delta-1}\Big].
\end{equation}
This contributes to the energy through $C_0(\alpha_*)$ as seen above, entering as a term $\propto\alpha_*^{2\hat{\Delta}-1}$ in the anomalous dimension,
\begin{equation}\label{eq:Girr}
    \Gamma_{1+q,1}(\alpha_*)\supset\frac{a(G)}{\pi}\,q\,\alpha_*^{2\hat\Delta-1}\big(1+\mathcal O(q)\big).
\end{equation}
Among neutral operators on the fused line, $\hat{\phi}^\dagger\hat{\phi}$ is the lowest with a non-vanishing one-point function. The displacement operator \eqref{eq:displacement} has dimension exactly two but carries transverse spin, so its one-point function vanishes in this $SO(3)$-invariant state, and it gives no contribution at $\mathcal{O}(\alpha_*)$.
This allows us to identify $2\hat{\Delta}=\Delta_{\rm irr}$, and expanding in $q$,
\begin{equation}\label{eq:Dirr}
    \Delta_{\rm irr}=2\hat{\Delta}=1+\sqrt{1-\frac{G^2q^2}{4\pi^2}}\sim 2-\frac{q^2G^2}{8\pi^2}+\dots<2.
\end{equation}
This is the dimension of the operator whose marginality marks the merger of fixed points on the Wilson line as observed in~\cite{Aharony:2022ntz,Aharony:2023amq}. In terms of exponents, we see
\begin{equation}\label{eq:Dirrlog}
    \alpha_*^{\Delta_{\rm irr}-1}=\alpha_*-\frac{G^2q^2}{8\pi^2}\alpha_*\log\alpha_* +\mathcal{O}(q^4).
\end{equation}
This logarithmic piece cannot be seen in the energy at $\mathcal{O}(q)$ nor $\mathcal{O}(G^2)$ but rather at $\mathcal{O}(q^3G^4)$, entering into the energy in the piece
\begin{equation}\label{eq:alphalog}
    \alpha_{q_1q_2}\,\alpha_*^{\Delta_{\rm irr}-1}=\frac{a(G)}{\pi}\,q\,\alpha_*\Big(1-\frac{G^2q^2}{8\pi^2}\log\alpha_*+\dots\Big),
\end{equation}
where the expansion of $a(G)$ starts at $G^2$. This means that one would naively deduce the wrong lowest irrelevant operator given the first few perturbative terms. Furthermore, since this dimension is computed semiclassically, additional logarithms could occur at $\mathcal{O}(Q^0)$. This supersedes the statement $\Delta_{\rm irr}=2+\mathcal{O}(G^3)$ in~\cite{Bersini:2026tpq}. There, the curvature term and leading coefficient of the irrelevant operator term are combined, and the logarithm is not observed yet in the perturbative regime\footnote{Note: eq.~(5.10) of \cite{Bersini:2026tpq} has an incorrect prefactor: the correct one can be obtained by expanding their eq.~(B.2) around small cusp angles, and taking the opposite sign.}.

\section{Discussion and Comparison to the Literature}\label{sec:discussion}

In this work, we have studied cusped Wilson lines with a defect-changing impurity in the abelian Higgs model, in the double scaling limit at leading order in $Q$. We took the charges $q_1=1+q$ and $q_2=1$, and expanded in the charge $q$ of the impurity. This allowed us to work exactly in $G$, throughout the subcritical regime $G<2\pi$. Our main result is the scaling dimension of the impurity on the straight line, through to order $\mathcal{O}(q^3)$,
\begin{equation}\label{eq:summary}
\boxed{\begin{aligned}
  \Delta_{1+q,1}={}&q\,\nu+q^2\left(P\left[\frac{\kappa}{4\pi}-\frac{\tau(2\nu+1)}{2(2\nu-1)}\right]-\frac{\tau^2}{2\nu-1}\right)\\
    &+q^3\left(-2\pi R^2\int_0^\pi\varrho^2\Big[\big(y'\big)^2+\big(R B^{(1)}\big)^2\Big]d\alpha\right)+\mathcal O(q^4),
\end{aligned}}
\end{equation}
where
\begin{equation}\label{eq:summarydefs}
    \tau=\frac{G}{4\pi},\quad \nu=\frac12+\frac12\sqrt{1-\frac{G^2}{4\pi^2}},\quad P=\frac{\Gamma\!\left(2\nu-\tfrac12\right)|\Gamma(\nu+i\tau)|^2}{4\sqrt\pi\,\Gamma(2\nu)\,|\Gamma(\nu+\tfrac12+i\tau)|^2},
\end{equation}
and
\begin{multline}
    \varrho=k_1\sin^\nu\!\alpha\,e^{-\tau\alpha},\qquad k_1=\frac{2^\nu e^{\pi\tau/2}|\Gamma(1+\nu+i\tau)|}{2\pi R\sqrt{\Gamma(1+2\nu)}},\nonumber\\
    RB^{(1)}=GR^2k_1^2\Big(\frac{\tau\cot\alpha+\nu}{2\nu(2\nu-1)}\sin^{2\nu}\!\alpha\,e^{-2\tau\alpha}+\Big(\cot\alpha-\frac{2\tau}{2\nu-1}\Big)\int_\alpha^\pi\sin^{2\nu}\!a\,e^{-2\tau a}\,da\Big)+C_0^{(1)},\nonumber\\
    C_0^{(1)}=P\Big[\frac{\kappa}{2\pi}-\frac{\tau(\nu+\frac12)}{\nu-\frac12}\Big],\qquad
    \varrho^2y'=\int_0^\alpha\Big[\kappa R^2\frac{\varrho^4}{\sin^2a}-2\big(\tau\cot a+\nu\big)\big(RB^{(1)}\big)\varrho^2\Big]da.\nonumber
\end{multline}
The key new results are the $\mathcal{O}(q^2)$ and $\mathcal{O}(q^3)$ parts, exact in $G$ and $\kappa$, as well as the manifest negativity of $\Gamma^{(3)}_{1+q,1}(\alpha_*)$. The expansion in $q$ is limited by the arm criticality at $q_*=\frac{1}{2\tau}-1$, and is generally also decreased by $\kappa$ away from the CFT fixed point.
Each term in the expansion \eqref{eq:summary} is the summation of infinitely many Feynman diagrams for the perturbative problem. The Catalan numbers occur in these expansions because the defect dimension solves a quadratic equation, and thus enters the energy via the state-operator correspondence. Near the critical coupling $G\sim 2\pi$, each coefficient beyond $\mathcal{O}(q)$ is singular, with leading singular behaviour
\begin{equation}\label{eq:Dncrit}
    \Delta^{(n)}_{1+q,1}\simeq-\frac{c_{n-2}}{2^{3n-4}(1+e^{-\pi})}\,\mu^{3-2n}\quad(n\ge2),\quad \mu=\tfrac12\sqrt{1-\tfrac{G^2}{4\pi^2}},
\end{equation}
where $c_m$ are the Catalan numbers. This behaviour is exactly resummable and is given by the asymptotic expression \eqref{eq:critresum}, which we repeat:
\begin{equation}\label{eq:critresum2}
   \boxed{\Delta_{1+q,1}=\frac{q}{2}+\mu^3\frac{x(\sqrt{1-x/2}+e^{-\pi})}{1+e^{-\pi}}+\mathcal{O}(\mu^4),}
\end{equation}
where $x=\frac{q}{\mu^2}\in[0,2]$, with $\mu=\nu-1/2$.
This gives us a window of control around the semiclassical approach to the loss of conformality. We also compute the energy of the excited states at leading order in $q$. For the straight line, the result is exact in $G$. At general cusp angle, we analytically find the splitting at leading order in $G$, for which the first few values are seen in Table \ref{tab:descsplit}. These results are verified numerically in Figure \ref{fig:excited}.

Away from the straight line, we also computed the Bremsstrahlung function to $\mathcal{O}(q^2)$
\begin{equation}\label{eq:betasummary}
    \beta_{1+q,1}=-\frac{G}{6\pi^2}+\big(2c(G)-\frac{G}{6\pi^2}\big)q+\beta^{(2)}_{1+q,1}q^2+\mathcal O(q^3).
\end{equation}
At $\mathcal{O}(q)$ it is given exactly in terms of the nontrivial function $c(G)$ of \eqref{eq:c(G)}, 
\begin{equation}
    \boxed{\begin{aligned}\label{eq:c(G)2}
c(G)={}&\frac{2\tau}{3\pi^2}\,\mathrm{Im}\,\psi(\nu+i\tau)-\frac{\tau^2}{6\pi^2}\bigg[\frac{3-5\nu}{\nu}\\
&+\frac{\Gamma\big(\frac32+\nu-\rt\big)\,\Gamma\big(\frac32+\nu+\rt\big)}{\Gamma(1+2\nu)}\\
&\qquad\times\sum_{n=0}^\infty\Big(f(n)-f\big(n+\rt-\nu-\tfrac12\big)\Big)\bigg],
\end{aligned}}
\end{equation} where $f(x)$ is given by \eqref{eq:fx},
and at $\mathcal{O}(q^2)$ as an integral $\beta^{(2)}_{1+q,1}$ with no known closed form \eqref{eq:beta2}, which nonetheless, when expanded as a series in $G$, can be evaluated to arbitrary orders \eqref{eq:beta2series}. In the fusion limit of the cusp, $\alpha_*\to 0$, we are also able to compute the associated data. We argue that the Casimir energy, found at leading order in $G$ in~\cite{Bersini:2026tpq}, is exact semiclassically for subcritical lines. We are also able to compute the dimension of the fused line to $\mathcal{O}(q^3)$ to all orders in $G$ and show that it is polynomial in $G$ in \eqref{eq:fuseddim}, which follows from rescaling. The leading irrelevant operator on the fused line is shown to have dimension $\Delta_{\rm irr}<2$, \eqref{eq:Dirr}, as is expected in the case of straight Wilson lines~\cite{Aharony:2022ntz,Aharony:2023amq}. We determine its coefficient in terms of the scattering length $a(G)$
\begin{equation}\label{eq:scattering2}
\boxed{
    a(G)=-\frac{G^2}{16\pi^2}-\frac{(16-\pi^2)\,G^4}{1024\pi^4}-\frac{\big(96-4\pi^2-63\zeta(3)+6\pi^2\log2\big)\,G^6}{24576\pi^6}+\mathcal O(G^8),}
\end{equation}
and predict that the first logarithm is then at $\mathcal{O}(q^3G^4)$.

We can compare our results in the regions of perturbative overlap, as well as show how they satisfy various bounds on the cusp, including in the Lorentzian limit.

\subsection{Regions of Perturbative Overlap}\label{sec:overlap}
We can compare our result to the work of~\cite{Bersini:2026tpq}, which calculates the cusp anomalous dimension as a perturbative series in $G$, where each order is exact in $q$. This is a complementary perturbative regime to that in which we work. Thus, we can compare our calculation expanded in small $G$ with their result expanded at small $q$, which is the regime of overlap. Note that this is a test of the same semiclassics, but in a different expansion. However, the calculations in~\cite{Bersini:2026tpq} were also checked against the available terms of the explicit one-loop $\epsilon$ expansion, for example in~\cite{Antipin:2022hfe,Kleinert_2003} and were in agreement. Additionally, our line dimensions match those of~\cite{Aharony:2022ntz,Aharony:2023amq}, and in the $G=0$ limit, the results of the global model of~\cite{Badel:2019oxl}. Our results are also in excellent agreement with numerics.

\begin{figure}[H]
    \centering
    \resizebox{\textwidth}{!}{\input{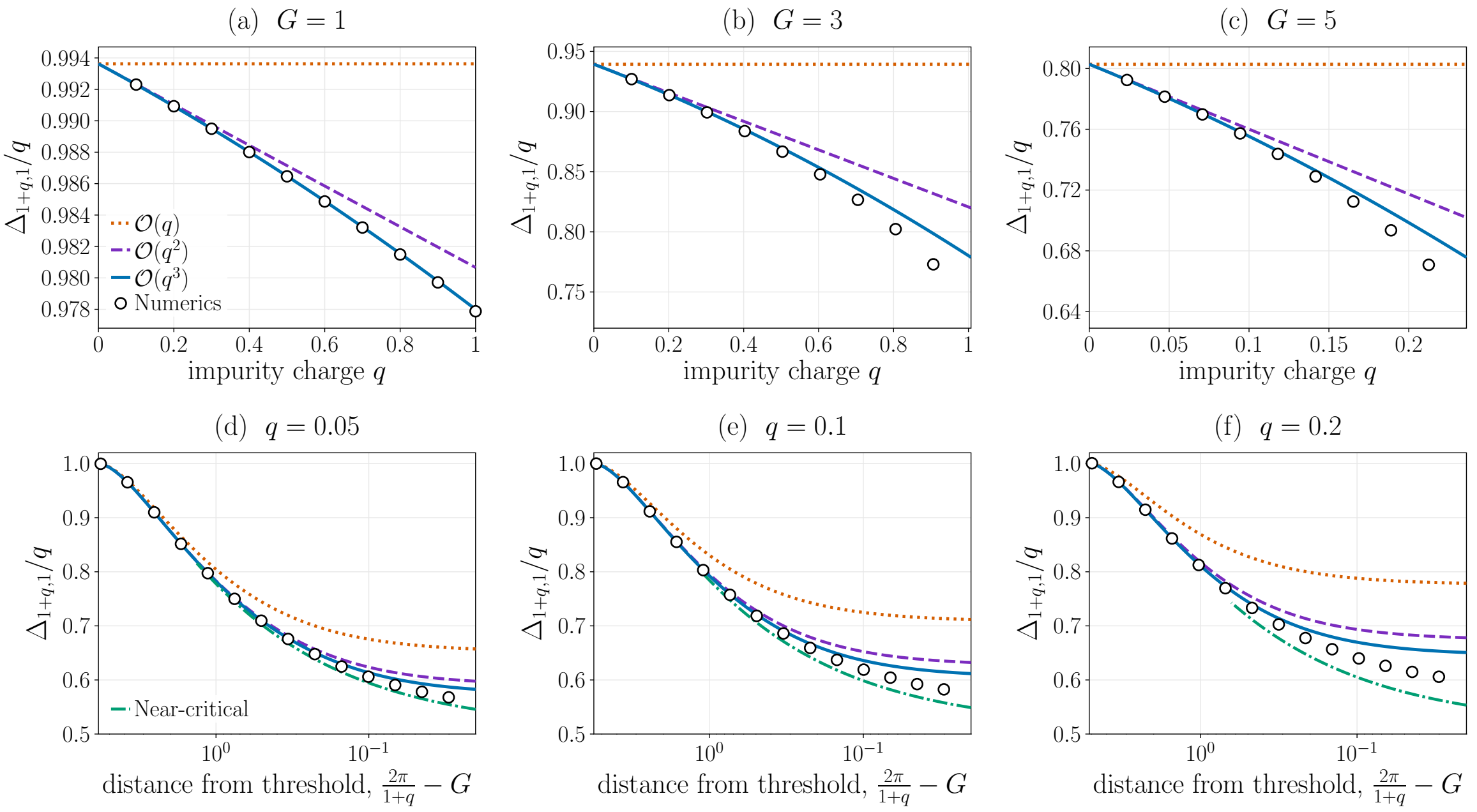}}
    \caption{The impurity dimension $\Delta_{1+q,1}=\Gamma_{1+q,1}(\pi)$ per unit impurity charge at $\kappa=1$. Open points are the numerical solution of the full nonlinear straight-line saddle at finite $q$. Lines are the truncations of the expansion in $q$ at $\mathcal{O}(q)$, $\mathcal{O}(q^2)$ and $\mathcal{O}(q^3)$. In the top row $G$ is fixed and $q$ varies, up to $q=1$ at $G=1$ and up to $q_*$ otherwise, with $q_*=\frac{1}{2\tau}-1$. In the bottom row, $q$ is fixed and $G$ increases, with a logarithmic axis measuring the distance from the critical threshold. We also show the near-critical result \eqref{eq:Gcrit}.}
    \label{fig:direct}
\end{figure}

In the perturbative regime, our present result is
\begin{align*}
\Delta_{1+q,1}={}&\begin{multlined}[t]
q\left(\rbox{1-\frac{G^2}{16\pi^2}}-\frac{G^4}{256\pi^4}-\frac{G^6}{2048\pi^6}-\dots\right)\\
+q^2\left(\bbox{\frac{\kappa}{8\pi^2}-\frac{3G}{16\pi^2}+G^2\Big(\frac{(\pi^2-6)\kappa}{384\pi^4}-\frac{1}{16\pi^2}\Big)}-\frac{(\pi^2-2)G^3}{256\pi^4}+\dots\right)\\
+q^3\left(\gbox{-\frac{\kappa^2}{32\pi^4}+G^2\Big(\frac{3-8\pi^2}{384\pi^4}+\frac{(\pi^2-6)\kappa}{384\pi^4}+\frac{(\pi^2-18)\kappa^2}{1536\pi^6}\Big)}\right.\\
\left.-G^3\Big(\frac{\pi^2-2}{256\pi^4}+\frac{(\pi^4+60\pi^2-480-210\zeta(3))\kappa}{7680\pi^6}\Big)+\dots\right).
\end{multlined}
\end{align*}
whereas the result of~\cite{Bersini:2026tpq} is
\begin{align*}
\frac{\Gamma_{(1+q)Q,Q}(\pi)}{Q}={}&\begin{multlined}[t]
\Big(\rbox{q}+\bbox{\frac{\kappa\,q^2}{8\pi^2}}\gbox{-\frac{\kappa^2q^3}{32\pi^4}}+\dots\Big)\;\bbox{-\frac{3q^2}{16\pi^2}}\;G\\
+\bigg(\rbox{-\frac{q}{16\pi^2}}\;\bbox{-\frac{q^2}{16\pi^2}+\frac{(\pi^2-6)\kappa\,q^2}{384\pi^4}}\\
+\gbox{\Big(\frac{3-8\pi^2}{384\pi^4}+\frac{(\pi^2-6)\kappa}{384\pi^4}+\frac{(\pi^2-18)\kappa^2}{1536\pi^6}\Big)q^3}\\
+\Big(\frac{(45-15\pi^2+2\pi^4)\kappa}{2880\pi^6}+\frac{(\pi^2-18)\kappa^2}{1536\pi^6}\Big)q^4\\
+\frac{(765-720\pi^2+32\pi^4)\kappa^2}{184320\pi^8}\,q^5+\dots\bigg)G^2+\mathcal O(G^3)+\mathcal O(1/Q).
\end{multlined}
\end{align*}
We highlight the $\mathcal{O}(q)$, $\mathcal{O}(q^2)$, and $\mathcal{O}(q^3)$ coefficients up to order $\mathcal{O}(G^2)$ of each result with red, blue, and green, respectively, as these are the orders of the expansion which are common to both results. In doing so, we see exact agreement in the coefficients, even though the series are organised differently. Thus, the unboxed terms of our result, the $G^3$ and higher terms at each order in $q$, are predictions for the $\mathcal{O}(G^3)$ and higher orders of~\cite{Bersini:2026tpq}. Conversely, the unboxed $q^4$ and $q^5$ terms of~\cite{Bersini:2026tpq} are predictions for the $G^2$ parts of $\Delta^{(4)}_{1+q,1}$ and $\Delta^{(5)}_{1+q,1}$. A comparison of our computation to numerics can be seen in Figure \ref{fig:direct}.

In our present work, we also computed the Bremsstrahlung function. This quantity is also extractable from the cusp anomalous dimension computed in~\cite{Bersini:2026tpq}. Our result, expanded as a series in $G$, is
\begin{align*}
\beta_{1+q,1}={}&\begin{multlined}[t]
\rbox{-\frac{G}{6\pi^2}}
+q\left(\bbox{-\frac{G}{6\pi^2}+\frac{(\pi^2+3)G^2}{72\pi^4}}-\frac{\big(45+15\pi^2+2\pi^4-360\zeta(3)\big)G^4}{34560\,\pi^6}+\dots\right)\\
+q^2\left(\gbox{\frac{(3+\pi^2)G^2}{72\pi^4}-\frac{\kappa\,G^2}{32\pi^6}}+\frac{\big(135-30\pi^2-8\pi^4+270\zeta(3)\big)G^3}{8640\,\pi^6}\right.\\
-\frac{\big(45+15\pi^2+2\pi^4-360\zeta(3)\big)G^4}{17280\,\pi^6}\\
\left.+\frac{\big(11\pi^4-360\pi^2+3330+240\pi^2\zeta(3)-3600\zeta(5)\big)\kappa\,G^4}{138240\,\pi^8}+\dots\right)+\mathcal O(q^3).
\end{multlined}
\end{align*}
The analogous result in~\cite{Bersini:2026tpq} is
\begin{align*}
\frac{\beta_{(1+q)Q,Q}}{Q}={}&\begin{multlined}[t]
\rbox{-\frac{G}{6\pi^2}}\;\bbox{-\frac{G\,q}{6\pi^2}}
+\bigg(\bbox{\frac{(3+\pi^2)\,q}{72\pi^4}}+\gbox{\Big(\frac{3+\pi^2}{72\pi^4}-\frac{\kappa}{32\pi^6}\Big)q^2}\\
+\Big(-\frac{\kappa}{32\pi^6}+\frac{(180-105\pi^2+14\pi^4)\kappa^2}{34560\pi^8}\Big)q^3+\dots\bigg)G^2+\mathcal O(G^3)+\mathcal{O}(1/Q),
\end{multlined}
\end{align*}
where we again box the terms with comparable perturbative orders. We see that these once more match exactly. Since $c(G)$ is even in $G$ by the exchange of the arms, we predict no $\mathcal{O}(q)$ terms will occur at $\mathcal{O}(G^3)$ in the expansion of~\cite{Bersini:2026tpq}.

We can also compare the dimension of the fused line. We obtain
\begin{equation*}
\Delta_{q0}=\rbox{q}+\bbox{\Big(\frac{\kappa}{8\pi^2}-\frac{3G}{16\pi^2}\Big)q^2}+\gbox{\Big(\frac{(3-8\pi^2)G^2}{384\pi^4}-\frac{\kappa^2}{32\pi^4}\Big)q^3}+\mathcal O(q^4).
\end{equation*}
We can compare this to the result in~\cite{Bersini:2026tpq}, in which they find
\begin{align*}
\frac{\Delta_{(Qq)0}}{Q}={}&\begin{multlined}[t]
\Big(\rbox{q}+\bbox{\frac{\kappa\,q^2}{8\pi^2}}\gbox{-\frac{\kappa^2q^3}{32\pi^4}}+\frac{\kappa^3q^4}{64\pi^6}-\frac{21\kappa^4q^5}{2048\pi^8}+\dots\Big)\;\bbox{-\frac{3q^2}{16\pi^2}}\;G\\
+\bigg(\gbox{\frac{(3-8\pi^2)q^3}{384\pi^4}}+\frac{(45-15\pi^2+2\pi^4)\kappa\,q^4}{2880\pi^6}\\
+\frac{(765-720\pi^2+32\pi^4)\kappa^2q^5}{184320\pi^8}+\dots\bigg)G^2+\mathcal O(G^3)+\mathcal{O}(1/Q).
\end{multlined}
\end{align*}
Here, the overlap is complete. Since our result is exact in $G$, we predict that there are no contributions below $\mathcal{O}(q^4)$ for all orders $\mathcal{O}(G^3)$ and above in~\cite{Bersini:2026tpq}. This can also be seen by rescaling $Q\to qQ$.

\subsection{Bounds on Cusps}\label{sec:bounds}
Recently, there has been much progress on general constraints on cusps~\cite{Lanzetta:2026pvr,Lanzetta:2026ogn,Chandra:2026wnf,Cuomo:2026mop,Bianchi:2026dhj}. Many of these follow from reflection positivity. Furthermore, many of the bounds hold only for cusps without impurities or impurities that fuse to the trivial line, while some are generic. As we are in an abelian theory, fusion to the trivial line implies that our charges are equal, which is the $q=0$ limit of the analysis. In this case, we obtain the pure Maxwell result for the cusp anomalous dimension which is not so interesting. We can nonetheless examine the bounds that are non-trivial in our regime. Our data are generally not close to saturating these bounds, as it is dominated by the Maxwell piece, so this functions primarily as a consistency test.

\subsubsection{Conformal Concavity Bounds}\label{sec:concavity}
In~\cite{Lanzetta:2026pvr}, the authors derive bounds for line defects from conformal concavity,
\begin{equation}\label{eq:concavity}
    \mathcal{E}_{q_1q_2}''(u)\leq 0, \quad \mathcal{E}_{q_1q_2}(u)=\sqrt{u}\Gamma_{q_1q_2}(u), \quad u=\tan^2\frac{\alpha_*}{4},
\end{equation}
where $\mathcal{E}_{q_1q_2}$ is the affine cusp anomalous dimension. This bound is a geometric one, coming from the energy of a cusp in an ``eye'' configuration which is nontrivially deformed by conformal transformations. From this, in terms of the cusp angle, one obtains a bound
\begin{equation}\label{eq:concavitybound}
    \Gamma_{q_1q_2}(\alpha_*)-\sin\alpha_*\Gamma'_{q_1q_2}(\alpha_*)-4\sin^2\frac{\alpha_*}{2}   \Gamma''_{q_1q_2}(\alpha_*)\geq 0.
\end{equation}
In the straight line limit, this gives
\begin{equation}\label{eq:Dm4bbound}
    \Delta_{q_1q_2}-4\beta_{q_1q_2}\geq 0.
\end{equation}
For our data, we find
\begin{equation}\label{eq:Dm4b}
    \Delta_{1+q,1}-4\beta_{1+q,1}=\frac{2G}{3\pi^2}+q\left(\nu+\frac{2G}{3\pi^2}-8c(G)\right)+\mathcal{O}(q^2).
\end{equation}
At $\mathcal{O}(q^0)$, this is strictly positive. The bracketed term at $\mathcal{O}(q)$ is also positive for all $G<2\pi$. In fact, $\beta_{1+q,1}<0$ in our case, which is not always the situation for generic theories.

Via concavity, one can use the fact that concave functions lie below their tangent lines to further derive~\cite{Lanzetta:2026pvr},
\begin{equation}\label{eq:tangentbound}
    \Gamma_{q_1q_2}(\alpha_*)+\sin\alpha_*\,\Gamma'_{q_1q_2}(\alpha_*)- \Delta_{q_1q_2}\sin\frac{\alpha_*}{2}\geq 0.
\end{equation}
For equal charges, $\Delta_{q_1q_1}=0$, this is stronger than ordinary concavity. For our cusp, the left-hand-side is
\begin{multline*}
  (1+q)\,\frac{G}{2\pi^2}\sin^2\frac{\alpha_*}{2}\Big(\frac{\pi-\alpha_*}{\sin\alpha_*}-1\Big)
    +q\Big(C_0(\alpha_*)+\sin\alpha_*\,C'_0(\alpha_*)-\nu\sin\frac{\alpha_*}{2}\Big)+\mathcal O(q^2).
\end{multline*}
The leading dominant term comes from the Maxwell cusp. Since $\pi-\alpha_*>\sin\alpha_*$, it is strictly positive for $0<\alpha_*<\pi$, and it vanishes at both ends, as $\frac{G}{12\pi^2}\varphi^2$ near the straight line and as $\frac{G}{8\pi}\alpha_*$ in the fusion limit.  The sign of the second term matters in these limits. Near the straight line, this reduces to the previous bound \eqref{eq:Dm4bbound}. In the fusion limit, the second term tends to $\Delta_{q0}$ and thus the bound reduces to $\Delta_{q0}>0$. We can also evaluate the second term numerically for generic cusp angles, using the $C_0(\alpha_*)$ from Figure \ref{fig:excited}, to see that this term is also positive at all angles we scanned, thus satisfying the bound on its own.

There are also bounds on fusion~\cite{Lanzetta:2026pvr},
\begin{equation}\label{eq:fusionbound}
    \Delta_{q_1q_2}\geq \frac{\mathcal{C}_{q_1(-q_2)(\overline{q_1-q_2})}}{2}+2\Gamma'_{q_1q_2}(\pi),
\end{equation}
which is stronger than ordinary unitarity $\Delta_{q_1q_2}\geq 0$ for a repulsive Casimir energy. In our case $\Gamma'_{1+q,1}(\pi)$ vanishes and this becomes $q\nu+\mathcal{O}(q^2)\geq -\frac{G(1+q)}{8\pi}$ which holds.

Ordinary concavity, together with monotonicity, also follows from reflection positivity of the displacement operators integrated along the two rays~\cite{Chandra:2026wnf}, where the reflection exchanges the rays, so it requires equal charges; \cite{Chandra:2026wnf} further observes numerically for free scalars that $-\Gamma_{q_1q_1}(\alpha_*)$ is completely monotonic in $\alpha_*$, which the Maxwell cusp also satisfies. For our impurity no such argument applies, yet our $\Gamma_{1+q,1}(\alpha_*)$ appears to be increasing and concave at $\mathcal O(q)$ nonetheless. At the two ends this can be seen analytically: in the fusion limit, the cusp is dominated by the Casimir energy $\mathcal{C}_{(1+q)(-1)\bar q}=-\frac{G(1+q)}{4\pi}<0$ and so the bound is trivial as the Casimir is attractive, and near the straight line, $\beta_{1+q,1}<0$ for all $G<2\pi$, since $2c(G)-\frac{G}{6\pi^2}<0$. In between, we find numerically $\Gamma^{(1)\prime\prime}_{1+q,1}(\alpha_*)\leq-0.0151,-0.0340,-0.0370$ is always negative for $G=1,3,5$, scanning over the cusp angles, where the maximum is obtained by the straight line. This is due to the Maxwell term, as the cloud is convex near the straight line with $C_0''(\pi)=2c(G)>0$.

\subsubsection{Right-angle Bounds}\label{sec:rightangle}
In~\cite{Lanzetta:2026ogn}, the authors derive a bound of the anomalous dimension at right angles versus the defect Casimir energy, for equal charges
\begin{equation}\label{eq:rightangle}
      \Gamma_{q_1q_1}\Big(\frac\pi2\Big)\geq\frac{3}{2\pi}\,\mathcal{C}_{q_1(-q_1)\bar0},
\end{equation}
which is saturated by two-dimensional boundary CFTs. At $q=0$ our cusp is the Maxwell cusp, $\Gamma_{1,1}(\frac\pi2)=-\frac{G}{4\pi^2}$, while $\frac{3}{2\pi}\mathcal{C}_{1(-1)\bar0}=-\frac{3G}{8\pi^2}$ which satisfies the bound as noted in~\cite{Lanzetta:2026ogn}. For generic defects, including excited states labelled by quantum numbers $[n]$, they obtain the bound for the shifted cusp
\begin{equation}\label{eq:shifted}
    \Gamma^{[n]}_{q_1q_2}(\alpha_*)\geq\frac12\Big[\Gamma_{q_1q_1}(\alpha_*-\alpha')+\Gamma_{q_2q_2}(\alpha_*+\alpha')\Big].
\end{equation}
At $q_1=q_2$, this is midpoint concavity, and so the small $q$ perturbation at $\alpha'=0$ tests the impurity, stating that our energy levels should be non-negative. Indeed, in Section \ref{sec:excited} we showed $C_{Nm}(\alpha_*)\geq \nu>\frac{1}{2}$ numerically in Figure \ref{fig:excited}. Observe also that for $q_1\neq q_2$ the lowest cusp need not have trivial quantum numbers. In our case, it however does for every angle as seen in our numerics. At $\alpha_*=\frac{\pi}{2}$, $\alpha'=0$, combining these bounds gives
\begin{equation}\label{eq:rightangleq}
    \Gamma_{1+q,1}\Big(\frac\pi2\Big)\geq\frac{3}{4\pi}\big[\mathcal{C}_{(1+q)(-1-q)\bar0}+\mathcal{C}_{1(-1)\bar0}\big].
\end{equation} 
With $\mathcal{C}_{q_1(-q_1)\bar0}=-\frac{Gq_1^2}{4\pi}$, the difference of the two sides is $\frac{G}{8\pi^2}+q\big[C_0(\frac\pi2)+\frac{G}{8\pi^2}\big]+\mathcal O(q^2)$, which is positive for all $G<2\pi$, since $C_0(\frac\pi2)>\frac12$.

\subsection{The Lorentzian Limit}\label{sec:lorentzian}
In~\cite{Cuomo:2026mop}, the cusp is analytically continued to an imaginary deflection angle $\varphi=i\eta$ (i.e.\ $\alpha_*=\pi-i\eta$), for which the cusp is Lorentzian and describes two lines boosted relative to one another with rapidity $\eta$. Physically, this describes a quench process where a particle of charge $q_1$ turns into another of $q_2$ with a kick in its velocity. The cusp anomalous dimension describes the infrared suppression of this process. The authors derive the bound, assuming Rindler positivity for defects,
\begin{equation}\label{eq:lorbound}
    \Gamma_{q_1q_2}(\pi-i\eta)\geq 0.
\end{equation}
At $\mathcal{O}(q)$, we can perform a small $\eta$ expansion of our result for the cusp anomalous dimension to obtain
\begin{equation}\label{eq:Gsmalleta}
      \Gamma_{1+q,1}(\pi-i\eta)=q\nu+\left((1+q)\frac{G}{12\pi^2}-qc(G)\right)\eta^2+\mathcal{O}(\eta^4,q^2).
\end{equation}
The coefficient of $\eta^2$ is set by $-\frac{1}{2}\beta_{1+q,1}>0$ and so the bound holds to this order of our perturbative solution. 

\begin{figure}
    \centering
    \resizebox{0.8\textwidth}{!}{\input{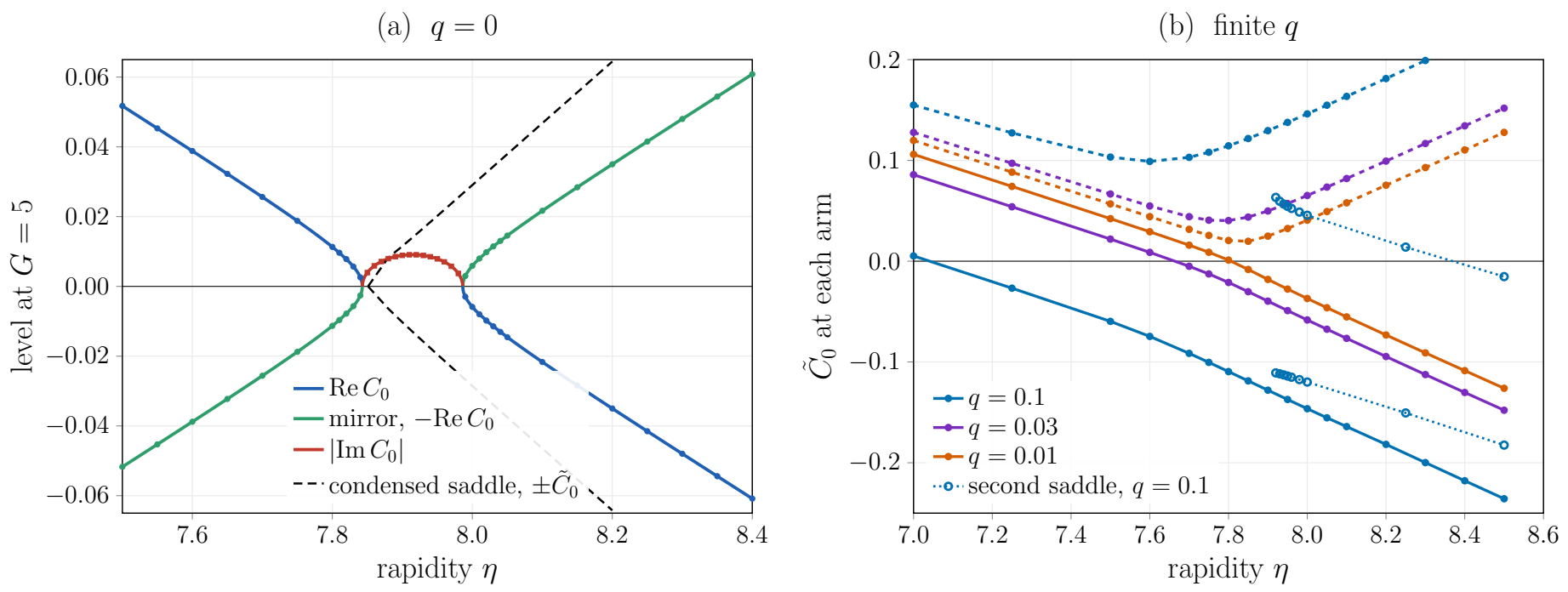}}
    \caption{The level near $\eta_c$ at $G=5$. In (a), at $q=0$, the level $C_0$ on the first arm and its mirror $-C_0$ on the second. They meet at $\eta=7.843$, and are imaginary until $\eta=7.986$, after which $C_0<0$. Dashed is the real saddle that branches off at the lower end of the window, at $\kappa=0.5$, with $q=0.001$. In (b), at $\kappa=0.5$, $\tilde{C}_0=RB-\frac{G}{4\pi^2}\big[q_1(\pi-\alpha)\cot\alpha-q_2(\pi-\gamma)\cot\gamma\big]$, which equals $C_0$ as $q\to0$ below the window, at the first arm (solid) and second arm (dashed). At finite $q$ the saddle stays real, and the transition is smooth. Dotted is a second real saddle at $q=0.1$, which continues the level on the first arm, and has higher energy. It starts at a fold near $\eta=7.91$, below which the solution is a complex pair.}
    \label{fig:c0imag}
\end{figure}

The large rapidity limit can also be considered. In~\cite{Cuomo:2026mop}, it is argued that for lines that can end on local operators, and if the bulk operators coupling the two lines have a gap in twist, then at large rapidity  the lines decouple and the cusp approaches 
\begin{equation}\label{eq:Glargeeta}
    \Gamma_{(1+q)Q,Q}(\pi-i\eta)\to Q\big(\Delta_{(1+q)0}+\Delta_{01}\big)+
    \mathcal{O}\big(\eta^{\delta_{J,0}}\,e^{-\tau_{\rm exch}\eta}\big),
\end{equation}
where $\tau_{\rm exch}=\Delta-J$ is the twist of the lowest operator exchanged between the lines. The physical interpretation is that it is energetically cheaper to end the lines on local operators, which is independent of $\eta$, rather than radiate soft photons, which grows in cost linearly in $\eta$. Indeed, in pure Maxwell theory,
\begin{equation}\label{eq:maxwelleta}
    \Gamma_{1,1}(\pi-i\eta)=\frac{G}{4\pi^2}(\eta\coth\eta-1),
\end{equation}
and one can clearly see the energy grows linearly like $\frac{G}{4\pi^2}\eta$ at large $\eta$. At leading order for $q=0$, our saddle coincides with this Maxwell cusp. This suggests a breakdown of our result at large $\eta$ as this linear growth should be suppressed by matter. This would happen due to the onset of condensation of the scalar. When the cusp has a defect changing operator, the $\mathcal{O}(q)$ part of the result of~\cite{Bersini:2026tpq} at $\mathcal{O}(G^2)$, analytically continued, gives the energy of a scalar carrying charge $q$,
\begin{equation}\label{eq:Ceta}
    C_0(\pi-i\eta)=1-\frac{G^2}{48\pi^4}\left(2\pi^2+\eta(\pi^2+\eta^2)\coth\eta\right)+\mathcal{O}(G^4).
\end{equation}
Expression \eqref{eq:Ceta} is valid only for $G\eta^2\ll4\pi^2$. Its zero $\eta_*\sim\big(\frac{48\pi^4}{G^2}\big)^{1/3}$ lies outside of this range and scales differently to the true threshold $\eta_c$ which is the onset of condensation. At large $\eta$ the second arm shifts the potential seen by the cloud of the first arm by a constant, $(\pi-\gamma)\cot\gamma\to-\eta-\log(\cos\alpha-i\sin\alpha\cos\theta)$, and so
\begin{equation}\label{eq:Clargeeta}
C_0(\pi-i\eta)=b(G)-\frac{G}{4\pi^2}\eta+\dots,
\end{equation}
where numerically $b(3)\approx1.112$ and $b(5)\approx1.002$ (see Appendix \ref{sec:appF}). For $q=0$, at a critical rapidity $\eta_c\approx4\pi^2b(G)/G$, the level $C_0$ merges with $-C_0$ of the other arm, becomes imaginary for a small window, before becoming real again with the levels exchanged. This can be seen numerically in Figure \ref{fig:c0imag}, which indicates the beginning of the formation of a condensate. The expansion in $q$ is not uniform there. Indeed, at $q>0$, there is no such window and the saddle stays real with a smooth transition. Since \eqref{eq:Ceta} is even in each as an expansion in $\eta$, whereas \eqref{eq:Clargeeta} is not, this suggests an order-of-limits issue between $G\to0$ and $\eta\to \infty$.

Let us derive the semiclassical behaviour at large rapidity, assuming that the Maxwell saddle is overtaken by a screened one. At $q=0$ this is an assumption, but for $q>0$ the numerics suggest a smooth transition between the Maxwell behaviour and screening, where the screened saddle lies below the other numerical saddle we find. Consider first a half line of charge $\q$ ending on a local operator, with solutions $\rho_{\q}$ and $RB_{\q}$. Along $\alpha=\pi-i\vartheta$, $RB_{\q}$ tends at large $\vartheta$ to a constant $RB_{\q}(\pi-i\infty)<1$, fixed by the endpoint saddle, and so $\rho_{\q}\propto e^{-(1-RB_{\q}(\pi-i\infty))\vartheta}$. This suggests the scalar splits up into parts localised around each arm, exponentially small in between, which screen the charge. The effective charge we observe after screening is thus
\begin{equation}\label{eq:qsdef}
    q_s=q_1-R^3\int_{\rm arm\,1}\rho^2 B=q_2+R^3\int_{\rm arm\,2}\rho^2 B.
\end{equation}
Here, we integrate over points that are closer to either arm, $|\cos\alpha|<|\cos\gamma|$ and, respectively, $|\cos\alpha|>|\cos\gamma|$. The screening is complete when $q_s=0$. We can take the integrands to be the solutions localised to each arm, neglecting the tails from the other arm, up to exponentially small corrections. These come from the overlap of the two tails between the arms, of $\mathcal{O}(\eta e^{-(1-RB(\pi-i\infty))\eta})$ with $RB(\pi-i\infty)$ the larger of the two constants, and from the neighbourhood of the circle $\alpha=\theta=\frac\pi2$ where both clouds are of order one, which has radius $e^{-\eta/2}$ and contributes $\mathcal O(e^{-\eta})$.

Considering an ansatz of the form
\begin{equation}\label{eq:rhoansatz}
    \rho=\rho_{q_1}(\alpha)+\rho_{q_2}(\gamma)+\mathcal O(q_s\eta^0),
\end{equation}
and
\begin{multline}\label{eq:ansatz}
  RB={\Big(1-\frac{q_s}{q_1}\Big)}RB_{q_1}(\alpha)-{\Big(1-\frac{q_s}{q_2}\Big)}RB_{q_2}(\gamma)+K\\
+\frac{Gq_s}{4\pi^2}\Big((\pi-\alpha)\cot\alpha-(\pi-\gamma)\cot\gamma\Big)+\mathcal O(q_s\eta^0),
\end{multline}
the saddle equations become
\begin{gather}\label{eq:ansatzcond}
-RB_{q_2}(\pi-i\infty)+K+\frac{G}{4\pi^2}q_{s}\eta=\mathcal{O}(q_s\eta^0),\\
RB_{q_1}(\pi-i\infty)+K-\frac{G}{4\pi^2}q_{s}\eta=\mathcal{O}(q_s\eta^0),\label{eq:ansatzcond2}
\end{gather}
and are solved provided that the constant $K$ is
\begin{equation}\label{eq:Kconst}
K=\frac12\big(RB_{q_2}(\pi-i\infty)-RB_{q_1}(\pi-i\infty)\big)+\mathcal O(q_s\eta^0),
\end{equation}
while the screened charge is
\begin{equation}\label{eq:qu}
q_{s}(\eta)=\frac{\pi}{2\tau\eta}\Big(RB_{q_1}(\pi-i\infty)+RB_{q_2}(\pi-i\infty)\Big)\Big(1+\mathcal O(\eta^{-1})\Big).
\end{equation}
Thus $q_s=\mathcal{O}(\eta^{-1})$. Only in the large boost limit $\eta\to\infty$ are the endpoints totally screened, with $q_s=0$.

\begin{figure}[H]
    \centering
    \resizebox{\textwidth}{!}{\definecolor{cmax}{HTML}{7A7A7A}\definecolor{cnum}{HTML}{1F5FB4}%
\definecolor{cend}{HTML}{C0392B}\definecolor{cstep}{HTML}{D35400}\definecolor{cpert}{HTML}{2E8B57}%
\begin{tikzpicture}[
  mpldash/.style={line width=#1pt, dash pattern=on {3.7*#1}pt off {1.6*#1}pt},
  mpldashdot/.style={line width=#1pt, dash pattern=on {6.4*#1}pt off {1.6*#1}pt on {1*#1}pt off {1.6*#1}pt},
  mpldot/.style={line width=#1pt, dash pattern=on {1*#1}pt off {1.65*#1}pt}]
\pgfplotsset{every axis/.append style={scale only axis, height=10.2cm, width=13.5cm, clip mode=individual,
  tick align=outside, tick pos=left, major tick length=3.5pt, every tick/.style={black, line width=0.8pt},
  axis line style={line width=0.8pt}, grid=major, grid style={black!8, line width=0.8pt},
  tick label style={font=\fontsize{16}{16}\selectfont}, label style={font=\fontsize{22}{16}\selectfont}, title style={font=\fontsize{24}{16}\selectfont},
  legend style={draw=none, fill=none, font=\fontsize{14}{16}\selectfont, row sep=1.5pt, cells={anchor=west}}, legend cell align=left,
  legend image code/.code={\draw[##1] (0cm,0cm) -- (0.36cm,0cm) (0.72cm,0cm) -- (0.36cm,0cm);
                           \draw[##1, solid] plot coordinates {(0.36cm,0cm)};},
  yticklabel style={/pgf/number format/fixed, /pgf/number format/fixed zerofill, /pgf/number format/precision=2}}}
% ---------------- (a) q = 0 ----------------
\begin{axis}[name=pa, xmin=0, xmax=40, ymin=0, ymax=2.0, xtick={0,5,...,40}, ytick={0,0.25,...,2.0},
  xlabel={rapidity $\eta$}, ylabel={$\Gamma_{1+q,1}(\pi-i\eta)$ per unit $Q$}, title={(a)\ \ $q=0$},
  legend style={at={(0.98,0.02)}, anchor=south east}]
\addplot[cmax, line width=1.8pt, domain=0.01:7.9, samples=100] {0.126651*(x/tanh(x)-1)};
\addlegendentry{Maxwell ($\rho=0$)}
\addplot[cmax, mpldot=1.0, domain=7.9:40, samples=150, forget plot, restrict y to domain=-1:3] {0.126651*(x/tanh(x)-1)};
\addplot[cnum, line width=2pt, mark=*, mark size=1.75pt, mark options={solid}] table {
  7.95 0.88007
  8.05 0.89228
  8.2 0.91008
  8.4 0.93287
  8.7 0.96515
  9 0.99528
  9.5 1.04118
  10 1.08228
  11 1.15239
  12 1.20955
  13 1.25673
  14 1.29614
  15 1.32942
  16.5 1.37048
  18 1.40351
  19.5 1.43055
  21 1.45303
  22.5 1.47197
  24 1.48812
  26 1.50623
  28 1.52130
  30 1.53400
  33 1.54966
  36 1.56226
  40 1.57561
};
\addlegendentry{Numerical saddle}
\addplot[cnum, only marks, mark=*, mark size=1.75pt, forget plot] table {
  0.05 0.00011
  1 0.03965
  2 0.13610
  3 0.25519
  4 0.38029
  5 0.50666
  6 0.63326
  7 0.75991
};
\addplot[cnum, mpldash=1.2, domain=0.3:40, samples=300, restrict y to domain=-12:3] {1.68471-3.566790/x};
\addlegendentry{Analytic large-$\eta$}
\addplot[cend, mpldashdot=1.2, domain=0:40] {1.68471};
\addlegendentry{Endpoint $2\Delta_{10}$, numerical}
\draw[black, mpldot=0.8] (axis cs:7.9,0) -- (axis cs:7.9,2);
\node[anchor=base west, font=\fontsize{16}{16}\selectfont, inner sep=0pt] at (axis cs:8.3,0.06) {$\eta_c$};
\end{axis}
% ---------------- (b) q = 0.1 ----------------
\begin{axis}[name=pb, at={(pa.south east)}, anchor=south west, xshift=3.4cm, xmin=0, xmax=40, ymin=0, ymax=2.0,
  xtick={0,5,...,40}, ytick={0,0.25,...,2.0},
  xlabel={rapidity $\eta$}, ylabel={$\Gamma_{1+q,1}(\pi-i\eta)$ per unit $Q$}, title={(b)\ \ $q=0.1$},
  legend style={at={(0.98,0.02)}, anchor=south east}]
\addplot[cpert, mpldot=1.8, domain=0.01:25, samples=250, restrict y to domain=-1:3]
  {(1+0.1)*0.126651*(x/tanh(x)-1)+0.1*(1-0.0053469*(2*pi^2+x*(pi^2+x^2)/tanh(x)))};
\addlegendentry{Small $G$ at $\mathcal{O}(q)$}
\addplot[cnum, line width=2pt, mark=*, mark size=1.75pt, mark options={solid}] table {
  0.05 0.07559
  0.5 0.08625
  1 0.11660
  1.5 0.16146
  2 0.21554
  3 0.33615
  4 0.46188
  5 0.58846
  6 0.71508
  6.5 0.77839
  7 0.84169
  7.3 0.87967
  7.6 0.91752
  7.8 0.94228
  8 0.96613
  8.2 0.98900
  8.5 1.02135
  9 1.07057
  9.5 1.11452
  10 1.15395
  11 1.22113
  12 1.27588
  13 1.32111
  14 1.35900
  15 1.39101
  16.5 1.43046
  18 1.46231
  20 1.49608
  22 1.52261
  24 1.54412
  26 1.56161
  28 1.57628
  30 1.58864
  33 1.60387
  36 1.61621
  40 1.62926
};
\addlegendentry{Numerical saddle}
\addplot[cnum, mpldash=1.2, domain=0.3:40, samples=300, restrict y to domain=-12:3] {1.73266-3.495293/x};
\addlegendentry{Analytic large-$\eta$}
\addplot[cend, mpldashdot=1.2, domain=0:40] {1.73266};
\addlegendentry{Endpoint $\Delta_{(1+q)0}+\Delta_{01}$, numerical}
\draw[black, mpldot=0.8] (axis cs:7.9,0) -- (axis cs:7.9,2);
\node[anchor=base west, font=\fontsize{16}{16}\selectfont, inner sep=0pt] at (axis cs:8.3,0.06) {$\eta_c$};
\draw[cpert, mpldot=0.8] (axis cs:4.915,0) -- (axis cs:4.915,2);
\node[cpert, anchor=base east, font=\fontsize{16}{16}\selectfont, inner sep=0pt] at (axis cs:4.515,0.06) {$\eta_*$};
\end{axis}
% ---------------- (c) q = 0.1, finite Q ----------------
\begin{axis}[at={(pb.south east)}, anchor=south west, xshift=3.4cm, xmin=10, xmax=131.33, ymin=1.27, ymax=1.76,
  xtick={20,40,...,120}, ytick={1.30,1.35,...,1.76},
  xlabel={rapidity $\eta$}, ylabel={$\Gamma_{1+q,1}(\pi-i\eta)$ per unit $Q$}, title={(c)\ \ $q=0.1$, large $\eta$, finite $Q$},
  legend style={at={(0.5,0.008)}, anchor=south, fill=white}]
\addplot[cstep, line width=2.4pt] table {
  10.000 1.38394
  11.674 1.43695
  15.010 1.50454
  21.013 1.57298
  35.022 1.64395
  105.067 1.73266
  131.334 1.73266
};
\addlegendentry{Analytic large-$\eta$, integer $\mathfrak{Q}=Q\,q_s$ ($Q=10$)}
\addplot[cnum, mpldash=1.4, domain=10:131.33, samples=200] {1.73266-3.495293/x};
\addlegendentry{Analytic large-$\eta$, continuous charge}
\addplot[cend, mpldashdot=1.2, domain=10:131.33] {1.73266};
\addlegendentry{Endpoint $\Delta_{(1+q)0}+\Delta_{01}$, numerical}
\draw[black, mpldot=0.8] (axis cs:105.07,1.27) -- (axis cs:105.07,1.76);
\node[anchor=base west, font=\fontsize{16}{16}\selectfont, inner sep=0pt] at (axis cs:106.57,1.47) {$\eta_s$};
\node[cstep, anchor=north, font=\fontsize{16}{16}\selectfont, inner sep=1pt] at (axis cs:118.20,1.7273) {$\mathfrak{Q}=0$};
\node[cstep, anchor=south east, font=\fontsize{16}{16}\selectfont, inner sep=1pt] at (axis cs:68.79,1.6950) {$\mathfrak{Q}=1$};
\node[cstep, anchor=south east, font=\fontsize{16}{16}\selectfont, inner sep=1pt] at (axis cs:34.12,1.6486) {$\mathfrak{Q}=2$};
\end{axis}
\end{tikzpicture}}
    \caption{Energies of the cusp per unit $Q$ at $G=5$, $\kappa=1$. In panel (a), with no impurity, $q_1=q_2=1$, the saddle is the free Maxwell cusp \eqref{eq:maxwelleta} up to the critical value $\eta_c$, computed numerically, where the scalar condenses. It then approaches the sum of endpoint dimensions as described by our large $\eta$ formula \eqref{eq:GammaEta}. Above $\eta_c$, we show the screened saddle, whose dominance we assume. In panel (b), with an impurity, $q_1=1.1$, $q_2=1$, we also show the small $G$ result at $\mathcal{O}(q)$ from \eqref{eq:Ceta}, which is zero at $\eta_*$, where the scalar energy vanishes perturbatively. In panel (c), with the same charges, we show the effect of continuous versus quantised charge at $Q=10$, for which the line charges are the integers $11$ and $10$. The energy \eqref{eq:Efinite} steps through integer $\mathfrak{Q}$, reaching $\mathfrak{Q}=0$ at $\eta_s$. This is an illustration of the difference in order of limits in taking large $Q$ (for which the charge is effectively continuous) versus large $\eta$ at fixed $Q$. Corrections of order $Q^0$ are not included, and need not be small on this scale.}
    \label{fig:largeeta}
\end{figure}
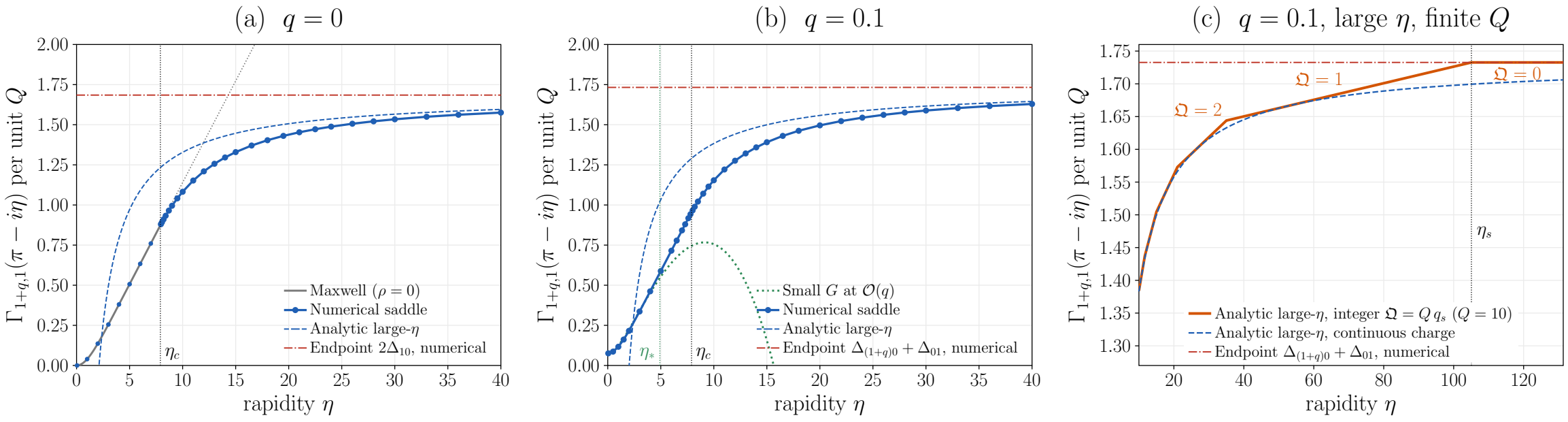

For the cusp anomalous dimension, the $\eta$-dependence enters only via the explicit dependence in the potential. In terms of the ansatz for $RB$ in \eqref{eq:ansatz}, the cusp anomalous dimension as a function of $q_s$ is
\begin{equation}\label{eq:Gqs}
    \Gamma_{q_1q_2}(q_s)=\Gamma_{q_1q_2}(0)-\Big(RB_{q_1}(\pi-i\infty)+RB_{q_2}(\pi-i\infty)\Big)q_s+\frac{G\eta}{4\pi^2}q_s^2+\mathcal O(q_s^2\eta^0),
\end{equation}
and using the expression for the screened charge \eqref{eq:qu} we get
\begin{equation}\label{eq:GammaEta}
\Gamma_{1+q,1}(\pi-i\eta)=\Delta_{(1+q)0}+\Delta_{01}-\frac{\pi}{4\tau\eta}\Big(RB_{1+q}(\pi-i\infty)+RB_{1}(\pi-i\infty)\Big)^2+\mathcal O(\eta^{-2}).
\end{equation}
Corrections to this are still sizeable at the rapidities we reach numerically, about a fifth of the $1/\eta$ term at $\eta=40$, but its coefficient can be tested without a fit. At stationary $q_s$, \eqref{eq:Gqs} gives $\del_\eta\Gamma=\frac{G}{4\pi^2}q_s^2$ and $\Gamma-\Gamma(0)=-\frac12\big(RB_{q_1}(\pi-i\infty)+RB_{q_2}(\pi-i\infty)\big)q_s+\mathcal{O}(q_s^3)$, so the numerics of Figure \ref{fig:largeeta} determine $q_s(\eta)$, independently of any constant offset in $\Gamma$. At $G=5$, $\kappa=1$, the slope of $1/q_s$ for $28\le\eta\le38$ agrees with \eqref{eq:qu} to about $2\%$, and the $\Gamma-\Gamma(0)$ relation holds to about $5\%$ for $\eta\ge21$.

Note that this is $\frac{1}{\eta}$ decay to the endpoint dimensions, rather than exponential suppression. This is a statement about working in the large $Q$ limit, where charge is effectively continuous. To see this, note that in reality, the charge seen after screening is quantised in units of $e$. This determines the range of validity of the formula \eqref{eq:GammaEta}. The total charge is actually then
\begin{equation}\label{eq:frakQ}
   Qq_s\equiv\mathfrak{Q} \in\mathbb{Z}.
\end{equation}
Thus, using \eqref{eq:Gqs}, the energy is
\begin{multline}\label{eq:Efinite}
\Gamma_{(1+q)Q,Q}(\pi-i\eta)=Q\big(\Delta_{(1+q)0}+\Delta_{01}\big)-\Big(RB_{1+q}(\pi-i\infty)+RB_{1}(\pi-i\infty)\Big)\mathfrak{Q} +\frac{G\eta}{4\pi^2Q}\mathfrak{Q} ^2\\+\mathcal O(\mathfrak Q^2\eta^0/Q)+\mathcal O(Q^0).
\end{multline}
At imaginary deflection the dilatation operator is not Hermitian, so excited levels can be complex, as in \eqref{eq:CN1} with $\sin\alpha_*\to i\sinh\eta$. We take the lowest level appearing in $Z_{Q_1Q_2}=\sum_nc_ne^{-\Gamma_n\log(L/a)}$, with $c_n\neq0$, to be among \eqref{eq:Efinite}. This level should minimise this energy. As $\eta$ grows, the screened charge must decrease in integer unit steps in order for the energy to be minimised. This reaches criticality at some $\eta_s$ yielding a fully screened ($\mathfrak{Q} =0$) configuration when
\begin{equation}\label{eq:etas}
\eta>\eta_s\equiv\frac{4\pi^2Q}{G}\Big(RB_{1+q}(\pi-i\infty)+RB_{1}(\pi-i\infty)\Big).
\end{equation}
For rapidities large with respect to $Q/G=1/e^2$, we reach a breakdown of the semiclassical result, and the lowest of these levels becomes the one with $\mathfrak{Q}=0$, and we expect to approach the endpoint energy exponentially. Thus, \eqref{eq:GammaEta} describes the regime $\eta_c\ll\eta\ll\eta_s$, which is the saddle where $q_s\neq0$ at any finite $\eta$. At large $Q$, the partial ending of the lines has lower energy until a critical $\eta_s$. This ignores the $\mathcal{O}(Q^0)$ corrections and tunnelling between different values of $\mathfrak{Q}$ which should be of order $Qe^{-(1-RB(\pi-i\infty))\eta}$ or smaller, so the step behaviour in the final panel of Figure \ref{fig:largeeta} should be treated as a sketch of the breakdown.  This extrapolation is consistent with the string-breaking crossover anticipated in~\cite{Cuomo:2026mop}. The approach to the endpoint dimensions for $\eta>\eta_s$ requires a separate analysis of the $\mathfrak Q=0$ state, which we do not attempt here.

\section{Outlook}\label{sec:outlook}
In this work, we have studied the cusp of the abelian Higgs model in $d=4-\epsilon$ dimensions in the semiclassical large charge limit. We have allowed for defect-changing operators, and computed the conformal data of the cusp perturbatively in the defect charge, exactly in the gauge and quartic 't Hooft couplings, at leading order in semiclassics. This is a complementary regime to~\cite{Bersini:2026tpq}. Our results here represent a resummation of perturbation theory in $\kappa$ and $G$, at each order in $q$ up to $\mathcal{O}(q^3)$, yielding systematic control of the expansion. We have also resummed the leading order singular behaviour near the critical gauge coupling to all orders in $q$, demonstrating a semiclassical window of control around the loss of conformality. The Casimir energy has been shown to receive no modifications from its classical value at leading order in semiclassics. We finally illustrate the interplay between the semiclassical and large rapidity limits.

Several questions remain open. Beyond $\mathcal O(q^3)$, the fields needed to compute $\Delta_{1+q,1}$ to higher orders are analytically inaccessible in closed form. It would be interesting to compare further numerics to this work and to~\cite{Bersini:2026tpq}. The excited cusp states were computed here at $\mathcal O(q)$, where the condensate does not yet affect the photon. At $\mathcal O(q^2)$ the Higgs mechanism lifts the photon modes and shifts the scalar levels, and this is where the fate of the descendants discussed in~\cite{Bianchi:2026dhj} would be decided. At large rapidity and finite $Q$ we see behaviour suggestive of string-breaking transitions, ending in full screening at $\eta_s\sim Q/G$. It would be interesting to study this crossover beyond the semiclassical limit. It would also be important to establish the dominance of the screened saddle, by a fluctuation analysis around it, and to test the coefficient of its $1/\eta$ approach. The data obtained here, namely the spectrum of cusp states, the exact Casimir energy and the fusion data, provide an interacting example with which to test the cusp bootstrap of~\cite{Bianchi:2026dhj,Lanzetta:2026ogn}. It is also known that the Wilson line defect admits an alternate choice of scalar boundary conditions~\cite{Aharony:2022ntz,Aharony:2023amq}, corresponding to a different conformal fixed point, rather than the one connected perturbatively to the global model we explore here. It would be interesting to further study this other choice.

It would be further interesting to consider the cusp in non-abelian theories in a similar semiclassical regime. The physics of non-abelian gauge symmetries is quite different and much richer, and could lead to interesting comparisons to cusp calculations in supersymmetric theories, which are often determined by integrability. Our work could also be applied to other higher dimensional defects, where they can be systematically studied as semiclassical objects in the limit of large quantum number.

\section*{Acknowledgements}
The author would like to thank Susanne Reffert, Domenico Orlando, and Jahmall Bersini for comments and discussions on this work. The author is supported by the Swiss National Science Foundation under grant number 200021\_219267. The author would like to thank the Ramsay Centre for Western Civilisation for its support.

\appendix
\section{Derivation of $\Gamma^{(2)}$}\label{sec:appA}
At $\mathcal{O}(q^{1/2})$ and $\mathcal{O}(q^{3/2})$, the matter equations are
\begin{gather}
\Big[-\Laplacian_{S^3}+1-\big(RB^{(0)}\big)^2\Big]\rho^{(1/2)}=0,\label{eq:rho12eq}\\
\Big[-\Laplacian_{S^3}+1-\big(RB^{(0)}\big)^2\Big]\rho^{(3/2)}=2\,RB^{(0)}\,RB^{(1)}\rho^{(1/2)}-\kappa R^2\big(\rho^{(1/2)}\big)^3.\label{eq:rho32eq}
\end{gather}
Multiplying \eqref{eq:rho32eq} by $\rho^{(1/2)}$, integrating over the sphere, using \eqref{eq:rho12eq}, and dropping the vanishing boundary term, gives the solvability condition
\begin{equation}\label{eq:solvgen}
\int_0^\pi\sin(\theta)\dd{\theta}\int_0^\pi\sin^2(\alpha)\dd{\alpha}\,\big(\rho^{(1/2)}\big)^2\Big[\,RB^{(0)}\,RB^{(1)}-\frac{\kappa}{2} R^2\big(\rho^{(1/2)}\big)^2\Big]=0 .
\end{equation}
For the straight line, the $\theta$ integral is trivial. The first term is linear in $B^{(1)}$ and thus $C_0^{(1)}$, so it can be isolated.
By recalling the charge condition
\begin{equation}\label{eq:charge1App}
    \int_0^\pi\sin^2\alpha\,B^{(0)}(\rho^{(1/2)})^2d\alpha=\frac{1}{4\pi R^3}
\end{equation}
we can express
\begin{equation}\label{eq:C01int}
        C_0^{(1)}=4\pi R^4\left[\frac{\kappa}{2}\int_0^\pi\sin^2\alpha\,(\rho^{(1/2)})^4d\alpha-\int_0^\pi\sin^2\alpha\,B^{(0)}\big(B^{(1)}-C_0^{(1)}/R\big)(\rho^{(1/2)})^2d\alpha\right].
\end{equation}
For the first term in the right-hand-side of \eqref{eq:C01int}, integrating by parts and using $\tau^2+\nu^2=\nu$ gives
\begin{equation}\label{eq:rho4identity}
    \int_0^\pi\sin^2\alpha\,(\rho^{(1/2)})^4d\alpha=k_1^4\int_0^\pi\sin^{4\nu-2}(\alpha)e^{-4\tau\alpha}d\alpha=\frac{4k_1^4}{4\nu-1}\int_0^\pi\sin^{4\nu}(\alpha)e^{-4\tau\alpha}d\alpha .
\end{equation}
For the other piece, defining the angular fraction of the charge (seen also in \eqref{eq:Fdef})
\begin{equation}\label{eq:FdefApp}
    F(\alpha)=4\pi R^3\int_0^\alpha \sin^2a B^{(0)}(\rho^{(1/2)})^2da
\end{equation}
and since
\begin{equation}\label{eq:Fprime}
    F'(\alpha)=4\pi R^3 \sin^2\alpha B^{(0)}(\rho^{(1/2)})^2
\end{equation}
remaining term with $C_0^{(1)}$ on the right-hand-side of \eqref{eq:C01int} is
\begin{equation}\label{eq:C01F}
\int_0^\pi\sin^2\alpha\, B^{(0)}\big( B^{(1)}-C_0^{(1)}/R\big)(\rho^{(1/2)})^2d\alpha=-\frac{1}{4\pi R^3}\int_0^\pi B^{(1)}{}'(\alpha)\,F(\alpha)\,d\alpha
\end{equation}
after an integration by parts. This can be expressed entirely in terms of $F(\alpha)$ by Gauss' law, which, when using the definition \eqref{eq:FdefApp}, can be integrated directly using identities
\begin{gather}\label{eq:identities}
    \int_0^\pi\frac{d\alpha}{\sin^2\alpha}\Big(\int_0^\alpha\sin^{2\nu}e^{-2\tau a}da\Big)\Big(\int_\alpha^\pi\sin^{2\nu}a e^{-2\tau a}da\Big)=\frac{1}{\nu}\int_0^\pi\sin^{4\nu}(\alpha)e^{-4\tau\alpha}d\alpha,\\
    \int_0^\pi\frac{\sin^{2\nu}(\alpha)e^{-2\tau\alpha}}{\sin^2\alpha}\Big(\int_\alpha^\pi-\int_0^\alpha\Big)\sin^{2\nu}a e^{-2\tau a}da\;d\alpha=\frac{4\tau}{\nu(2\nu-1)}\int_0^\pi\sin^{4\nu}(\alpha)e^{-4\tau\alpha}d\alpha
\end{gather}
and \eqref{eq:rho4identity} to get
\begin{multline}\label{eq:C01F2}
\int_0^\pi\sin^2\alpha\, B^{(0)}\big( B^{(1)}-C_0^{(1)}/R\big)(\rho^{(1/2)})^2d\alpha=\frac{\tau}{4\pi R^4}\int_0^\pi\frac{F(1-F)}{\sin^2\alpha}d\alpha\\=\frac{\tau}{4\pi R^4}(4\pi R^2k_1^2)^2\,\frac{2\nu+1}{(2\nu-1)(4\nu-1)}\int_0^\pi\sin^{4\nu}(\alpha)e^{-4\tau\alpha}d\alpha.
\end{multline}
Together, this allows $C_0^{(1)}$ to be isolated and removes $k_1$ by the charge condition $F(\pi)=1$
\begin{equation}\label{eq:C01isolated}
C_0^{(1)}=\frac{1}{2\pi(4\nu-1)}\left[\kappa-2\pi\tau\,\frac{2\nu+1}{2\nu-1}\right]\frac{\int_0^\pi\sin^{4\nu}(\alpha)\,e^{-4\tau\alpha}\,d\alpha}{\Big(\int_0^\pi\sin^{2\nu}(\alpha)\,e^{-2\tau\alpha}\,d\alpha\Big)^2}.
\end{equation}
The remaining integrals are evaluated using our master identity \eqref{eq:masterint}, which yields the expression \eqref{eq:C01} in the text. To obtain the near arm behaviour \eqref{eq:B1arm}, Gauss' law implies as $\alpha\to 0$
\begin{equation}
    RB^{(1)}=\frac\tau\alpha+C_0^{(1)}-\tau\int_0^\pi\cot\alpha\,F'\,d\alpha+\dots
\end{equation}
This final piece evaluates to $\tau\int_0^\pi\cot\alpha\,F'\,d\alpha=\frac{2\tau^2}{2\nu-1}$.

\section{Derivation of $\Gamma^{(3)}$}\label{sec:appB}
By \eqref{eq:GnBreg}, $\Gamma^{(3)}_{1+q,1}(\pi)=\frac13RB^{(2)}|_{\alpha=0}$, where $B^{(2)}$ is finite at the arm, so we need its value there. Consider first the $\rho^{(3/2)}$ equation. With $\rho^{(3/2)}=y\,\rho^{(1/2)}$ and \eqref{eq:rho12eq}, in terms of $\varrho$ equation \eqref{eq:rho32eq} becomes
\begin{equation}\label{eq:ypsi}
    \big(\varrho^2y'\big)'=\kappa R^2\frac{\varrho^4}{\sin^2\alpha}-2\big(RB^{(0)}\big)\big(RB^{(1)}\big)\varrho^2 .
\end{equation}
Integrating with no constant at $\alpha=0$ gives \eqref{eq:yprime}, which also vanishes at $\alpha=\pi$ by \eqref{eq:solvgen}. At the next order, the Maxwell equation is
\begin{equation}\label{eq:gaussB2}
    \big(\sin^2\alpha\,RB^{(2)}{}'\big)'=GR^2\varrho^2\big(RB^{(1)}+2y\,RB^{(0)}\big),
\end{equation}
whose right-hand side integrates to zero by the charge condition at $\mathcal{O}(q^2)$. The matter equation at $\mathcal{O}(q^{5/2})$, multiplied by $\rho^{(1/2)}$ and integrated as for \eqref{eq:solvgen}, gives the solvability condition
\begin{multline}\label{eq:solv52}
    2\int_0^\pi\big(RB^{(0)}\big)\big(RB^{(2)}\big)\varrho^2d\alpha\\=\int_0^\pi\Big[-\big(RB^{(1)}\big)^2\varrho^2-2\big(RB^{(0)}\big)\big(RB^{(1)}\big)\varrho^2y+3\kappa R^2\frac{\varrho^4}{\sin^2\alpha}y\Big]d\alpha .
\end{multline}
Using the charge fraction \eqref{eq:Fprime}, the behaviour near the arm is
\begin{equation}\label{eq:RB20F}
    RB^{(2)}\big|_{\alpha=0}=4\pi R^2\int_0^\pi\big(RB^{(0)}\big)\big(RB^{(2)}\big)\varrho^2d\alpha-\int_0^\pi RB^{(2)}{}'\,(1-F)\,d\alpha.
\end{equation}
Integrating the final term by parts gives
\begin{equation}\label{eq:RB20gauss}
    -\int_0^\pi RB^{(2)}{}'\,(1-F)\,d\alpha=-4\pi R^2\int_0^\pi\Big[\big(RB^{(1)}\big)^2\varrho^2+2\big(RB^{(0)}\big)\big(RB^{(1)}\big)\varrho^2y\Big]d\alpha .
\end{equation}
Substituting this \eqref{eq:RB20gauss} and the solvability condition \eqref{eq:solv52} into the near arm behaviour gives
\begin{equation}\label{eq:RB2y}
    RB^{(2)}\big|_{\alpha=0}=6\pi R^2\int_0^\pi\Big[-\big(RB^{(1)}\big)^2\varrho^2+\kappa R^2\frac{\varrho^4}{\sin^2\alpha}y-2\big(RB^{(0)}\big)\big(RB^{(1)}\big)\varrho^2y\Big]d\alpha .
\end{equation}
We must eliminate $y$ from this. The terms involving $y$ are proportional to the matter equation \eqref{eq:ypsi}, so using this and integrating by parts gives \eqref{eq:gamm3}. 

The same form in fact holds for any cusp angle. The same logic allows us to write
\begin{equation}\label{eq:G3general}
\Gamma^{(3)}_{1+q,1}(\alpha_*)=-\pi R^2\int_0^\pi\sin(\theta)\dd{\theta}\int_0^\pi\sin^2(\alpha)\dd{\alpha}\,\big(\rho^{(1/2)}\big)^2\Big[(\nabla y)^2+\big(RB^{(1)}\big)^2\Big],
\end{equation}
which shows that $\Gamma^{(3)}_{1+q,1}(\alpha_*)\le0$ at every angle.

\section{The Bremsstrahlung Function at $\mathcal{O}(q)$}\label{sec:appD}
Here we derive $c(G)$ of \eqref{eq:Cphi}, which yields \eqref{eq:c(G)}. To proceed, we expand the background gauge field solution in $\varphi$,
\begin{equation}\label{eq:B0bent}
\begin{aligned}
RB^{(0)}={}&\tau\cot\alpha+\nu+\varphi\,\frac\tau\pi\bo\cos\theta\\
&+\varphi^2\bigg[c(G)+\frac\tau\pi\Big(\frac12\cot\alpha\bo\sin^2\!\theta+\frac{\alpha\cot\alpha-1}{\sin^2\!\alpha}\cos^2\!\theta\Big)\bigg]+\mathcal{O}(\varphi^3).
\end{aligned}
\end{equation}
We define $f(\alpha,\theta)$ as the ratio of the full matter solution to the straight line matter solution,
\begin{equation}\label{eq:fdef}
    f(\alpha,\theta)=\frac{\rho(\alpha,\theta)}{k_1\sin^{\nu-1}\alpha\, e^{-\tau\alpha}}.
\end{equation}
We expand this in terms of the deflection angle,
\begin{equation}\label{eq:fexp}
    f=1+\varphi f_1+\varphi^2f_2+\mathcal{O}(\varphi^3).
\end{equation}
For $f_1$, via the matter equation, we then obtain
\begin{equation}\label{eq:f1eq}
    \nabla^2f_1+2\big[(\nu-1)\cot\alpha-\tau\big]\partial_\alpha f_1+2\Bh\frac\tau\pi\bo\cos\theta=0 .
\end{equation}
This implies a solution of the form $f_1=U(\alpha)\cos\theta$, where $U$ satisfies
\begin{equation}\label{eq:Uode}
U''+2(\nu\cot\alpha-\tau)\,U'-\frac{2U}{\sin^2\!\alpha}=-2\Bh\frac\tau\pi\bo .
\end{equation}
The solution is regular at $\alpha=0$, and at the far arm it is the rigid displacement of the core,
\begin{equation}\label{eq:Uarm}
    U=\frac{1-\nu}{\pi-\alpha}-\tau +\mathcal{O}(\pi-\alpha).
\end{equation}

To isolate $c(G)$, let us consider $f_2$. The relevant equation is
\begin{multline}\label{eq:f2eq}
\nabla^2f_2+2\big[(\nu-1)\cot\alpha-\tau\big]\partial_\alpha f_2\\
+2\Bh\Big[c(G)+\frac\tau\pi\Big(\frac12\cot\alpha\bo\sin^2\!\theta+\frac{\alpha\cot\alpha-1}{\sin^2\!\alpha}\cos^2\!\theta\Big)\Big]\\
+\frac{\tau}{\pi}\bo\Big[\frac\tau\pi\bo+2\Bh U\Big]\cos^2\!\theta=0 .
\end{multline}
By averaging over $\theta$, with $\bar f_2$ the average of $f_2$,
\begin{multline}\label{eq:f2bar}
\big(\wt\,\bar f_2'\big)'=-\wt\Big\{2\Bh\Big(c(G)-\frac\tau{3\pi}\Big)\\
+\frac13\,\frac{\tau}{\pi}\bo\Big[\frac\tau\pi\bo+2\Bh U\Big]\Big\}.
\end{multline}
Integrating then over $\alpha$ eliminates $\bar{f}_2$. Using $\int_0^\pi\wt\,\Bh\,\dd\alpha=\int_0^\pi\wt\,\dd\alpha$, this allows us to isolate $c(G)$,
\begin{multline}\label{eq:solv}
c(G)=\frac{\tau}{3\pi}-\frac{1}{6\int_0^\pi\wt\;\dd\alpha}\\
\times\FP\!\!\int_0^\pi\wt\;\frac\tau\pi\bo\Big[\frac\tau\pi\bo+2\Bh U\Big]\dd\alpha,
\end{multline}
where FP denotes the finite part. The integral diverges as $(\pi-\alpha)^{2\nu-4}$ at $\alpha=\pi$, which is an artefact of expanding in $\varphi$ at fixed $\alpha$, since the coefficients hold only for $\pi-\alpha\gg\varphi$. We define it as a finite part, by analytically continuing in the power of $\pi-\alpha$, which is non-integer for $0<G<2\pi$. Integrating by parts against the rotation zero mode $(\nu-1)\cot\alpha-\tau$ shows that the term $\frac{\tau}{3\pi}$ in \eqref{eq:solv} is the same integral with $\bo$ replaced by $\pi/\sin^2\!\alpha$, and combining the two gives an integral that converges at both arms, so selecting the finite part removes only this artefact. Doing this gives
\begin{multline}\label{eq:cforce}
c(G)=-\frac{\tau}{6\pi\int_0^\pi\wt\,\dd\alpha}\\
\times\int_0^\pi\wt\,\Big(\cot\alpha+\frac{\pi-\alpha}{\sin^2\!\alpha}\Big)\Big[\frac\tau\pi\bo+2\Bh U\Big]\dd\alpha.
\end{multline}

To evaluate \eqref{eq:solv}, we write the source in \eqref{eq:Uode} as
\begin{equation}\label{eq:sourcesplit}
    \frac\tau\pi\bo=\Big[\frac\tau\pi\bo-\frac{1-\nu}{\pi}\Big]+\frac{1-\nu}{\pi},
\end{equation}
and split $U$ into two parts,
\begin{equation}\label{eq:Uelnew}
    U=U_1+\frac{1-\nu}{\pi}U_2,\qquad U_1=\frac{1-\nu}{\pi}\big(1-\alpha\cot\alpha\big)-\frac{\tau\alpha}{\pi}.
\end{equation}
Here $U_1$ is the elementary solution with the bracketed term of the source, and carries the behaviour \eqref{eq:Uarm}. $U_2$ is the response to the uniform dipole source,
\begin{equation}\label{eq:U2ode}
U_2''+2(\nu\cot\alpha-\tau)U_2'-\frac{2U_2}{\sin^2\!\alpha}=-2\Bh,\qquad U_2(0)=U_2(\pi)=0 .
\end{equation}
Inserting this into \eqref{eq:solv}, integrating total derivatives reduces the elementary integrals to the $\wt$-weighted averages of $\alpha$ and $\alpha^2$. The latter cancel, and differentiating \eqref{eq:masterint} gives $\langle\alpha\rangle=\frac\pi2+\frac\tau\nu-\mathrm{Im}\,\psi(\nu+i\tau)$, so that
\begin{equation}\label{eq:cU}
c(G)=\frac{1}{6\pi^2}\Big[4\tau\,\mathrm{Im}\,\psi(\nu+i\tau)-(1-\nu)(3-5\nu)-2(1-\nu)^2\,\hat{\mathcal U}\Big],
\end{equation}
where the non-trivial remaining part is
\begin{equation}\label{eq:Uhatdef}
\hat{\mathcal U}=\frac{\int_0^\pi\wt\,\Bh\,U_2\,\dd\alpha}{\int_0^\pi\wt\,\dd\alpha}=\frac{\int_0^\pi\wt\big[U_2'^2+2U_2^2/\sin^2\!\alpha\big]\dd\alpha}{2\int_0^\pi\wt\,\dd\alpha}.
\end{equation}
In the variable $z=1-e^{2i\alpha}$, \eqref{eq:U2ode} is an inhomogeneous hypergeometric equation, with solution
\begin{multline}\label{eq:U2}
U_2=\mathrm{Re}\bigg\{\frac{\nu+i\tau}{2(1-\nu)}\bigg[z\;{}_3F_2\Big(1,1,1+\nu+i\tau;\;\tfrac32+\nu-\rt,\,\tfrac32+\nu+\rt;\;z\Big)\\
-\frac{\Gamma\big(\frac12-\nu+\rt\big)\,\Gamma\big(\frac32+\nu-\rt\big)\,\Gamma\big(\frac32+\nu+\rt\big)\,\Gamma\big(\frac12+\rt+i\tau\big)}{\Gamma(1+\nu+i\tau)\;\Gamma\big(1+2\rt\big)}\\
\times z^{\frac12-\nu+\rt}\,{}_2F_1\Big(\tfrac12-\nu+\rt,\,\tfrac12+\rt+i\tau;\;1+2\rt;\;z\Big)\bigg]\bigg\}.
\end{multline}
Using contour integration around the circle $|1-z|=1$, which runs around the branch cut along $z\in[0,1]$, we deform the contour around the cut and take the difference between the two sides of the cut. Everything converges, and we obtain
\begin{equation}\label{eq:Uhatsum}
\hat{\mathcal U}=\frac{\Gamma\big(\frac32+\nu-\rt\big)\,\Gamma\big(\frac12+\nu+\rt\big)}{2\,\Gamma(2\nu)}\sum_{n=0}^\infty\frac{f(n)-f\big(n+\rt-\nu-\frac12\big)}{\rt-\nu-\frac12},
\end{equation}
with $f(x)$ as in \eqref{eq:fx}. Finally, $\big(\rt-\nu-\frac12\big)\big(\rt+\nu+\frac12\big)=2(1-\nu)$ and $\nu(1-\nu)=\tau^2$ turn \eqref{eq:cU} into \eqref{eq:c(G)}.

\section{The Bremsstrahlung Function at $\mathcal{O}(q^2)$}\label{sec:appE}
To compute the $\mathcal{O}(q^2)$ Bremsstrahlung function to arbitrary orders of $G$, the full expression is
\begin{multline}\label{eq:beta2}
\tfrac12\beta^{(2)}_{1+q,1}=2\pi\kappa R^4k_1^4\int_0^\pi\frac{(\wt)^2}{\sin^2\alpha}\,U^2\,d\alpha\\
-4\pi R^2k_1^2\int_0^\pi\wt\;RB^{(1)}\Big[c(G)-\frac{\tau}{3\pi}\\
+\frac13U\Big(\frac{2\tau}{\pi}\bo+\Bh U\Big)\Big]d\alpha\\
+4\pi R^2k_1^2\int_0^\pi d\alpha\,y'(\alpha)\int_0^\alpha da\,\sin^{2\nu}\!a\,e^{-2\tau a}\bigg[2(\tau\cot a+\nu)\Big(c(G)-\frac{\tau}{3\pi}\Big)\\
+\frac{\tau}{3\pi}\Big(\cot a-\frac{a}{\sin^2a}\Big)s(a)\bigg]\\
-\frac{8\pi\tau}{3}R^4k_1^4\int_0^\pi d\alpha\,\wt\;s(\alpha)\Big(\cot\alpha+\frac{\pi-\alpha}{\sin^2\alpha}\Big)\\
\times\int_0^\alpha da\,\sin^{2\nu}\!a\,e^{-2\tau a}\,s(a)\Big(\cot a-\frac{a}{\sin^2a}\Big).
\end{multline}
Here, $s=\frac\tau\pi\bo+2\Bh U$, where $U$ is the function of \eqref{eq:Uode}, $RB^{(1)}$ the $\mathcal{O}(q)$ straight line gauge field, and $y'$ is given by \eqref{eq:yprime}. Individual integrals diverge at $\alpha=\pi$ and are defined as finite parts, as in \eqref{eq:solv}. These integrals can be expanded in $\tau$ to give \eqref{eq:beta2series}.

\section{Numerical Methods}\label{sec:appF}
The straight-line saddle of Figure~\ref{fig:direct} is a one-dimensional boundary value problem in $s=\log\tan\frac\alpha2$, which we solve by Newton iteration with Chebyshev collocation. Integrating $RB_{\rm reg}|_{\alpha=0}$ in $q$ then gives $\Gamma_{1+q,1}(\pi)$. The endpoint dimensions $\Delta_{q0}$ follow in the same way with $q_2=0$, and the constants $RB_{\q}(\pi-i\infty)$ by continuing the half-line solutions along $\alpha=\pi-i\vartheta$.

The levels of Figure~\ref{fig:excited} solve the matter equation at $\mathcal{O}(q)$ as an eigenvalue problem in $C_0$ at real $\alpha_*$. We use finite elements on $S^3$, factoring out the singular behaviour $\alpha^{\nu-1}\gamma^{\nu-1}$ at the arms for $m=0$, and imposing that $\rho$ vanishes at the arms for $m\neq0$.

At imaginary deflection, for $b(G)$ and the numerical saddle of Figure~\ref{fig:largeeta}, we use the same finite elements on a deformation of the real $S^3$ with the arms held fixed. Let $t$ be the angle along the great circle through the arms, which sit at $t=0$ and $t=\alpha_*$. We take $t\to t-i\eta\,h(t)$, with $h$ a smoothed step from $0$ near $t=0$ to $1$ near $t=\pi$, so that there is a real hemisphere around each arm, joined by a Lorentzian tube. The nonlinear saddle is then found by Newton iteration, with continuation in $\eta$ and $q$.

As checks, the straight-line solver reproduces \eqref{eq:G2} and \eqref{eq:gamm3} at small $q$, and the finite elements reproduce $C_{nl}$ on the straight line and the entries of Table \ref{tab:descsplit} at small $G$. Comparing meshes, the saddle of Figure \ref{fig:largeeta} carries an uncertainty of up to about $3\times10^{-3}$.

The levels of Figure \ref{fig:c0imag}(a) are computed from the quadratic pencil in $C_0$, on the conformal map of $S^3$ to $\mathbb{R}\times S^2$, with a spectral discretisation. The ends of the window are stable to $10^{-5}$ under changes of the mesh, box and grid map.

\printbibliography

\end{document}